\documentclass{article}

\usepackage{algorithmic, algorithm, xcolor, wrapfig, fancyvrb}
     \usepackage[final, nonatbib]{neurips_2020}

\usepackage[utf8]{inputenc}

\usepackage{amsmath, amsfonts, hyperref, url, booktabs, nicefrac, microtype, amssymb, accents, enumerate}
\usepackage[T1]{fontenc}
\newcommand{\RR}{\mathbb{R}}
\usepackage[pdftex]{graphicx}
\newcommand{\PL}{{P\L}}
\newcommand{\cX}{\mathcal{X}}
\newcommand{\cS}{\mathcal{S}}
\newcommand{\cB}{\mathcal{B}}
\newcommand{\cD}{\mathcal{D}}
\newcommand{\cP}{\mathcal{P}}
\newcommand{\dom}{{\rm dom}\, }
\newcommand{\prox}{{\rm Prox}}
\newcommand{\diag}{{\rm Diag}}
\newcommand{\interior}{{\rm int}}
\newcommand{\rank}{{\rm rank}}
\newcommand{\qed}{\hfill $\square$}
\newcommand{\ones}{\mathbf{1}}
\newcommand{\ubar}[1]{\underaccent{\bar}{#1}}
\newcommand{\nnnl}{\nonumber \\}
\newcommand{\st}{\ \ \textnormal{s.t.}\ \ }

\newtheorem{theorem}{Theorem}
\newtheorem{lemma}{Lemma}
\newtheorem{corollary}{Corollary}

\DeclareMathOperator*{\argmax}{arg\,max}
\DeclareMathOperator*{\argmin}{arg\,min}

\usepackage{xcolor}

\newcommand\update[1]{\textcolor{black}{#1}}
\newcommand\highlight[1]{\textcolor{black}{#1}}

\title{First-Order Methods for Large-Scale Market Equilibrium Computation \\}
\author{%
	Yuan Gao \\
	Department of IEOR, Columbia University \\
	New York, NY, 10027 \\
	\texttt{gao.yuan@columbia.edu} \\
	\And
	Christian Kroer \\
	Department of IEOR, Columbia University \\
	New York, NY, 10027 \\
	\texttt{christian.kroer@columbia.edu}
}

\begin{document}

\maketitle

\begin{abstract} 
	Market equilibrium is a solution concept with many applications such as digital ad markets, fair division, and resource sharing. For many classes of utility functions, equilibria can be captured by convex programs. We develop simple first-order methods suitable for solving these convex programs for large-scale markets. We focus on three practically-relevant utility classes: linear, quasilinear, and Leontief utilities. Using structural properties of market equilibria under each utility class, we show that the corresponding convex programs can be reformulated as optimization of a structured smooth convex function over a polyhedral set, for which projected gradient achieves linear convergence. To do so, we utilize recent linear convergence results under weakened strong-convexity conditions, and further refine the relevant constants in existing convergence results. 
	Then, we show that proximal gradient (a generalization of projected gradient) with a practical linesearch scheme achieves linear convergence under the \textit{Proximal-\PL} condition, a recently developed error bound condition for convex composite problems. For quasilinear utilities, we show that Mirror Descent applied to a new convex program achieves sublinear last-iterate convergence and yields a form of Proportional Response dynamics, an elegant, \update{interpretable} algorithm for computing market equilibria originally developed for linear utilities. Numerical experiments show that Proportional Response dynamics is highly efficient for computing approximate market equilibria, while projected gradient with linesearch can be much faster when higher-accuracy solutions are needed.
\end{abstract}

% intro

\section{Introduction}

Market equilibrium is a classical model from economics, where the decision of who gets what and why is based on finding a set of market-clearing prices such that every market participant gets allocated an optimal bundle given the prices and their budgets.
In this paper, we study the \emph{Fisher market} model, where a set of $n$ buyers compete for $m$ items. In the \emph{competitive equilibrium from equal incomes} (CEEI) mechanism for fair division, a set of $m$ divisible items are to be fairly divided among $n$ agents. The mechanism works by giving every agent a budget of one unit of fake money. 
Then, a market equilibrium is computed, and the allocation from the equilibrium is used as the fair division of the items, while the prices are discarded since they are for fake money \cite{varian1974equity}. An allocation given by CEEI can be shown to satisfy various fairness desiderata, e.g., the allocation makes all agents envy-free, Pareto optimal and get at least their proportional share of utility \cite{budish2011combinatorial,nisan2007algorithmic}.
CEEI has recently been suggested for \emph{fair recommender systems} \cite{kroer2019scalable,kroer2019computing}. 
Divisible Fisher market equilibrium has also been shown equivalent to the \emph{pacing equilibrium} solution concept for budget smoothing in Internet ad auctions \cite{conitzer2018multiplicative,conitzer2019pacing}. When items are indivisible, (approximate) CEEI has been used for allocating courses to students~\cite{budish2011combinatorial,budish2016course}, and the related max Nash welfare solution for fairly dividing estates and other goods (see, e.g., \cite{caragiannis2019unreasonable} and also the online fair division service platform \texttt{spliddit.org}). 
The indivisible setting can be related to the divisible setting via lotteries, though the question of when lotteries can be resolved satisfactorily is intricate~\cite{budish2013designing,akbarpour2019approximate}.

\paragraph{Market equilibrium.} In this paper we focus on computing Fisher market equilibrium in the divisible setting. 
Let the market consists of $n$ buyers and $m$ goods. Buyer $i$ has budget $B_i > 0$ and utility function $u_i: \RR^m_+ \rightarrow \RR_+$. As mentioned above, $B_i = 1$ for all $i$ corresponds to the CEEI mechanism. Without loss of generality, assume each good $j$ has \textit{unit} supply. 
An (aggregate) \textit{allocation} is a matrix $x = [x_1^\top; \dots; x_n^\top]\in \RR^{n\times m}_+$, where $x_{ij}$ is the amount of item $j$ purchased by buyer $i$. Given prices $p\in \RR^m_+$ of the goods, the \emph{demand set} of buyer $i$ is defined as the set of utility-maximizing allocations:
\[ D_i(p) = \argmax \left\{u_i(x_i)\mid  x_i\in \RR^n_+,\, \langle p, x_i\rangle \leq B_i \right\}. \] 
A \textit{competitive equilibrium} is a pair of prices and allocations $(p, x)$, $p\in \RR_+^m$ that satisfy the following conditions \cite{eisenberg1959consensus, jain2010eisenberg}: 
\begin{itemize}
	\item \textit{Buyer optimality}: $x_i \in D_i(p)$ for all $i\in [n]$, 
	\item \textit{Market clearance}: $\sum_{i=1}^n x_{ij} \leq 1$ for all $j\in [m]$, and equality must hold if $p_j > 0$. 
\end{itemize}

We say that $u_i$ is \textit{homogeneous} (with degree $1$) if it satisfies $u_i(\alpha x_i) = \alpha u_i(x_i)$ for any $x_i \geq 0$ and $\alpha>0$ \cite[\S 6.2]{nisan2007algorithmic}. We assume that $u_i$ are concave, continuous, nonnegative, and homogeneous utilities (CCNH). This captures many widely used utilities, such as linear, Leontief, Cobb-Douglas, and general \textit{Constant Elasticity of Substitution} (CES) utilities (see, e.g., \cite[\S 6.1.5]{nisan2007algorithmic} and \cite{branzei2014fisher}).
%Below are some widely used CCNH utilities (see, e.g., \cite{branzei2014fisher}).
%\begin{itemize}
%	\item \textit{Constant Elasticity of Substitution} (CES): $u_i(x_i) = \left( \sum_{j=1}^m v_{ij} x_{ij}^\rho \right)^{1/\rho}$, where $v\geq 0$, $\rho \in (- \infty, 1]\setminus\{0\}$. In particular, $\rho = 1$ gives \textit{linear} utilities $u_i(x_i) = \langle v_i, x_i \rangle$.
%	\item Leontief: $u_i(x_i) = \min_{j\in J_i} \frac{x_i}{a_{ij}}$ for some $ a = (a_{ij})$, $a_{ij}>0$ $J_i \subseteq [m]$, $J_i \neq \emptyset$. It corresponds to CES with $\rho \rightarrow -\infty$ and models complementary items.
%	\item Cobb-Douglas: $u_i(x_i) = \prod_{j=1}^m x_{ij}^{\lambda_{ij}}$, where $\lambda_{ij} \geq 0$, $\sum_{j}\lambda_{ij} = 1$. It corresponds to CES with $\rho \rightarrow 0$ and models substitutive items.
%\end{itemize} 
For CCNH utilities, a market equilibrium can be computed using the following \emph{Eisenberg-Gale convex program} (EG): \begin{align}
\max \sum_{i=1}^n B_i \log u_i(x_i)\quad {\rm s.t.} \ \sum_{i=1}^n x_{ij} \leq 1, \ x\geq 0.
\label{eq:eisenberg-gale-primal}
\end{align}
More precisely, we state the following theorem, which is well-known in various forms in the literature: see, e.g., \cite[\S 6.2]{nisan2007algorithmic} for the case of differentiable $u_i$ and Eisenberg's original work \cite[Theorem 4]{eisenberg1961aggregation} for the convex program with a maximization objective of the Nash social welfare). 
For completeness, we present an elementary, self-contained proof (all proofs are in the Appendix). Note that the assumption essentially means that everyone can at least get positive utility.

\begin{theorem}
	Let $u_i$ be concave, continuous, nonnegative, and homogeneous (CCNH). Assume $u_i(\ones) >0$ for all $i$. Then, (i) \eqref{eq:eisenberg-gale-primal} has an optimal solution and (ii) any optimal solution $x^*$ to \eqref{eq:eisenberg-gale-primal} together with its optimal Lagrangian multipliers $p^*\in \RR_+^m$ constitute a market equilibrium, up to arbitrary assignment of zero-price items. Furthermore, $\langle p^*, x^*_i \rangle = B_i$ for all $i$.
	\label{thm:eg-gives-me-for-certain-ui}
\end{theorem}
%\todo{we should comment on the existence assumption (which typically holds)}
%There are other convex programs that also capture market equilibriums, such as Shmyrev's \cite{shmyrev2009algorithm}. In fact, it can be derived from \eqref{eq:eisenberg-gale-primal} via convex programming duality. More details can be found in Appendix \ref{app:shmyrev-cp}.
We focus on linear, quasilinear (QL) and Leontief utilities. Under linear utilities, the utility of buyer $i$ is $u_i(x_i) = \langle v_i, x_i \rangle$. As mentioned above, linear utilities are used in the CEEI mechanism in fair division and fair recommender systems. QL utilities have the form $u_i(x_i) = \sum_j (v_{ij} - p_j) x_{ij}$, which depends on the prices \cite{chen2007note, cole2017convex}.  It captures budget-smoothing problems in first- and second-price auction markets~\cite{conitzer2018multiplicative,conitzer2019pacing}. Leontief utilities have the form $u_i(x_i) = \min_{j\in J_i} \frac{x_{ij}}{a_{ij}}$ (see \S \ref{sec:leontief-utilities}). They model perfectly complementary goods and are suitable in resource sharing where an agent's utility is capped by its \emph{dominant resource}, such as allocating different types of compute resources to computing tasks \cite{ghodsi2011dominant, kash2014no}.

Another notable convex program that captures market equilibrium under linear utilities is Shmyrev's \cite{shmyrev2009algorithm} (see Appendix \ref{app:shmyrev-cp} and \eqref{eq:shmyrev} there for more details). \cite{birnbaum2011distributed} shows that the well-known Proportional Response dynamics (PR) \cite{zhang2011proportional} for computing market equilibrium under linear utilities is in fact applying Mirror Descent to Shmyrev's convex program \eqref{eq:shmyrev}. In \S \ref{sec:ql-utilities}, we will show that both Shmyrev's convex program and PR generalize elegantly to QL utilities.

In principle, \eqref{eq:eisenberg-gale-primal} can be solved using an interior-point method (IPM) that handles exponential cones \cite{skajaa2015homogeneous, dahl2019primal}. However, it does not scale to large markets since IPM requires solving a very large linear system per iteration. This linear system can often be ill-conditioned and relatively dense, even when the problem itself is sparse. This makes IPM impractical for applications such as fair recommender systems and ad markets, where the numbers of buyers and items are typically extremely large. In this paper, we investigate iterative (gradient-based) first-order methods (FOMs) for computing equilibria of Fisher markets. Because each FOM iteration has low cost, which usually scales only with the number of nonzeros when the problem is sparse, these methods are suitable for scaling up to very large markets.
\highlight{Furthermore, there have been variants of FOMs that allow parallel and distributed updates, making them even more scalable \cite{chen2012fast, shi2015proximal, wang2016parallel, lin2014accelerated}.}
This is analogous to other equilibrium-computation settings such as zero-sum Nash equilibrium, where iterative first-order methods are also the state-of-the-art~\cite{brown2018superhuman,moravvcik2017deepstack,kroer2020faster}. We will focus on the important classes of linear, quasi-linear and Leontief utilities, which are discussed in \S \ref{sec:linear-utilities}, \ref{sec:ql-utilities} and \ref{sec:leontief-utilities}, respectively. 
\highlight{We also note that under CES utilities, PR yields linearly convergent prices and utilities \cite[Theorem 4]{zhang2011proportional}. Meanwhile, market equilibrium under Cobb-Douglas utilities can be computed explicitly. See Appendix \ref{app:CD-and-CES} for more details.} 
We note that there have been highly nontrivial algorithms that find market equilibria in time polynomial in $m, n,\log \frac{1}{\epsilon}$ in theory (where $\epsilon$ is a desired upper bound on error in prices) \cite{devanur2002market,vazirani2007combinatorial,bei2019ascending}. 
However, none of these are as easily implementable as the FOMs we consider here, which also achieves $\epsilon$-equilibrium prices in $O\left(\log \frac{1}{\epsilon}\right)$ time thanks to the linear convergence guarantees. Secondly, these methods all have prohibitively-expensive per-iteration costs from a large-scale perspective.

%In particular, we will address the following ones.
%\begin{itemize}
%	\item Buyers have additional \textit{spending constraints} \cite{vazirani2010spending, birnbaum2011distributed}. This leads to additional simple bound constraints in the formulation \eqref{eq:shmyrev}.
%	\item Buyers have \textit{quasi-linear} utilities of the form $\sum_j (v_{ij} - p_j) x_{ij}$. In this case, a modified convex program and its dual capture the the corresponding equilibrium \cite{cole2017convex}.
%	\item In an auction market and the fair recommender system setting, buyers have \textit{at-most-one} (AMO) constraints $x_{ij} \leq 1$.
%\end{itemize}
%%% Local Variables:
%%% mode: latex
%%% TeX-master: "../main"
%%% End:

\paragraph{First-order methods.}
Consider optimization problems of the form
\begin{align}
f^* = \min_{x\in \cX} f(x) = h(Ax) + \langle q, x\rangle,
\label{eq:std-form-proj-grad-desc}
\end{align}
where $\cX$ is a bounded polyhedral set, $h:\RR^r \rightarrow \RR$ is $\mu$-strongly convex with a $L$-Lipschitz continuous gradient on $\cX$ (or $(\mu, L)$-s.c. for short), $A\in \RR^{d\times r}$ and $q\in \RR^d$. We say that an algorithm for \eqref{eq:std-form-proj-grad-desc} converges \textit{linearly} (in objective value) with rate $\rho\in (0,1)$ if its iterates $x^t$  satisfy $f(x^t) - f^* \leq \rho^t (f(x^0) - f^*)$ for all $t$, where $x^0\in \dom \cX$ is an initial iterate and $f^*$ is the minimum objective value. 
Unless otherwise stated, $\cX^*$ denotes the set of optimal solutions to \eqref{eq:std-form-proj-grad-desc}, which is always a bounded polyhedral set \cite[Lemma 14]{wang2014iteration}. 

Various FOMs, such as the following, are naturally suitable for \eqref{eq:std-form-proj-grad-desc}.
\begin{itemize}
	\item Projected gradient (PG): $x^{t+1} = \Pi_{\cX} (x^t - \gamma_t \nabla f(x) ) $, where $\gamma_t$ is the stepsize.
	\item Frank-Wolfe (FW): $x^{t+1} = x^t + \gamma_t (w^t - x^t)$, where $w^t \in \argmin_{w\in \cX} \langle \nabla f(x^t), w\rangle $ \cite{frank1956algorithm, bubeck2015convex, beck2017linearly, lacoste2015global}. Unless otherwise stated, $w^t$ is chosen from the set of vertices of $\cX$. 
	\item Mirror Descent: $x^{t+1} = \argmin_{x\in\cX} \langle \nabla f(x^t), x - x^t\rangle + \gamma D(x \| x^t)$ \cite{nemirovski1983information, beck2003mirror}. Here, $D$ is the \textit{Bregman divergence} of a differentiable convex function $d$ (see, e.g., \cite[\S 5.3]{ben2019lectures}).
\end{itemize}

As is well-known, for a broad class of convex optimization problems, these FOMs and their variants achieve \textit{sublinear} convergence, that is, $f(\tilde{x}^t) - f^* = O\left( t^{-k} \right)$ for some $k>0$, where $\tilde{x}^t$ is either $x^t$ or a weighted average of $x^\tau$, $\tau\leq t$ \cite{bubeck2015convex, ben2019lectures, beck2017first}. 
When $f$ is strongly convex, linear convergence can be derived. However, strong convexity is highly restrictive and relaxed sufficient conditions for linear convergence of FOMs have been considered. 
For example, the classical Polyak-\L ojasiewicz (\PL) condition ensures linear convergence of gradient descent \cite{polyak1963gradient, lojasiewicz1963topological}. Also have been extensively studied are various \textit{error bound} (EB) conditions, which, roughly speaking, say that the distance from a feasible solution $x$ (possibly required to be close to $\cX^*$) to $\cX^*$ is bounded by a constant multiple of a computable ``residual'' of $x$ \cite{luo1993error, tseng2010approximation, wang2014iteration}. Another notable example is the \textit{quadratic growth} (QG) condition, which essentially means that the objective grows at least quadratically in the distance to $\cX^*$ \cite{rockafellar1976monotone, anitescu2000degenerate, drusvyatskiy2018error}. Furthermore, it can be seen to be equivalent to an EB under further assumptions on the problem \cite{drusvyatskiy2018error}. Recently, \cite{karimi2016linear} shows that, for convex composite ``$f+g$'' problems, a so-called \textit{Proximal-\PL} condition, which generalizes the {\PL} condition, is sufficient for linear convergence of proximal gradient. It is also shwon to be equivalent to a few existing conditions under further assumptions on the problem.

%\todo{discuss related work: various iterative methods with linear or sublinear rate}

%%% Local Variables:
%%% mode: latex
%%% TeX-master: "../main"
%%% End:

\noindent\textbf{Summary of contributions.} 
We show that PG, FW and MD are suitable for computing various market equilibria via their convex optimization formulations. In terms of first-order methods, we prove that proximal gradient with a non-standard practical linesearch scheme (Algorithm \ref{alg:ls-subroutine} in the appendix) converges linearly, with a bounded number of backtracking steps per iteration, for the more general class of problems satisfying the {Proximal-\PL} condition (Theorem \ref{thm:pg-ls-conv}). In terms of market equilibria under different utility classes, we establish simple bounds on equilibrium quantities by exploiting properties of market equilibria (Lemmas \ref{lemma:eg-linear-u_min-u_max}, \ref{lemma:ql-shmyrev-upper-lower-bounds}, \ref{lemma:r-ubar-r-bar}).
Through various problem reformulations exploiting convex optimization duality and using these bounds, we show that the convex programs for ME can be reformulated into ones with highly structured objectives over simple polyhedral sets \eqref{eq:std-form-proj-grad-desc}. 
Then, we derive linear convergence of PG for these convex programs through establishing the {Proximal-\PL} condition (Theorems \ref{thm:pg-for-eg-lin-conv},\ref{thm:pg-ql-shmyrev-lin-conv},\ref{thm:pg-Leontief}). Specifically, for linear utilities, we show that PG for the EG convex program \eqref{eq:eisenberg-gale-primal} converges linearly. 
For QL utilities, based on the relation between EG and Shmrev's convex programs, we derive a new ``QL-Shmyrev'' convex program \eqref{eq:ql-shmyrev} whose optimal solutions give equilibrium prices and bids. Similarly, PG for this convex program also achieves linear convergence.
We also show that Mirror Descent for the same convex program \eqref{eq:ql-shmyrev} achieves sublinear last-iterate convergence with a small constant. MD for this convex program also leads to a form of Proportional Response dynamics (PR), a scalable and interpretable algorithm for computing market equilibrium, extending the a series of results for linear utilities \cite{birnbaum2011distributed, zhang2011proportional}. 
For Leontief utilities, we show that PG for a reformulated dual of \eqref{eq:eisenberg-gale-primal}, with variables being the prices, achieves linear convergence. 
%In all cases, FW also achieves the classical sublinear convergence. 
For all utility classes, linear convergence of running iterates (e.g., prices $p^t$) to their corresponding equilibrium quantities ($p^*$) can be easily derived using linear convergence of the objective values. 
Extensive numerical experiments demonstrate that PR (and sometimes FW) can quickly compute approximate equilibrium allocations and prices, while PG with linesearch is more efficient for computing a higher-accuracy solution.

%%% Local Variables:
%%% mode: latex
%%% TeX-master: "main"
%%% End:

\noindent\textbf{Notation.} The $j$-th unit vector and vector of $1$'s in $\RR^d$ are $\mathbf{e}^{j,(d)}$ and $\ones^{(d)}$, respectively, where the superscript $(d)$ is omitted when $d$ is clear form the context. The $d$-dimensional simplex is $\Delta_d = \{ x\in \RR^d: \ones^\top x = 1,\, x\geq 0 \}$. For any vector $x$, $x_+$ denotes the vector with entries $\max\{x_i, 0 \}$ and $\verb|nnz|(x)$ denote its number of nonzeros. All unsubscripted vector norms and matrix norms are Euclidean $2$-norms and matrix $2$-norms (largest singular value), respectively. For a closed convex set $\cX\subseteq \RR^d$, $\interior(\cX)$ denotes the interior of $\cX$, $\Pi_\cX(x)$ denotes the Euclidean projection of $x\in \RR^d$ onto $\cX$ and ${\rm Diam}(\cX) = \sup_{x, y\in \cX}\|x-y\|$. For $p, q\in \RR^d_+$, the (generalized)  Kullback–Leibler (KL) divergence of $p$ w.r.t. $q$ is $D(p\|q) = \sum_i p_i \log \frac{p_i}{q_i} - \sum_i p_i + \sum_i q_i$. Denote the maximum and minimum \textit{nonzero} singular values of $A$ as $\sigma_{\max}(A)$ and $\sigma_{\min}(A)$, respectively.
% main sections

\section{Linear convergence of first-order methods}
For PG, it has been shown that linear convergence can be achieved for non-strongly convex objectives under various relaxed conditions \cite{luo1993error, drusvyatskiy2018error, karimi2016linear}. In particular, \eqref{eq:std-form-proj-grad-desc} can be shown to satisfy several such conditions and therefore guarantees linear convergence under PG. One notion of central significance in many such results is the \textit{Hoffman constant}. For $A\in \RR^{d\times r }$ and polyhedral set $\cX\subseteq \RR^d$, denote the (relative) Hoffman constant of $A$ w.r.t. $\cX$ as $H_\cX(A)$, which is the smallest $H>0$ such that, for any $z\in {\rm range}(A)$, $\cS = \{x: Ax=z \}$, it holds that $\|x - \Pi_{\cX\cap \cS}(x)\| \leq H \|Ax-z\|$ for all $x\in \cX$ (the Hoffman inequality). Note that $H_\cX(A)$ is always well-defined, finite, and depends only on $A$ and $\cX$ \cite{hoffman1952approximate, pena2020new}. 
%Intuitively, it can be viewed as a ``condition number'' of $\cX\cap \cS$ that controls the distance from $x$ to $\cX \cap \cS$ by the residual of the linear constraints $\|Ax-z\|$. 
In the optimization context, the intuition is as follows: the set of optimal solutions $\cX^*$ can often be expressed as $\cX^* = \cX \cap \cS$, which means the distance to optimality can be bounded by a constant multiple of the residual $\|Ax-z\|$. Therefore, the Hoffman inequality can be viewed as an EB condition. Further details on this definition, as well as its connection to the classical explicit expression, can be found in Appendix \ref{app:char-hoffman-const}.
%\todo{the above is too dense. Also, we probably need to try to give some intuition on Hoffman constants and say that they may be hard to evaluate}
Using the Hoffman inequality, we have the following linear convergence guarantee. It is a special case of \cite[Theorem 5]{karimi2016linear}. A proof is in Appendix \ref{app:proof-h(Ax)+poly_ind-conv}, which is essentially the same as \cite[Appendix F]{karimi2016linear}. However, our proof is slightly refined to allow a weakened condition, provide more details on the relevant constants and conclude iterate convergence.
\begin{theorem}
	For \eqref{eq:std-form-proj-grad-desc} with $q = 0$, PG with a constant stepsize $\gamma_t = \frac{1}{L \|A\|^2}$ generates $x^t$ such that $ \frac{\mu}{2 H_\cX(A)} \|x^t - \Pi_{\cX^*}(x^t)\|^2 \leq f(x^t) - f^* \leq \left( 1 - \frac{\mu}{\max\{ \mu, LH_\cX(A)^2 \|A\|^2  \}} \right)^t \left(f(x^0) - f^* \right) $ for all $t$. 
	 \label{thm:h(Ax)+poly_ind}
\end{theorem}
%In Theorem \ref{thm:h(Ax)+poly_ind}, intuitively, we can think of $\frac{\mu}{H^2}$ as the effective strong convexity modulus of $f$, while $L \|A\|^2$ is a Lipschitz constant. 
For the case of $q\neq 0$, we have the following instead, which relies heavily on the boundedness of $\cX$. The proof is in Appendix \ref{app:proof-h(Ax)+<q,x>+poly_ind-conv}. The key is to make use of the polyhedral structure of the set of optimal solutions and establish a QG condition similar to the one in \cite[Lemma 2.3]{beck2017linearly}. Here, we slightly refine the constants using the monotonicity of $f(x^t)$. 
We note that linear convergence with different rates can also be established through invoking different error bound conditions that also hold for \eqref{eq:std-form-proj-grad-desc} (see, e.g., \cite{wang2014iteration, drusvyatskiy2018error}). 
%\todo{is it really equivalent if the rate is only similar and not the same?}
\begin{theorem}
	 There exists unique $z^*\in \RR^r$ and $t^*\in \RR$ such that $Ax^* = z^*$ and $\langle q, x^*\rangle = t^*$ for any $x^* \in \cX^*$. Furthermore, PG for \eqref{eq:std-form-proj-grad-desc} with constant stepsize $\gamma_t = \frac{1}{L \|A\|^2}$, starting from $x^0\in \cX$, generates $x^t$ such that $ \frac{1}{\kappa} \|x^t - \Pi_{\cX^*}(x^t)\| \leq f(x^t) - f^* \leq C \rho^t $, where\newline $\rho = 1 - \frac{1}{\max\{ 1, 2\kappa L \|A\|^2\}}$, $\kappa = H_\cX(A)^2 \left( C + 2 G D_A + \frac{2(G^2 + 1)}{\mu} \right)$, 
%	 $H = \max\left\{ H_\cX\left( \begin{bmatrix} A \\ q^\top \end{bmatrix} \right), \frac{1}{2 \|A\|} \right\}$,
	 $C = f(x^0) - f^*$, $G = \|\nabla h(z^*)\|$, $D_A = \sup_{x, x'\in \cX} \|A(x - x')\|$.
	 \label{thm:h(Ax)+<q,x>-lin-conv}
\end{theorem}

For FW, the classical sublinear convergence holds. Specifically, for \eqref{eq:std-form-proj-grad-desc}, using FW with static stepsizes or exact linesearch, that is, $\gamma_t = \frac{2}{2+t}$ or  $\gamma_t \in \argmin_{\gamma\in [0,1]} f(x^t + \gamma (w^t - x^t))$, it holds that $f(x^t) - f^* \leq \frac{{\rm Diam}(\cX)^2 L\|A\|^2}{k+2}$ for all $t$ (see, e.g., \cite[Theorem 1]{jaggi2013revisiting} and \cite[Theorem 2.3]{clarkson2010coresets}). The advantage of FW is that it can maintain highly sparse iterates when $\cX$ has sparse vertices.
In fact, if we start with $x^0$ being a vertex, then $x^t$ is a convex combination of $(t+1)$ vertices. For ME convex programs such as \eqref{eq:eisenberg-gale-primal}, the constraint set $\cX$ is a product of simplexes $\Delta_n$. Therefore, any vertex of $\cX$ has the form $x\in \{0,1\}^{n\times m}$, where each $(x_{1j}, \dots, x_{nj})$, $i\in [m]$ is a unit vector. Hence, $x^t$ is a linear combination of $(t+1)$ vertices and has at most $(t+1)m$ nonzeros. 
There have been recent theories on linear convergence of an away-step variant of FW for \eqref{eq:std-form-proj-grad-desc} \cite{beck2017linearly, lacoste2015global}, in which a succinct ``vertex representation'' \cite[\S 3.2]{beck2017first} must be maintained through an additional, potentially costly reduction procedure. Initial trials suggest that it does not work well for our specific convex programs and are therefore not considered further.

\paragraph{PG with Linesearch.} Linesearch is often essential for numerical efficiency of gradient-based methods, especially when the Lipschitz constant cannot be computed. %\todo
Here, we consider a modified version of backtracking linesearch for PG (see, e.g., \cite[\S 10.4.2]{beck2017first} and \cite[\S 1.4.3]{beck2009gradient}). Specifically, given increment factor $\alpha\geq 1$, decrement factor $\beta\in (0,1)$, and stepsize upper bound $\Gamma >0$, at each iteration, set the (candidate) stepsize $\gamma_t$ to either the previous one $\gamma_{t-1}$ or $\min\{\Gamma, \alpha \gamma_{t-1}\}$, depending on whether backtracking occurs at all in the previous iteration. Then, perform the usual backtracking with factor $\beta$ until a sufficient decrease condition is met. denote the linesearch subroutine as $(x^{t+1}, \gamma_t, k_t) \leftarrow \mathcal{LS}_{\alpha, \beta, \Gamma}(x^t, \gamma_{t-1}, k_{t-1})$, where $k_{t-1}$ is the number of backtracking steps in the previous iteration, $\gamma_{-1} = \Gamma$, $k_{-1}=0$. 
%\todo{I think we have to explain the LS here}
More details, as well as the proof of the following convergence result (in fact, a more general one that holds for the proximal gradient method), can be found in Appendix \ref{app:pg-ls-details}. 
Let $L^\cX_f = \sup_{x,y\in \cX,\, x\neq y}\frac{\|\nabla f(x) - \nabla f(y)\|}{x-y} \leq L \|A\|^2$ be the (effective) Lipschitz constant of $\nabla f$ on $\cX$. 
Note that, in addition to ensure linear convergence, linesearch can potentially improve the constant in the rate, since $L_f^\cX$ can be much smaller than $L\|A\|^2$.
\begin{theorem}
	PG with linesearch $\mathcal{LS}_{\alpha,\beta, \Gamma}$ for \eqref{eq:std-form-proj-grad-desc} with $q=0$ and $q\neq 0$ converges linearly with rate $1 - \frac{\mu }{\max\{\mu, H^2/\Gamma, H^2 L_f^\cX / \beta \}}$ (where $H = H_\cX(A)$) and $1 - \frac{1}{\max \{1, 2\kappa/\Gamma, 2\kappa L_f^\cX/\beta \}}$, respectively.  For both, the number of projections $\Pi_\cX$ per iteration is at most $1 + \frac{\log \frac{\Gamma}{\tilde{\gamma}}}{\log \frac{1}{\beta}}$, where $\tilde{\gamma} :=  \min\left\{ \Gamma, \beta/L^\cX_f\right\}$.
	\label{thm:pg-ls-conv}
\end{theorem}

%\todo{Is this the first linesearch-based proximal gradient result using PL setup? We should maybe make a slightly bigger deal out of it if so}
%%% Local Variables:
%%% mode: latex
%%% TeX-master: "../main"
%%% End:

\section{Linear utilities} \label{sec:linear-utilities}
Let $u_i(x_i) = \langle v_i, x_i\rangle$ and $v = [v_1, \dots, v_n]\in \RR^{n\times m}$ be the matrix of all buyers' valuations. Without loss of generality, from now on, we assume the following \textit{nondegeneracy} condition, that is, \textit{$v$ does not contain any zero row or column}. 
Then, the inequality constraints in \eqref{eq:eisenberg-gale-primal} can be replaced by equalities without affecting any optimal solution.
%\begin{align}
%	\cX:= \left\{ x\in \RR^{n \times m}: \sum_{i=1}^n x_{ij} = 1,\, j\in [m],\, x\geq 0 \right\}, \ \ A x = \left(\langle v_1, x_i \rangle, \dots, \langle v_n, x_n \rangle\right). \label{eq:eg-linear-X-and-A-def}
%\end{align}
Subsequently, EG with linear utilities can be written as \eqref{eq:std-form-proj-grad-desc} with $\cX = (\Delta_n)^m$, $f(x) = h(A x)$, where $x$ can be viewed as a $(nm)$-dimensional vector, $A\in \RR^{n\times nm}$ is a block-diagonal matrix with $i$-th block being $v_i^\top$, $h(u) = \sum_{i=1}^n h_i(u_i)$, $h_i(u_i) := -B_i \log u_i$. Clearly, $\|A\| = \max_i \|v_i\|$. However, in this way, $h$ does not have a Lipschitz continuous gradient on the interior of $\mathcal{U} = \{ Ax:x\in \cX \}$: for $u\in \interior (\mathcal{U})$, each $u_i$ can be arbitrarily close to $0$. This can be circumvented by the following bounds on equilibrium utilities. 
\begin{lemma}
	Let $\ubar{u}_i =  \frac{B_i\|v_i\|_1}{\|B\|_1}$ and $\bar{u}_i = \|v_i\|_1$, $i\in [n]$. Any feasible allocation $x\in \cX$ satisfies $\langle v_i, x_i \rangle \leq \bar{u}_i$ for all $i$. For any equilibrium allocation $x^*$, we have $\langle v_i , x^*_i \rangle \geq \ubar{u}_i$ for all $i$. \label{lemma:eg-linear-u_min-u_max}
\end{lemma}
%For the Lipschitz constant $L$ of $\nabla h$, since $u_i = \langle v_i, x_i\rangle$ may be arbitrarily small for $x\in \cX$, it is not \textit{a priori} clear how it can be bounded.
Using Lemma \ref{lemma:eg-linear-u_min-u_max}, we can replace each $h_i$ by its ``quadratic extrapolation'' when $\langle v_i, x_i\rangle \notin [\ubar{u}_i, \bar{u}_i]$. Specifically, let $\tilde{h}(z):= \sum_{i=1}^n \tilde{h}_i(u_i)$, where
\begin{align}
\tilde{h}_i(u_i) := \begin{cases}
\frac{1}{2}h_i''(\ubar{u}_i) (u_i-\ubar{u}_i)^2 + h_i'(\ubar{u}_i)(u_i - \ubar{u}_i) + h_i(\ubar{u}_i) & {\rm if}\ u_i \leq \ubar{u}_i, \\
h_i(u_i) = -B_i\log u_i & {\rm otherwise.}
\end{cases}
%\label{eq:h-tilde-EG}
\end{align}
In the above, clearly, we have $h''(\ubar{u}_i) = \frac{B_i}{\ubar{u}_i^2} > 0$, $h'(\ubar{u}_i) = - \frac{B_i}{\ubar{u}_i}$ and $h_i(\ubar{u}_i) = - B_i \log \ubar{u}_i$. Therefore, $\tilde{h}$ is $\mu$-strongly convex and smooth with $L$-Lipschitz continuous gradient on $\mathcal{U}$, where 
$\mu = \min_i \frac{B_i} {\|v_i\|_1^2}$ and $L =  \max_i \frac{B_i}{\ubar{u}_i^2} = \max_i \frac{\|B\|_1^2 \|v_i\|_1}{B_i}$. We can then use the minimization objective $f(x) = \tilde{h}(Ax)$ in \eqref{eq:std-form-proj-grad-desc} instead, without affecting any optimal solution. 
A direct application of Theorem \ref{thm:h(Ax)+<q,x>-lin-conv} gives the following convergence guarantee regarding PG for the modified EG convex program. A similar conclusion can also be made using Theorem \ref{thm:pg-ls-conv} on PG with linesearch for the same problem.
\begin{theorem}
	Let $\cX$ and $A$ be as above. Then, PG for the problem $\min_{x\in \cX} \tilde{h}(Ax)$ with constant stepsize $\gamma_t = \frac{1}{L\|A\|^2}$ converges linearly with rate $1 - \frac{\mu}{\max\{\mu, LH_\cX(A)^2 \|A\|^2 \}}$, where $\|A\| = \max_i \|v_i\|$.
	\label{thm:pg-for-eg-lin-conv}
\end{theorem}

\noindent\textbf{Cost per iteration.\ } In every iteration (or linesearch step), PG requires computing the projection $\Pi_\cX$, which decomposes item-wise into $\Pi_{\Delta_n}$. This can be computed via an efficient $O(n\log n)$ sorting-based algorithm \cite{condat2016fast, chen2011projection, wang2013projection} (in fact, only nonzero elements need to be sorted), while $O(n)$ algorithms are also available \cite{duchi2008efficient, brucker1984n}. Furthermore, when $v$ is sparse, that is, only a small subset of buyers $I_j$$\subseteq [n]$ value each item $j$, the total cost for projection scales as $O(\sum_j |I_j|\log |I_j|) = O(\verb|nnz|(v)\log \verb|nnz|(v))$, without explicit dependence on $n,m$. Clearly, this is also true for computing the utilities $\langle v_i, x_i\rangle$, $i\in [n]$. Therefore, the cost per iteration scales with number of nonzeros $\verb|nnz|(v)$ without explicit dependence on $n,m$.
Furthermore, the gradient computation $\nabla f(x) = \sum_i \tilde{h}'_i(\langle v_i, x_i\rangle) v_i $ decomposes buyer-wise and $\Pi_\cX(x) = \left( \Pi_{\Delta_n}(x_{:, j})\right)$ decomposes item-wise, allowing straightforward parallelization of PG. In fact, the same holds for subsequent QL and Lenontief utilities.

\noindent\textbf{Convergent utilities and prices.\ } The equilibrium utilities $u^*$ are unique \cite[Theorem 1]{eisenberg1961aggregation}. So are the equilibrium prices $p^*$. Furthermore, $u^t_i = \langle v_i, x^t_i\rangle$ converges linearly to $u^*$ by strong convexity of $h$. 
In addition, the simplex projection algorithm in each iteration yields a linearly convergent sequence of prices $p^t$ in terms of relative price error $\eta^t := \max_j \frac{|p^t_j - p^*_j|}{p^*_j}$, a commonly used error measure in ME computation \cite{zhang2011proportional,birnbaum2011distributed}. See Appendix \ref{app:eg-convergence-of-simplex-prices} for details.

By Theorem \ref{thm:pg-for-eg-lin-conv} and the above discussion, we have the following, where $u^*_i$ is the (unique) equilibrum utility of buyer $i$. 
Similar statements can also be made following the subsequent convergence results, i.e., Theorem \ref{thm:pg-ql-shmyrev-lin-conv} and \ref{thm:pg-Leontief}.
\begin{corollary}
	PG for \eqref{eq:eisenberg-gale-primal} computes an feasible allocation $x^t$ such that $|\langle v_i, x^t_i\rangle -u^*_i| \leq \epsilon$ for all $i$ in $t = O( \tilde{n} \log \tilde{n} \log (1/\epsilon) )$ time, where $\tilde{n} = \verb|nnz|(v)$ and the constant only depends on $v_{ij}$ and $B_i$. 
\end{corollary}

\noindent\textbf{Handling additional constraints. } Although efficient computation of $\Pi_\cX$ depends on the structure of $\cX$, it is still arguably more flexible compared to combinatorial algorithms for computing market equilibria, which depend crucially on the specific market structures \cite{vazirani2010spending, vazirani2007combinatorial}. 
Here, for example, the simplex projection algorithms can be modified easily to handle box constraints like $\ubar{x}_{ij} \leq x_{ij} \leq \bar{x}_{ij}$ without affecting the time complexity \cite{pardalos1990algorithm, brucker1984n}. 
This allows additional at-most-one constraints $x_{ij} \leq 1$, useful in in fair division applications \cite{kroer2019scalable}.
% in the QL case (\S \ref{sec:ql-utilities}),
Similar bounds on the individual \textit{bids} $b_{ij}$ (which denote how much each buyer spends), such as \textit{spending constraints} (see, e.g., \cite[Eq. (2)]{birnbaum2011distributed} and \cite{vazirani2010spending}) can also be incorporated without incurring additional cost (this requires solving the Shmyrev convex program \eqref{eq:shmyrev} which we introduce in Appendix \ref{app:shmyrev-cp}).

\section{Quasi-linear utilities} \label{sec:ql-utilities}
In quasi-linear (QL) utilities, the money has value outside the market, which means each buyer's utility is the utility they derive from the items minus their payments, i.e., $u_i(x_i) = \sum_j (v_{ij} - p_{j})x_{ij} $, where $p_j$ is the price of item $j$. 
Note that QL utilities are not CCNH, as it depends on the prices. Nevertheless, based on EG and its dual for linear utilities, another pair of primal and dual convex programs can be derived to capture ME under QL utilities \cite[Lemma 5]{cole2017convex}. 
In order to derive Proportional Response dynamics for a market with QL utilities, we now introduce the following convex program, which we call QL-Shmyrev, for its similarity in structure to Shmyrev's convex program for linear utilities \eqref{eq:shmyrev} \cite{shmyrev2009algorithm}. It is the dual of a reformulation of a convex program in \cite[Lemma 5]{cole2017convex}.
%The dual, which minimizes over prices, is as follows:
%\begin{align}
%	\min \, \sum_j p_j - \sum_i B_i \log \beta_i \ \ 
%	{\rm s.t.} \ v_{ij}\beta_i \leq p_j,\ \forall\, i,j,\ \ 0 \leq \beta \leq 1. \label{eq:ql-dual-p-beta}
%\end{align}
%Under the same nondegeneracy assumption as in the case of linear utilities, by a change of variable and duality, we can derive the dual of \eqref{eq:ql-dual-p-beta}. 
Let the bids be $b = (b_1, \dots, b_n)$, where each $b_i = (b_{ij}) \in \RR^m$. Denote $p_j(b) = \sum_i b_{ij}$ and $p(b) = (p_1(b), \dots, p_m(b))$. Introduce slack variables $\delta = (\delta_1, \dots, \delta_n)$ representing buyers' leftover budgets. Let $\cB = \left\{(b, \delta) \in \RR^{n\times m} \times \RR^n: (b_i, \delta_i)\in \Delta_{n+1},\ i\in[n] \right\}$. The convex program is
\begin{align}
	\varphi^* = \min_{\bar{b} = (b, \delta)}\, \varphi(b) = -\sum_{i, j} (1+\log v_{ij}) b_{ij} + \sum_j p_j(b)\log p_j(b)  \st \bar{b} \in \cB. \label{eq:ql-shmyrev}
\end{align} 
The derivation is in Appendix \ref{app:ql-shmyrev-dual}. This convex program differs from Shmyrev's convex program \eqref{eq:shmyrev} for linear utilities in the coefficients of $b_{ij}$ and constraints on $b_{ij}$ (allowing $\sum_j b_{ij} \leq B_i$ instead of $=$). 
For notational brevity, we assume that $v>0$, although all results hold in the general case of nondegenerate $v$ (no zero row or column) with summations like $\sum_{ij} \log v_{ij}$ being over $(i,j): v_{ij}>0$ instead. 
\update{Here, $\delta_i$ is interpreted as the ``leftover'' budget of buyer $i$: since $u_i$ depends on the prices $p_j$, a buyer may not spend their entire budget $B_i$, if prices are too high relative to their valuations. In contrast, for CCNH $u_i$, budgets are always depleted at equilibrium.} 
In order to apply PG to solve \eqref{eq:ql-shmyrev}, we need a $(\mu, L)$-s.c. objective, while the function $ h(p) = \sum_j p_j \log p_j $ does not have Lipschitz continuous gradient on $\{p(b): (b, \delta)\in \cB \}$, since $p_j(b)$ can be arbitrarily small for $b\in \mathcal{B}$. TO address this, we establish the following bounds on equilibrium prices. We also show that they are unique and equal to the sum of buyers' bids at equilibrium.
\begin{lemma}
	For nondegenerate $v$, the equilibrium prices $p^*$ under QL utilities are unique. Let $\ubar{p}_j = \max_i \frac{v_{ij} B_i}{\|v_i\|_1 + B_i}$  and $\bar{p}_j = \max_i v_{ij}$ for all $j$. For any optimal solution $(b^*, \delta^*)$ to \eqref{eq:ql-shmyrev}, we have $\ubar{p}_j \leq p^*_j = p_j(b^*) \leq \bar{p}_j$.
	\label{lemma:ql-shmyrev-upper-lower-bounds}
\end{lemma}
To cast \eqref{eq:ql-shmyrev} into the standard form \eqref{eq:std-form-proj-grad-desc}, we take $\cX = \cB$, the linear map $A: (b, \delta) \mapsto (p_1(b), \dots, p_m(b))$, $h(p) = \sum_j p_j \log p_j$ and $q = (q_{ij})$, $q_{ij} = -1 - \log v_{ij}$. Viewing $(b, \delta)$ as a $n(m+1)$-dimensional vector concatenating each $b_i$ and $\delta$, we have $A := [I, \dots, I, 0]\in \RR^{ n \times (n(m+1)) }$ and $\|A\| = n$. Then, replace $h$ with a $(\mu, L)$-.s.c. function $\tilde{h}$ via a smooth extrapolation similar to that in \S \ref{sec:linear-utilities}, where $\mu = \frac{1}{\max_j \bar{p}_j}$ and $L = \frac{1}{\min_j \ubar{p}_j}$. Let $\tilde\varphi(b) = -\sum_{i,j} (1+\log v_{ij})b_{ij} + \sum_j \tilde{h}(p(b))$. Clearly, $\varphi(b) = \tilde{\varphi}(b)$ as long as $p(b)\in [\ubar{p}, \bar{p}]$, since $h = \tilde{h}$ on $[\ubar{p}, \bar{p}]$. 
Combining Theorem \ref{thm:h(Ax)+<q,x>-lin-conv} and Lemma \ref{lemma:ql-shmyrev-upper-lower-bounds}, we have the following.
\begin{theorem}
	Let $(b^0, \delta^0)$ satisfies $p(b^0)\in [\ubar{p}, \bar{p}]$. Then, PG with stepsize $\gamma_t = \frac{1}{L n^2}$ for the problem $\min_{(b,\delta)\in \cB} \tilde{\varphi}(b)$ converges linearly with rate $1 - \frac{1}{\max\{1, 2\kappa L n^2 \}}$, where \newline $\kappa = H_\cX(A)^2 \left( C + 2 G D + \frac{2(G^2 + 1)}{\mu} \right)$, 
%	$H = \max\left\{ H_\cB \left( \begin{bmatrix} A \\ q^\top  \end{bmatrix} \right), \frac{1}{2 n} \right\}$, \newline
	$C = \varphi(b^0) - \varphi^*$, \newline $G = \|\nabla \tilde{h}(p^*)\|$, $D = \sup_{(b, \delta), (b',\delta')\in \cB} \|p(b) - p(b')\|$.
	\label{thm:pg-ql-shmyrev-lin-conv}
\end{theorem}
\noindent\textbf{Remark.\ } Some constants can be bounded explicitly. Clearly, $D \leq \sqrt{\sum_j (\max_i v_{ij})^2}$. By Lemma \ref{lemma:ql-shmyrev-upper-lower-bounds}, $\tilde{h}(p^*) = h(p^*)$, since $p^*\in [\ubar{p}, \bar{p}]$. Therefore, $G = \sqrt{\sum_j (1+\log p^*_j)^2 } \leq \sqrt{\sum_j (1+\log \max_i v_{ij})^2}$. 

\noindent\textbf{MD for \eqref{eq:ql-shmyrev} as Proportional Response dynamics.\ } Similar to \cite{birnbaum2011distributed}, we can also apply MD to \eqref{eq:ql-shmyrev} and obtain a PR dynamics under QL utilities with $O(1/T)$ convergence in both objective value $\varphi(b^t) - \varphi^*$ and price error $D(p^t\|p^*)$. Recall that MD for \eqref{eq:ql-dual-p-beta} with unit stepsize $\gamma_t = 1$ performs the following update:
\begin{align}
(b^{t+1}, \delta^{t+1}) = \argmin_{(b, \delta)\in \cB}\, \langle \nabla \varphi(b^t), b - b^t \rangle + D(b, \delta\| b^t, \delta^t), \label{eq:md-udpate}
\end{align}
where $D(b',\delta'\|b, \delta) = \sum_{i,j} b'_{ij}\log \frac{b'_{ij}}{b_{ij}} + \sum_i \delta'_i\log \frac{\delta'_i}{\delta_i}$ is the usual KL divergence.  Extending (and slightly strengthening) \cite[Eq. (13) and (16)]{birnbaum2011distributed}, we first establish the following last-iterate convergence in objective value and price error. Its proof is in Appendix \ref{app:proof-MD-1/T}.
\begin{theorem}
	Let $b^0_{ij} = \frac{B_i}{m+1}$, $\delta^0_i = \frac{B_i}{m+1}$ for all $i,j$. Then, MD applied to \eqref{eq:ql-shmyrev} generates iterates $(b^t, \delta^t)$, $t = 1, 2, \dots$ such that $ D(p(b^t)\|p^*) \leq \varphi(b^t) - \varphi^* \leq \frac{\|B\|_1 \log(m+1)}{t}$ for all $t$. \label{thm:pr-ql-shmyrev-1/T}
\end{theorem}

The MD update \eqref{eq:md-udpate} leads to a form of PR dynamics \cite{zhang2011proportional, birnbaum2011distributed} as follows (see Appendix \ref{app:details-md-to-pr-ql} for the derivation). At time $t$, buyers first submit their bids $b^t = (b^t_{ij})$. Then, item prices are computed via $p^t_j = \sum_j b^t_{ij}$. Next, each buyer is allocated $x^t_{ij} = b^t_{ij}/p^t_j$ amount of item $j$. Finally, the bids and leftover budgets are updated via
\begin{align}
b^{t+1}_{ij} = B_i \cdot \frac{v_{ij}x^t_{ij}}{\sum_\ell v_{i\ell}x^t_{i\ell} + \delta^t_i},\ \ \delta^{t+1}_i = B_i \cdot \frac{\delta^t_i}{\sum_\ell v_{i\ell} x^t_{i\ell} + \delta^t_i}. \label{eq:pr-ql-shmyrev-final}
\end{align}

\noindent\textbf{Convergence of prices and a computable bound.} 
Similar to the linear case, PG for QL utilities also yields prices $p^t$ with relative error $\eta^t$ converging linearly to $0$. Furthermore, it can also be bounded explicitly by computable quantities. See Appendix \ref{app:ql-price-conv} for details.
%
%\[ \frac{\|p^t - p^*\|_1}{p_{\min}} \leq \frac{\sqrt{2\times {\rm dgap}}}{m} \]
%\section*{Other homogeneous utilities}
%\noindent\textbf{Cobb-Douglas} For Cobb-Douglas utilities, \eqref{eq:eisenberg-gale-primal} separates buyer-wise into simple minimization problems over $x_i$.
%\vspace{-5px}
\section{Leontief utilities} \label{sec:leontief-utilities}
%Leontief utilities represent complementary goods. Assume each buyer $i$ is characterized by $a_{ij}>0$, $j\in J_i\neq \emptyset$ such that, its utility given an allocation $x_i$ is
%\[ u_i(x_i) = \min_{j\in J_i}\, \frac{x_{ij}}{a_{ij}}. \]
Leontief utilities model perfectly complementary items and are suitable for resource sharing scenarios where utilities are capped by dominant resources. The utility function is $u_i(x_i) = \min_{j\in J_i}\frac{x_{ij}}{a_{ij}}$, where $a_{ij}>0$ for all $j\in J_i\neq \emptyset$. Denote $I_j = \{ i\in [n]: j\in J_i \}$ and assume that $I_j\neq \emptyset$ for all $j$, without loss of generality. Denote $a_i = (a_{i1}, \dots, a_{im})\in \RR_+^m$, where $a_{ij}:=0$ if $j\notin J_i$ and $a = [a_1^\top; \dots, a_n^\top]\in \RR^{n\times m}$. Under Leontief utilities, EG \eqref{eq:eisenberg-gale-primal} can be written in terms of the utilities $u\in \RR_+^n$ (see \eqref{eq:eg-Leontief} in Appendix \ref{app:leontief-dual}), whose dual, after reformulation, is
\begin{align}
\min\, f(p) = - \sum_i B_i \log \langle a_i, p \rangle \ \ {\rm s.t.}\ p\in \mathcal{P},	
\label{eq:eg-Leontief-dual-reform}
\end{align}
where $\cP = \{ p\in \RR^m_+: \sum_j p_j = \|B\|_1 \}$. The derivation is in Appendix \ref{app:leontief-dual}. In particular, we use the fact that $\|p^*\|_1 = \|B\|_1$ at equilibrium. Similar to the case of linear and QL utilities, we have the following bounds on $\langle a_i, p \rangle$ for Leontief utilities.
\begin{lemma}
	For any $p\in\cP$, it holds that $\langle a_i, p\rangle \leq \bar{r}_i := \|B\|_1 \|a_i\|_\infty$ for all $i$. Furthermore, for any equilibrium prices $p^*$, it holds that $\langle a_i, p^* \rangle \geq \ubar{r}_i := B_i \|a_i\|_\infty$ for all $i$. \label{lemma:r-ubar-r-bar}
\end{lemma}
%\begin{lemma}
%	For any $p\in \mathcal{P}$, it holds that $\langle a_i, p \rangle \leq \bar{r}_i:=  \|a_i\|_\infty \|B\|_1$. Furthermore, if $p$ is an optimal solution to \eqref{eq:eg-Leontief-dual-reform}, then it holds that, for all $i$, 
%	\[ \langle a_i, p\rangle \geq \ubar{r}_i := \left( \frac{\left(\frac{\|B\|_1}{m} \right)^{\|B\|_1}\prod_\ell \|a_\ell\|_1^{B_\ell}}{\left( \prod_{\ell\neq i} \|a_\ell\|_\infty^{B_\ell} \right)\|B\|_1^{ \sum_{\ell\neq i} B_\ell } } \right)^{\frac{1}{B_i}} > 0. \]
% \label{lemma:leontief-bounds-on-<a(i),p>}
%\end{lemma}
%The proof is in Appendix \ref{app:proof-leontief-bounds}.
%
%Therefore, at optimality, $p$ must satisfies $u_i := \langle a_i, p\rangle \geq \delta_i> 0$ for all $i$. If we assume, without loss of generality, that $\|B\|_1 = 1$, then we have $\ubar{r}_i = \left( \frac{\prod_\ell \|a_\ell\|_1^{B_\ell}}{m \prod_{\ell\neq i} \|a_\ell\|_\infty^{B_\ell}} \right)^{\frac{1}{B_i}}$. Furthermore, assuming without loss of generality that $\|a_i\|_\infty = 1$ for all $i$, then $\ubar{r}_i = \left( \frac{\prod_{\ell\neq i} \|a_\ell\|_1^{B_\ell}}{m} \right)^{\frac{1}{B_i}} \cdot \|a_i\|_1$.
Let $h(u) = -\sum_i B_i \log u_i$. Again, using Lemma \ref{lemma:r-ubar-r-bar}, we can use a simple quadratic extrapolation to construct a $(\mu, L)$-s.c. function $\tilde{h}$ with
	$\mu = \min_i \frac{B_i}{\bar{r}^2_i}$, $L = \max_i \frac{B_i}{\ubar{r}^2_i}$
without affecting any optimal solution. Clearly, a Lipschitz constant of the gradient of $f(p) = \tilde{h}(ap)$ is $L \|a\|^2$. Analogous to Theorem \ref{thm:pg-for-eg-lin-conv}, we have the following convergence guarantee for Leontief utilities. Here, equilibrium prices may not be unique, but the equilibrium utilities $u^*$ are (e.g., by \eqref{eq:eg-Leontief} in the appendix). We can also easily construct $u^t$ that converges linearly to $u^*$. See Appendix \ref{app:leontief-utilities-lin-conv} for details.
\begin{theorem}
	PG with fixed stepsize $\gamma_t = \frac{1}{L \|a\|^2}$ for the problem $\min_{p\in \cP} \tilde{h}(ap)$ converges linearly at a rate $ 1- \frac{\mu}{\max\{ \mu, L H_\cP(a)^2\|a\|^2 \}}$.
	\label{thm:pg-Leontief}
\end{theorem}

%\highlight{
%\noindent\textbf{Remark} }

%%% Local Variables:
%%% mode: latex
%%% TeX-master: "../main"
%%% End:

\section{Experiments} \label{sec:experiments}
We perform numerical experiments on market instances under all three utilities with various generated parameters. The algorithms are PGLS (PG with linesearch $\mathcal{LS}_{\alpha, \beta, \Gamma}$, see Appendix \ref{app:pg-ls-details}), PR and FW (with exact linesearch). 
For linear utilities, we generate market data $v = (v_{ij})$ where $v_{ij}$ are i.i.d. from standard Gaussian, uniform, exponential, or lognormal distribution. For each of the sizes $n=50, 100, 150, 200$ (on the horizontal axis) and $m=2n$, we generate $30$ instances with unit budgets $B_i=1$ and random budgets $B_i = 0.5 + \tilde{B}_i$ (where $\tilde{B}_i$ follows the same distribution as $v_{ij}$).
For QL utilities, we repeat the above (same random $v$, same sizes and termination conditions) using budgets $B_i = 5( 1+\tilde{B}_i)$.
The termination criterion is either (i) $\epsilon(p^t, p^*) = \max_i \frac{|p^t_i - p^*_i|}{p^*_i} \leq \eta$, where $p^*$ are the prices computed by CVXPY+Mosek \cite{diamond2016cvxpy, mosek2010mosek, dahl2019primal}, or (ii) average duality gap ${\rm dgap}_t/n \leq \eta$, for various thresholds values $\eta$. For PGLS, we report the number of linesearch iterations (that is, the total number of projection computations). 
For other algorithms, we report the number of iterations. \highlight{As a fair comparison, we use the same parameters $\alpha, \beta, \Gamma$ for PGLS throughout without handpicking.}
The plots report average numbers of iterations to reach the termination condition and their standard errors across $k=30$ repeats. In Appendix \ref{app:details-experiment} we show: more details on the setup, additional plots with different termination criteria and for Leontief utilities. Codes for the numerical experiments are available at \texttt{https://github.com/CoffeeAndConvexity/fom-for-me-codes}.

As can be seen, for linear utilities, PR is more efficient for obtaining an approximate solution (i.e., termination at $\epsilon(p^t, p^*)\leq 10^{-2}$ or ${\rm dgap}_t/n\leq 10^{-3}$). When higher accuracy is required, PGLS takes far fewer iterations. For QL utilities, PR is more efficient in most cases, except when very high accuracy is required (${\rm dgap}_t/n\leq 5 \times 10^{-6}$). We do not show FW for the QL case, as it performed extremely badly. In the appendix, we also see that for Leontief utilities, PGLS with linesearch terminates within tens of iterations in all cases.
%For \textbf{QL} utilities, we compare PG and PR for \eqref{eq:ql-shmyrev} through similar experiment setups.

\begin{center}
	\includegraphics[width=0.8\linewidth]{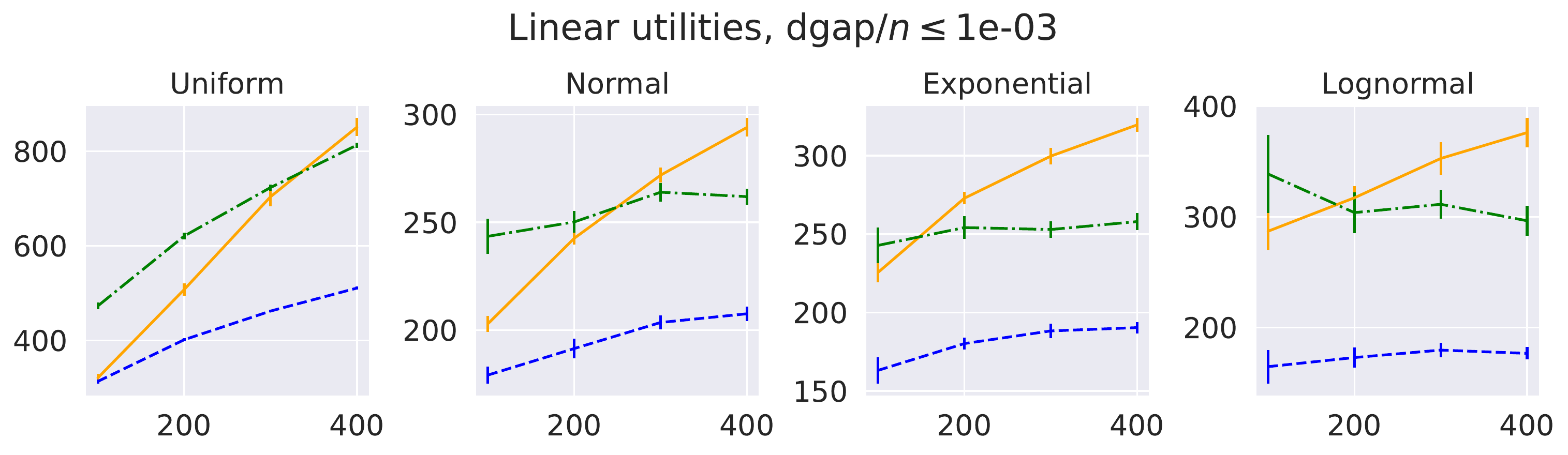}
	\includegraphics[width=0.8\linewidth]{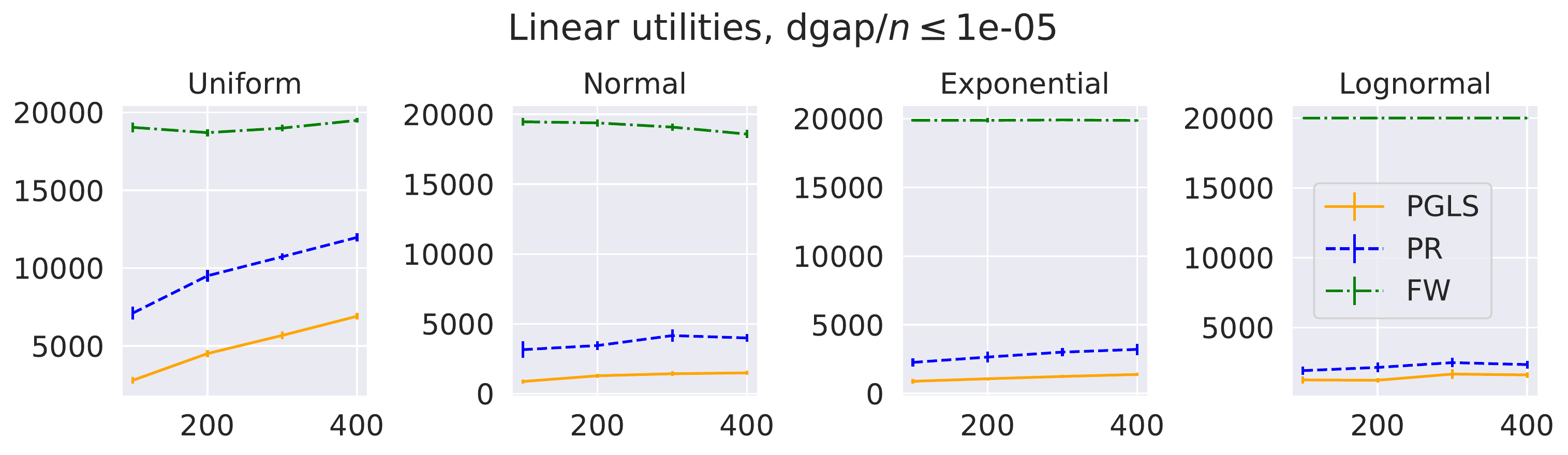} \quad
	\includegraphics[width=0.8\linewidth]{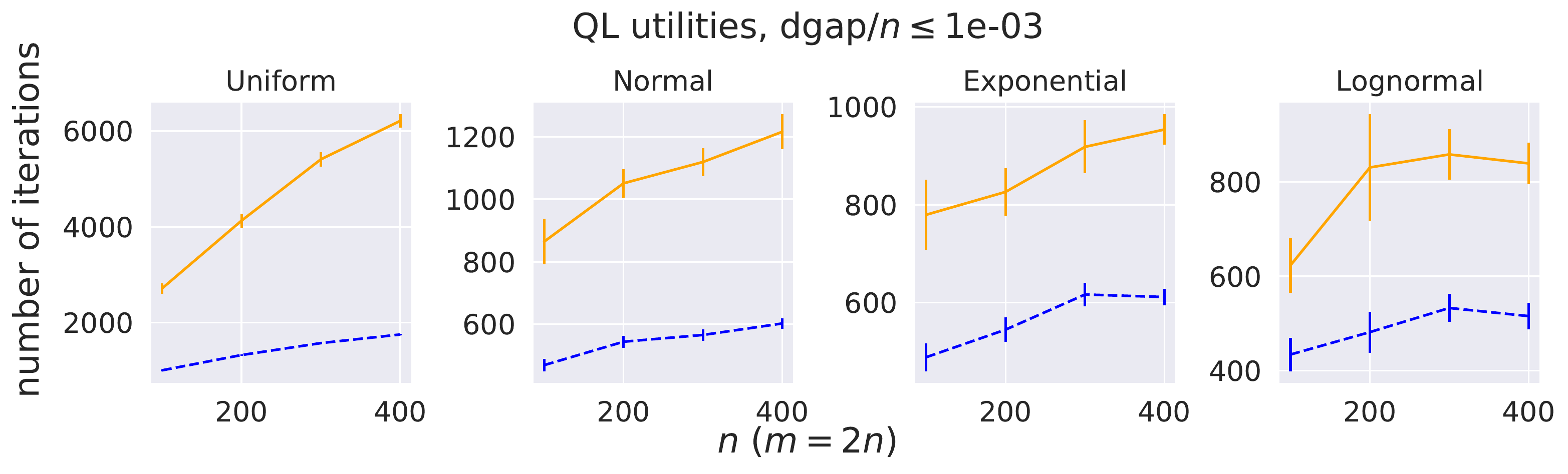}
	\includegraphics[width=0.8\linewidth]{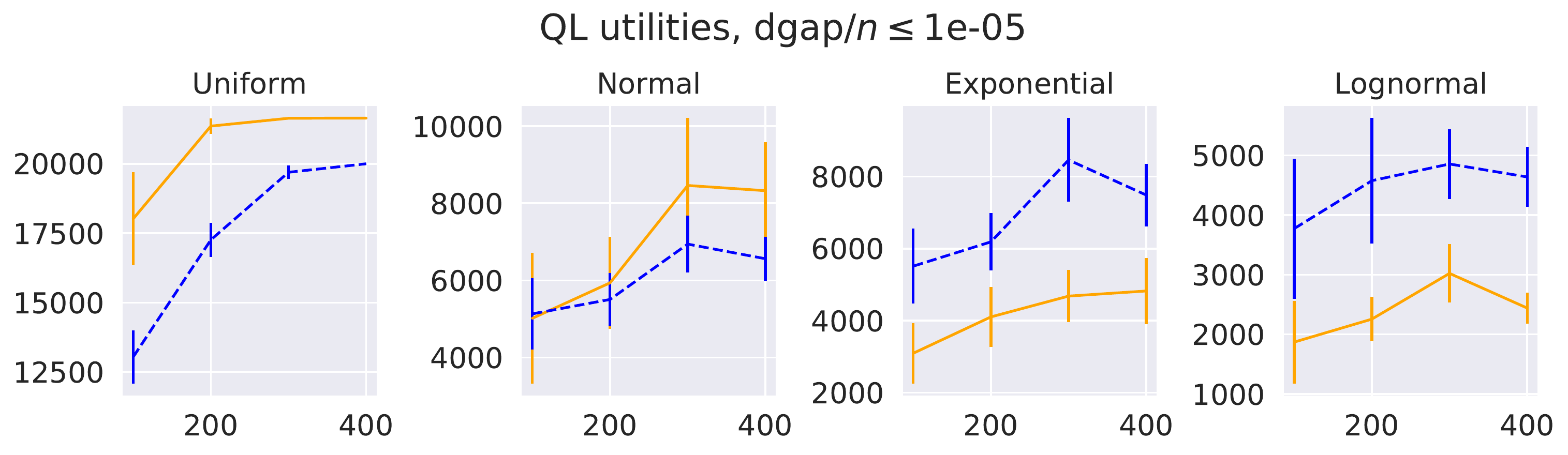}
\end{center}

%%% Local Variables:
%%% mode: latex
%%% TeX-master: "../main"
%%% End:

\section{Conclusions} \label{sec:conclu}
We investigate the computation of market equilibria under different buyer utility assumptions using simple first-order methods. Through convex optimization duality and properties of market equilibria, we show that the associated convex programs can be reformulated into structured forms suitable for FOMs. We show that projected gradient achieves linear convergence for these reformulations, through weakened strong-convexity conditions. As a technical contribution on FOMs, we also prove a more general linear convergence result of proximal gradient with linesearch under the {Proximal-\PL} condition. For QL utilities, we derive a form of Proportional Response through Mirror Descent on a convex optimization formulation, extending the classical convergence results for linear utilities. Finally, extensive numerical experiments compare the efficiency of the FOMs for various low- and high-accuracy termination criteria. %and demonstrate their general efficiency over an off-the-shelf approach.

%%% Local Variables:
%%% mode: latex
%%% TeX-master: "../main"
%%% End:

\newpage
\section{Broader Impact}
As mentioned in the introduction, large-scale market equilibrium computation problems arise in important applications such as Internet advertising markets, course assignment at universities, fair recommender systems and compute resource allocation. As such, this work has the following potential positive societal impact: Based on this work, resource allocation schemes, especially those with desirable fairness properties, that were previously deemed hard to solve at scale (and are thus simplified or disregarded), can be implemented in reasonable time through computing a market equilibrium by a first-order method. Progress of a FOM can be easily monitored via the duality gap while feasibility is guaranteed. Our work also enables greater scalability of certain Internet advertising market equilibrium models. This could be used for greater market efficiency or better monetization of such markets. Whether this is viewed as a positive or negative thing is beyond the scope of our paper.

\newpage
\bibliography{references}
\bibliographystyle{siam}

\appendix
\newpage

\section{Preliminaries}
We list and prove a few elementary lemmas used in subsequent proofs and discussions.

\begin{lemma}
	Let $n \geq m$, $A\in \RR^{m \times n}$ and $b\in \RR^m $ be such that $\mathcal{S} = \{ x\in \RR^n: Ax=b \} \neq \emptyset$. Let the singular value decomposition (SVD) of $A^\top$ be $A^\top = U\Sigma V^\top$, where $U\in \RR^{n\times r}$ and $V\in \RR^{m \times r}$ have orthonormal columns, $\Sigma = \diag(\sigma_1, \dots, \sigma_r)$, $\sigma_1\geq \dots \sigma_r >0$, $r = \rank(A)$. For any $x\in \RR^n$, the projection of $x$ onto $\mathcal{S}$ can be expressed as
	\[ \Pi_\cS(x) = (I - UU^\top) x + U \Sigma^{-1} V^\top b. \]
	In particular, when $A$ has full rank, 
	\[ \Pi_\cS(x) = (I - A^\top (AA^\top)^{-1}A)x + A^\top (AA^\top)^{-1}b. \]
	\label{lemma:proj-subspace-expression}
\end{lemma}
\textit{Proof.} The case of full rank $A$ is well-known, see, e.g., \cite[Eq. (1)]{plesnik2007finding}. For general $A$, $Ax = b \Leftrightarrow V\Sigma U^\top x = b$. Since $V\Sigma$ has full column rank, there exists a unique $w\in \RR^{r}$ such that $V\Sigma w = b$. In fact, $w = \Sigma^{-1}V^\top b$. Therefore, \[Ax = b \Leftrightarrow U^\top x = \Sigma^{-1} V^\top b.\] 
Since $U^\top$ has full row rank, the formula follows directly from the full rank case.
\qed

\begin{lemma}
	Under the same assumptions as Lemma \ref{lemma:proj-subspace-expression}, for any $x\in \RR^n$, it holds that \[ \sigma_r \cdot \| x - \Pi_\mathcal{S}(x) \| \leq \|Ax-b\|. \] \label{lemma:hoffman-const-subspace}
\end{lemma}
\textit{Proof.} 
By Lemma \eqref{lemma:proj-subspace-expression} and the fact that $U$ and $V$ have orthonormal columns, 
\begin{align*}
	\|x - \Pi_\cS(x) \| &= \| U(U^\top x - \Sigma^{-1}V^\top  b) \| = \| U^\top x - \Sigma^{-1}V^\top  b \| \\
	& = \|  \Sigma^{-1}V^\top  (Ax -b) \| \leq \|\Sigma^{-1}\| \|Ax-b\|,
\end{align*}
where $\|\Sigma^{-1}\| = \frac{1}{\sigma_r}$.\qed

\begin{lemma}
	For $B>0$ and $g\in \RR^d$, let
	\[ x^* = \argmin \left\{ \langle g, x\rangle + \sum_{i=1}^d x_i \log x_i : x\geq 0,\, \ones^\top x = B \right\}. \] \label{lemma:md-subproblem}
	Then, $x^*_i = B\cdot \frac{e^{-g_i}}{\sum_\ell e^{-g_\ell}}$, $i\in [d]$. 
\end{lemma}
\textit{Proof.} It can be easily verified via KKT optimality conditions. The Lagrangian is 
\[ L(x, \lambda) = \langle g, x \rangle + \sum_i x_i \log x_i - \lambda(\ones^\top x - B). \]
By the first-order condition, for any $\lambda$, the minimizer $x(\lambda)$ of $L(x, \lambda)$ has $x_i(\lambda) = e^{\lambda - 1 - g_i}$. Primal feasibility implies $\sum_i x_i(\lambda) = B \Rightarrow e^\lambda = \frac{B}{\sum_i e^{-(1+g_i)}}$. Therefore, 
\[ x^*_i = B\cdot \frac{e^{-g_i}}{\sum_\ell e^{-g_\ell}}. \]
\qed 

\subsection{Proof of Theorem \ref{thm:eg-gives-me-for-certain-ui}} \label{app:thm-eg-and-mkt-equi}
Existence of an optimal solution of \eqref{eq:eisenberg-gale-primal} can be shown via a standard argument. Specifically, since $u_i$ are CCNH, the objective function $ F(x) = \sum_i B_i \log u_i(x_i)$ is continuous and concave on the interior of the feasible region $\mathcal{X} = \left\{ x\in \RR_+^{n\times m}: \sum_i x_{ij} \leq 1,\, \forall\, j \right\}$. Since $u_i(\ones) >0$ for all $i$, by homogeneity, $u_i(\ones/n) >0$ for all $i$ for the feasible allocation $x^0_i = \ones/n$, $i\in [n]$. Hence, the optimal objective value $>-\infty$. Meanwhile, for any feasible $x$, 
\[ F(x) \leq \sum_i B_i \log u_i(\ones) < \infty.  \]
Hence, $F^* = \sup_{x\in \mathcal{X}} F(x)$ is finite. Furthermore, 
\[ \mathcal{X}' = \left\{ x\in \mathcal{X},\, F(x) \geq F(x^0) \right\} \]
is a compact set, on which $F$ is continuous. Therefore, $F^* = F(x^*)$ for some $X^* \in \mathcal{X}$. 

Next, we show that an optimal solution of \eqref{eq:eisenberg-gale-primal} gives a ME. 
Consider the minimization form of \eqref{eq:eisenberg-gale-primal}. Let $p_j\geq 0$ be the dual variable associated with constraint $\sum_i x_{ij} \leq 1$. The Lagrangian is
\[\mathcal{L}(x,p) = \left[ -\sum_i B_i \log u_i(x) + \left \langle p, \sum_i x_i\right \rangle - \sum_j p_j \right]. \]
The Lagrangian dual is 
\begin{align}
\max_{p\geq 0} \, g(p) := \min_{x\geq 0}\, \mathcal{L}(x,p). \label{eq:general-eg-dual}
\end{align}

Since $u_i$ are homogeneous, \eqref{eq:eisenberg-gale-primal} has a strictly feasible solution with finite objective value ($\sum_i x_{ij} < 1$ for all $j$), e.g., $x_i = \ones/(n+1)$ for all $i$. 
The dual \eqref{eq:general-eg-dual} clearly has a strictly feasible solution $p>0$ with finite $g(p) = \min_{x\geq 0} \mathcal{L}(x, p)$. Therefore, strong duality holds by Slater's condition and the KKT conditions are necessary and sufficient for (primal and dual) optimality of a solution pair $(x,p)$ (see, e.g., \cite[Appendix D]{ben2019lectures}). Let $x^*$ and $p^*$ be optimal solutions to the primal \eqref{eq:eisenberg-gale-primal} and dual \eqref{eq:general-eg-dual}, respectively. Clearly, $u_i(x^*_i)>0$ for all $i$ (since $F^*$ is finite). 
By Lagrange duality, we have
\[x^* \in \argmin_{x\geq 0} \mathcal{L}(x, p^*).\]
In other words, each $x^*_i$ maximizes \[ r_i(x_i, p^*) := B_i \log u_i(x_i) - \langle p^*, x_i\rangle\] 
on $x_i \geq 0$. We show that $(x^*, p^*)$ is a market equilibrium. 

\paragraph{Buyer optimality} First, we verify that $\langle p^*, x^*_i\rangle = B_i$ for all $i$. Assume that $\langle p^*, x^*_i\rangle > B_i$ for some $i$. Let $\tilde{x}_i = (1-\epsilon) x^*_i$, where $0\leq \epsilon<1$. Consider the smooth function
\[\phi(\epsilon) = B_i \log u_i((1-\epsilon)x^*_i) - \langle p^*, (1-\epsilon)x^*_i\rangle = B_i \log u_i(x^*_i) - \langle p^*, x^*_i\rangle + B_i \log (1-\epsilon) + \epsilon \langle p^*, x^*_i\rangle. \]
Note that $\phi$ is differentiable on $(0,1)$ and \[\phi'(\epsilon) = - \frac{B_i}{1-\epsilon} + \langle p, x^*_i\rangle.\] 
Since $\langle p^*, x^*_i\rangle >B_i$, we have $\phi'(0) >0$. In other words, replacing $x^*_i$ by $\tilde{x}_i$ with sufficiently small $\epsilon$ strictly decreases the value of $r_i(x_i)$, contradicting to the choice of $x^*$. Therefore, \[\langle x^*, x^*_i\rangle \leq B_i\] 
for all $i$. Completely analogously, we can also show that \[\langle p^*, x^*_i\rangle \geq B_i\] 
for all $i$. Therefore, for each buyer $i$, $x_i^*$ is feasible and depletes its budget $B_i$ under prices $p^*$. Hence, for any $x_i \in \RR^m_+$, $\langle p^*, x_i\rangle \leq B_i$, since $x^*_i$ maximizes $r_i(x_i, p^*)$, we have
\[ B_i \log u_i(x^*_i) -\langle p^*, x^*_i \rangle \geq B_i \log u_i(x_i) - \langle p^*, x_i\rangle. \]
Since $\langle p^*, x_i\rangle \leq B_i = \langle p^*, x^*_i\rangle$, the above implies
\[ B_i \log u_i(x^*_i) \geq B_i \log u_i(x_i). \]
Therefore, $u_i(x^*_i) \geq u_i(x_i)$. In other words, $x^*_i \in D_i(p^*)$ (buyer $i$ is optimal) for all $i$.
\paragraph{Market clearance} By the complementary slackness condition, for item $j$ such that $\sum_i x^*_{ij} < 1$, it must holds that $p^*_j = 0$, completing the proof.
\paragraph{Remark} We can also assign any leftover of item $j$ to any buyer $i$ without violating its budget constraint, in order to ``clear'' the market. Meanwhile, since $u_i$ is CCNH, it is also ``monotone'' in the following sense: for any $\alpha\geq 0$, 
\[ u_i(x^*_i + \alpha \mathbf{e}^j) \geq u_i(x^*_i) + \alpha u_i(\mathbf{e}^j) \geq 0. \]
In other words, buyer $i$'s optimality is not affected by the assignment of any leftover of any item whose price is zero.

\subsection{Characterizations of Hoffman constant}\label{app:char-hoffman-const}
We compare our definition of Hoffman constant and another common, explicit characterization. Recall that $H_\cX(A)$ is the smallest $H$ such that, for any $b$, $\cS = \{x: Ax=b\}$,
\[ \|x-\Pi_{\cX\cap \cS}(x)\|\leq H\|Ax-b\|,\ \forall\, x\in \cX. \]

For any matrix $M$, let $\cB(M)$ be the set of \textit{nonsingular} submatrices consisting of rows of $M$. Define
\begin{align}
H(M) = \max_{B\in \mathcal{B}(M)} \frac{1}{\sigma_{\min}(B)} < \infty.  \label{eq:hoffman-unconstrained}
\end{align}
The following fact is known (see, e.g., \cite[\S 11.8]{guler2010foundations} and \cite[\S 2.1]{beck2017linearly}).
\begin{lemma}
	Suppose the reference polyhedral set can be represented by inequality constraints $\cX = \{x: Cx\leq d\}$.  Then,
	\[ H_{\cX}(A) \leq H\left( \begin{bmatrix}
	A \\ C
	\end{bmatrix} \right). \]
\end{lemma}
Clearly, $H(M)$ is finite for any $M$. In fact, this is the most well-known characterization of Hoffman constant, and is tight in the following sense: let $\cS = \{x: Ax=b\}$ for some arbitrary right hand side $b$, then it is the smallest constant $H$ such that  
\[ \|x - \Pi_{\cX \cap \cS}(x)\| \leq H \left\| \begin{bmatrix}
Ax - b \\ (Cx-d)_+
\end{bmatrix} \right\|\]
for all $x$ (not necessarily $\in \cX$). However, for all of our purposes, that is, analysis of PG, $x$ is always restricted to be $\in \cX$. Therefore, we choose to define $H_\cX(A)$ as such, consistent with \cite{beck2003mirror} and \cite{pena2020new}.
Meanwhile, the following is clear. 
\begin{lemma}
	For any matrices $A\in \RR^{m\times n}$, $m\leq n$ and $C\in \RR^{\ell \times n} $, it holds that \[
	H\left(\begin{bmatrix}
	A \\ C
	\end{bmatrix} \right) \geq \max \left\{\frac{1}{\sigma_{\min}(A)}, H(A) \right\}. \label{lemma:compare-H-H(A)-sigma_min}\]
\end{lemma}
\textit{Proof.} By definition \eqref{eq:hoffman-unconstrained}, $H' := H\left(\begin{bmatrix}
A \\ C^\top
\end{bmatrix} \right) \geq H(A)$. If ${\rm rank}(A) = m$, then $H' \geq \frac{1}{\sigma_{\min}(A)}$ because $A\in \cB\left( \begin{bmatrix}
A \\ C
\end{bmatrix} \right)$. If $r = {\rm rank}(A) < m$, let the (nonzero) singular values of $A$ be $\sigma_1 \geq \dots \geq \sigma_r = \sigma_{\min}(A) > 0$. Consider any $B\in \cB(A) \subseteq \cB\left( \begin{bmatrix}
A \\ C
\end{bmatrix} \right)$ with rank $r$ (having exactly $r$ rows), let its nonzero singular values be $\sigma'_1\geq \dots \geq \sigma'_r = \sigma_{\min}(B) > 0$. Applying Cauhchy's Interlacing Theorem (see, e.g., \cite[Theorem 1]{fan1957imbedding}) on $AA^\top$ and its principal submatrix $BB^\top$, we have
\[ \sigma_1 \geq \sigma'_1 \geq \dots \geq \sigma_r \geq \sigma_r'. \]
Therefore, $H' \geq \frac{1}{\sigma_{\min}(B)} \geq \frac{1}{\sigma_{\min}(A)}$.
\qed

\subsection{Proof of Theorem \ref{thm:h(Ax)+poly_ind}} \label{app:proof-h(Ax)+poly_ind-conv}
We follow the development in \cite[\S 4 \& Appendix F]{karimi2016linear} and further articulate the constants. There, the authors show that proximal gradient achieves linear convergence under the so-called \textit{Proximal-\PL} inequality. Consider the following general nonsmooth problem
\begin{align}
F^* = \min_x F(x) = f(x) +g(x) \label{eq:min-F=f+g}
\end{align} where $f$ is smooth convex with $L_f$-Lipschitz continuous gradient, $g$ is simple closed proper convex and $\dom g \subseteq \dom f$.
One iteration of the proximal gradient method with stepsize $\gamma > 0$ is as follows:
\begin{align}
x^{t+1} = \prox_{g}\left( x^t - \gamma \nabla f(x^t) \right) = \argmin_x \left[ \langle \gamma \nabla f(x), x-x^t\rangle + \frac{1}{2}\|x - x^t\|^2 + g(x)\right]. \label{eq:pg-iter}
\end{align}

For any $\alpha>0$ and any $x\in \dom g$, define
\begin{align}
\cD(x, \alpha) =  -2 \alpha \min_{x'} \left[ \langle \nabla f(x), x' - x \rangle  + \frac{\alpha}{2}\|x'-x\|^2 + g(x') - g(x) \right].
\label{eq:def-D_g(x,a)}
\end{align}
Say that $F = f+g$ satisfies the {proximal-\PL} inequality at $x$ w.r.t. $\Lambda \geq \lambda > 0$ if
\begin{align}
\frac{1}{2} \cD(x, \Lambda) \geq \lambda(F(x) - F^*),  \label{eq:prox-PL}
\end{align}
%Say that an algorithm converges \textit{linearly} (in objective value) at a rate $\theta \in (0,1)$ if 
%\[F(x^t) - F^* \leq \theta^t \left( F(x^0) - F^* \right),\ \ t = 0, 1, 2, \dots \] 
%for all $t$. 
Below is essentially \cite[Theorem 5]{karimi2016linear}, which shows that the so-called {Proximal-\PL} condition is sufficient for linear convergence. Note that, different from \cite[Theorem 5]{karimi2016linear}, we only require \eqref{eq:prox-PL} to hold for $x\in \cX$ such that $F(x)\leq F(x^0)$ instead of all $x\in \cX$. In addition, we note that in some cases \eqref{eq:prox-PL} may hold with $\Lambda > L_f$, in which case the rate needs to be slightly adjusted. Since $\cD(x, \cdot)$ is monotone \cite[Lemma 1]{karimi2016linear}, \eqref{eq:prox-PL} holds when $\Gamma$ is replaced by $\Gamma' \geq \Gamma$. The statement and proof are the same as \cite[pp. 9]{karimi2016linear} otherwise.
\begin{theorem}
	Let $x^0 \in \dom g$. If $f$ and $g$ satisfies \eqref{eq:prox-PL} for all $x\in \dom g$ such that $F(x) \leq F(x^0)$, then $x^t$ defined by \eqref{eq:pg-iter} starting from $x^0$ with constant stepsize $\gamma = 1/L_f$ converges linearly with rate $1 - \frac{\lambda}{\bar{L}}$, where $\bar{L} = \max\{\Lambda, L_f \}$. In other words,
	\[ F(x^t) - F^* \leq \left( 1- \frac{\lambda}{\bar{L}}\right)^t (F(x^0) - F^*), \ t = 1, 2, \dots \]
	\label{thm:pg-lin-conv}
\end{theorem}
\textit{Proof.} By assumption, \eqref{eq:prox-PL} holds for all $x\in \dom g$, $x\leq F(x^0)$. In particular, it holds for $x^t$, $t, 1, 2, \dots$, since proximal gradient is a \textit{descent} method, i.e., $F(x^0) \geq F(x^1) \geq \dots$ (see, e.g., \cite[Corollary 10.18]{beck2017first}). 
%Since $\bar{L} \geq \Lambda$, by monotonicity of $\cD(x, \cdot)$ \cite[Lemma 1]{karimi2016linear}, 
%\[  \]
Therefore, by $L_f$-Lipschitz continuity of $\nabla f$, proximal gradient update \eqref{eq:pg-iter}, definition of $D(x, \cdot)$, its monotonicity, and \eqref{eq:prox-PL} for all $x^t$,
\begin{align*}
F(x^{t+1}) 
& \leq F(x^t) + \langle \nabla f(x^t), x^{t+1} - x^t\rangle  + \frac{L_f}{2} \|x^{t+1} - x^t\|^2 + g(x^{t+1}) - g(x^t) \\
&\leq F(x^t) + \left[\langle \nabla f(x^t), x^{t+1} - x^t\rangle  + \frac{\bar{L}}{2} \|x^{t+1} - x^t\|^2 + g(x^{t+1}) - g(x^t) \right] \\
& \leq F(x^t) - \frac{1}{2\bar{L}} \cD(x^t, \bar{L}) \\
& \leq F(x^t) - \frac{\lambda}{\bar{L}}\left( F(x^t) - F^* \right) \\
\Rightarrow\ \ 	& F(x^{t+1}) - F^* \leq \left( 1 - \frac{\lambda}{\bar{L}} \right)\left( F(x^t) - F^* \right).
\end{align*}
Repeatedly applying the above inequality completes the proof. \qed

% Since proximal gradient is a \textit{descent} method, that is, $F(x^0) \geq F(x^1)\geq F(x^2)\geq \dots$ (see, e.g., \cite[Corollary 10.18]{beck2017first}), this is sufficient for the third inequality in the proof in \cite[pp. 9]{karimi2016linear}.

% For $x\in \cX$ and $\alpha>0$, denote
% \[\cD(x, \alpha) = - 2\alpha \min_{y\in \cX}\left\{ \langle \nabla f(x), y-x\rangle + \frac{\alpha}{2}\|y-x\|^2 \right\} \geq 0.\] 
Then, we prove Theorem \ref{thm:h(Ax)+poly_ind}. Clearly, problem \eqref{eq:std-form-proj-grad-desc} is \eqref{eq:min-F=f+g} with $g(x)=\delta_\cX(x)$ and PG is a special case of proximal gradient. By Theorem \ref{thm:pg-lin-conv}, in order to prove Theorem \ref{thm:h(Ax)+poly_ind}, it suffices to establish the {Proximal-\PL} condition \eqref{eq:prox-PL} (for all $x\in \cX, f(x) \leq f(x^0)$ for some initial iterate $x^0$). Let $\cX^*$ be the set of optimal solutions to \eqref{eq:std-form-proj-grad-desc} and $f^*$ be the optimal objective value.  Since $h$ is $\mu$-strongly convex and $f(x) = h(Ax)$, there exists $z^*\in \dom f$ such that $\mathcal{S} = \{x: A x = z^*\}$ and $\cX^* = \cX \cap \mathcal{S}$. Therefore, for any $x\in \cX$, $x_p := \Pi_{\cX^*}(x)$, we have
\[ f(x_p) = h(A x_p) \geq h(A x) + \langle \nabla h(A x), A(x_p - x)\rangle + \frac{\mu}{2}\|A(x_p-x)\|^2. \]
Note that 
\[ \langle \nabla h(A x), A(x_p - x)\rangle =  \langle A^\top \nabla h(A x ), x_p - x\rangle = \langle \nabla f (x), x_p - x\rangle. \]
%\textbf{Question}: Is there any simple bound on $H$, or even a closed-form expression? In general, are there such results for Hoffman constants of simple polyhedral sets?

Hence, for any $x\in \cX$, by strong convexity of $h$ and definition of $H = H_\cX(A)$, we have
\begin{align*}
f(x_p) & \geq f(x) + \langle \nabla f(x), x_p - x\rangle + \frac{\mu}{2}\|A(x-x_p)\|^2 \\
& = f(x) + \langle \nabla f(x), x_p - x\rangle + \frac{\mu}{2}\|A x- z^*\|^2 \\
%& \geq f(x) + \langle \nabla f(x), x_p - x\rangle + \frac{\mu}{2 H_\cX(A)^2} \| x - x_p \|^2 \\
& \geq f(x) + \langle \nabla f(x), x_p - x\rangle + \frac{\mu}{2 H^2} \| x - x_p \|^2, 
\end{align*}
%where $H = \max\left\{H_\cX(A), \frac{1}{\|A\|} \right\}$. 
Therefore, 
\begin{align*}
f^* &\geq f(x) + \langle \nabla f(x), x_p - x\rangle + \frac{\mu}{2 H^2}\|x-x_p\|^2 \\
&\geq f(x) + \min_{y\in \cX} \left\{ \langle \nabla f(x), y-x\rangle + \frac{\mu}{2 H^2} \|y-x\|^2 \right\} \\
& \geq f(x) - \frac{H^2}{2\mu} \cD\left(x, \frac{\mu}{H^2}\right)  \\
\Rightarrow\ \ & \frac{1}{2} \cD\left(x, \frac{\mu}{H^2}\right) \geq \frac{\mu}{H^2}(f(x) - f^*).
\end{align*}
%\footnote{Here, $\frac{1}{H}$ can be viewed as a ``constrained'' singular value, as it differs from $\sigma_{\min}(A)$ due to the fact that $\cS$ is further constrained into $\cX^*$.} 
%Since $L \geq \mu$, by the choice of $H$, it holds that
%\[L_f = L \|A\|^2 \geq \frac{\mu}{H^2}. \]
%By monotonicity of $\cD(x, \cdot)$, 
%\[ \frac{1}{2}\cD(x, L_f) \geq \frac{1}{2}\cD\left(x, \frac{\mu}{H^2}\right) \geq \frac{\mu}{H^2}(f(x) - f^*). \]
Thus, \eqref{eq:prox-PL} holds for all $x\in \cX$, $f(x)\leq f(x^0)$ with \[\Lambda = \lambda = \frac{\mu}{H^2}.\] 
Since $\nabla f(x) = A^\top \nabla h(A x)$ and $h$ is $(\mu, L)$-s.c., its Lipschitz constant can be chosen as 
\[L_f = L \|A\|^2.\] 
%\[\bar{\mu} = \frac{\mu}{H^2},\ L_f = L \|A\|^2.\] 
By Theorem \ref{thm:pg-lin-conv}, PG with stepsize $\gamma = \frac{1}{L_f}$ converges linearly with rate

\[ 1 - \frac{\frac{\mu}{H^2}}{\max\left\{ \frac{\mu}{H^2}, L\|A\|^2 \right\}} = 1 - \frac{\mu}{\max\{ \mu, LH^2 \|A\|^2  \}}.  \]
Finally, convergence of the distance to optimality $\|x^t - \Pi_\cX(x^t)\|$ is straightforward: for any $x\in \cX$, by the strong convexity of $h$ and definition of $H$,
\[ f(x) - f^* = h(Ax) - h(Ax_p) \geq \frac{\mu}{2}\|Ax - Ax_p\|^2 = \frac{\mu}{2}\|Ax-z^*\|^2 \geq \frac{\mu}{2 H} \|x - x_p\|^2. \] \qed

\paragraph{Remark}
A special case is when $d\geq r$ (recall that $A\in \RR^{d\times r}$) and ${\rm rank}(A) =r$. In this case, $f(x) = h(Ax)$ itself is strongly convex with modulus $\mu \sigma_{\min}(A)^2$. In this case, classical analysis (e.g., \cite[\S 10.6]{beck2017first}) implies linear convergence with rate $1 - \frac{\mu \sigma_{\min}(A)^2}{L \|A\|^2}$. Meanwhile, in the above analysis, we have $\cX^* = \{x^*\} = \cS = \{x: Ax = z^*\} = \cX \cap \cS$ (since $x^*, z^*$ are unique and ${\rm rank}(A)=r$). By Lemma \ref{lemma:hoffman-const-subspace}, for any $x$, it holds that
\[\|x - \Pi_{\cX^*}(x) \|\leq \frac{1}{\sigma_{\min}(A)^2}\|Ax-z^*\|.\]
Therefore, by the definition of Hoffman constant, $H_\cX(A)\leq \frac{1}{\sigma_{\min}(A)^2}$ and the classical rate under strong convexity is recovered.

\subsection{Proof of Theorem \ref{thm:h(Ax)+<q,x>-lin-conv}} \label{app:proof-h(Ax)+<q,x>+poly_ind-conv}
Let $\cX^*$ be the set of optimal solutions to \eqref{eq:std-form-proj-grad-desc}. First, recall the following lemma \cite[Lemma 14]{wang2014iteration}, which ensures the first part of the theorem, that is, uniqueness of $Ax^*$ and $q^\top x^*$ for all $x^*\in \cX^*$.
\begin{lemma}
	There exist unique $z^*\in \RR^r$ and $w^*\in \RR$ such that for any $x^*\in \cX^*$, 
	\[ Ax^* = z^*,\ \langle q^*, x\rangle = w^*. \]
	\label{lemma:Ax^*=z^*,qTx^*=w^*}
\end{lemma}
The next lemma is essentially \cite[Lemma 2.5]{beck2017linearly}. Different from the statement of \cite[Lemma 2.5]{beck2017linearly}, we keep $\|\nabla h(z^*)\|$ instead of bounding it by $\sup_{x\in \cX} \|\nabla h(Ax)\|$. We also define $C = f(x^0) - f^*$ instead of $C = \sup_{x\in \cX} f(x) - f^*$, since subsequent application of the lemma only involves PG iterates $x^t$, which have monotone decreasing objective values $f(x^0)\geq f(x^1)\geq \dots$ The proof remains unchanged otherwise.
\begin{lemma}
	Let $z^*$ be as in Lemma \ref{lemma:Ax^*=z^*,qTx^*=w^*} and $x^0 \in \cX$. For any $x\in \cX$ such that $f(x) \leq f(x^0)$, it holds that
	\begin{align*}
	\| x - \Pi_{\cX^*}(x)\|^2 \leq \kappa \left( f(x) - f^* \right),
	% \label{eq:quad-growth-h(Ax)+<q,x>}
	\end{align*}
	where, same as in Theorem \ref{thm:h(Ax)+<q,x>-lin-conv},
	$\kappa = H_\cX(A)^2 \left( C + 2 G D_A + \frac{2(G^2 + 1)}{\mu} \right)$, 
	%		 $H =\max\left\{ H_\cX\left( \begin{bmatrix} A \\ q^\top \end{bmatrix} \right), \frac{1}{2 \|A\|} \right\}$, 
	$C = f(x^0) - f^*$,\newline 
	$G = \|\nabla h(z^*)\|$, $D_A = \sup_{x, y\in \cX} \|A(x - y)\|$.
	\label{lemma:quad-growth-h(Ax)+<q,x>}
\end{lemma}

%Let $s = \rank(A)$ and $\bar{A} = \begin{bmatrix} A \\ q^\top \end{bmatrix}$. We first assume that $ \rank(\bar{A}) = s+1$. 
Finally, take $L_f = L \|A\|^2$ as a Lipschitz constant of $\nabla f$. By Lemma \ref{lemma:quad-growth-h(Ax)+<q,x>} and \cite[\S 4.1]{karimi2016linear}, it holds that \eqref{eq:std-form-proj-grad-desc} satisfies the {proximal-\PL} inequality \eqref{eq:prox-PL} with \[\Lambda = \lambda = \frac{1}{2\kappa}\] 
for all $x\in \cX$ such that $f(x) \leq f(x^0)$ (in particular, for all $x^t$, $t = 1,2,\dots$). 
%Since $\kappa \geq \frac{2H^2}{\mu}$ and $H \geq \frac{1}{2\|A\|}$, we clearly have
%\[ L_f \geq \frac{L}{4H^2} \geq  \frac{1}{2\kappa} = \bar{\mu}. \]
By Theorem \ref{thm:pg-lin-conv}, PG converges linearly with rate $1 - \frac{\lambda}{\max\{ \Lambda, L_f \}} = 1 - \frac{1}{\max\{ 1, 2\kappa L \|A\|^2 \}}$.
%Let the nonzero singular values of $A$ be $\sigma_1 \geq \dots \geq \sigma_r = \sigma_{\min}(A) > 0$ and those of $\bar{A}$ be $\sigma'_1 \geq \dots \geq \sigma'_s \geq \sigma'_{s+1} = \sigma_{\min}(\bar{A}) > 0$. By Cauchy's Interlacing Theorem, it holds that 
%\[ \sigma'_1\geq \sigma_1 \geq \dots \geq \sigma'_s \geq \sigma_s \geq \sigma'_{r+1} > 0. \]
%Therefore, $\sigma_{\min}(A) \geq \sigma_{\min}(\bar{A})\geq \frac{1}{H}$. 
%Note that $\kappa \geq \frac{2 H^2}{\mu}$, which implies
%\[ L_f = L \|A\|^2 \geq \mu \sigma_{\min}^2(A) \geq \mu \sigma_{\min}^2(\bar{A}) \geq \frac{\mu}{H^2} \geq \frac{2}{\kappa} > \frac{1}{2\kappa},
%\]
%where the third inequality is by Lemma \ref{lemma:min-sigma-vs.-H(A)}.
%Therefore, by Theorem \ref{thm:pg-lin-conv}, PG converges at a rate $1 - \frac{\bar{\mu}}{L_f} = 1 - \frac{1}{2\kappa L \|A\|^2}$. 

%Next, assume $\rank(\bar{A}) = r$, that is, $q = A^\top v$ for some $v\in \RR^r$. In this case, consider $\hat{h}(z) = h(z) + \langle v, z \rangle$, which is $(\mu, L)$-s.c. Therefore, the objective is of the form $\hat{h}(Az)$. By Theorem \ref{thm:h(Ax)+poly_ind}, PG with stepsize $\frac{1}{L \|A\|^2}$ converges at a rate $1 - \frac{\mu}{H_\cX(A)^2 L \|A\|^2 }$

\paragraph{Remark} Lemma \ref{lemma:quad-growth-h(Ax)+<q,x>} shows that QG holds. Similar convergence guarantees can also be derived from other QG-based analysis, e.g., \cite[Corollary 3.7]{drusvyatskiy2018error}.

\subsection{Linear convergence of PG with linesearch} \label{app:pg-ls-details}
First, we consider the more general proximal gradient setup \eqref{eq:min-F=f+g}. Let $L_f$ be a Lipschitz constant of $\nabla f$ and the {Proximal-\PL} inequality \ref{eq:prox-PL} holds with $\Lambda \geq \lambda \geq 0$ for all $x\in \dom g$ such that $F(x)\leq F(x^0)$. 
Let $\alpha\geq 1$, $\beta\in (0,1)$, $\Gamma >0$ (increment factor, decrement factor, upper bound on stepsize, respectively). The linesearch subroutine $\mathcal{LS}_{\alpha, \beta, \Gamma}$ is defined in Algorithm \ref{alg:ls-subroutine}. 

\begin{algorithm}
	\caption{$x_{t+1}, \gamma_t, k_t\leftarrow \mathcal{LS}_{\alpha,\beta, \Gamma}(x, \gamma, k_{\rm prev})$ with parameters $\alpha\geq 1$, $\beta\in (0,1)$, $\Gamma>0$.}
	\label{alg:ls-subroutine}
	\begin{algorithmic}
		\STATE If $k_{\rm prev} = 0$, set $\gamma^{(0)} = \min\{ \alpha\gamma, \Gamma\}$. Otherwise, set $\gamma^{(0)} = \gamma$. 
		\STATE For $k=0, 1, 2, \dots$
		\begin{enumerate}
			\item Compute $x^{(k)} = \prox_{\lambda^{(k)} g}(x - \gamma^{(k)}\nabla f(x) )$.
			\item Break if 
			\begin{align}
			f(x^{(k)}) \leq f(x) + \langle \nabla f(x), x^{(k)} - x\rangle + \frac{1}{2 \gamma^{(k)}} \| x^{(k)} - x\|^2.  \label{eq:ls-break-condition}
			\end{align}
			\item Set $\gamma^{(k+1)} = \beta \gamma^{(k)}$ and continue to $k+1$. 
		\end{enumerate}
		Return $x_{t+1} = x^{(k)}$, $\gamma_t = \gamma^{(k)}$, $k_t = k$.
	\end{algorithmic}
\end{algorithm}

In this way, proximal gradient with linesearch can be described formally as follows: starting from $x^0 \in \dom f$, $\gamma_{-1} = \Gamma$, $k_{-1} = 0$, perform the following iterations
\begin{align*}
(x^{t+1}, \gamma_t, k_t) \leftarrow \mathcal{LS}_{\alpha, \beta, \Gamma}(x^t, \gamma_{t-1}, k_{t-1}), \ \ t = 1, 2, \dots 
%	\label{eq:prox-grad-with-ls}
\end{align*}

Note that \eqref{eq:ls-break-condition} holds for any $\gamma^{(k)} \leq \frac{1}{L_f}$ (see, e.g., \cite[Theorem 10.16]{beck2017first}). Therefore, Algorithm $1$ terminates when $ \gamma^{(0)} \beta^k \leq \frac{1}{L_f}$. This means 
\begin{align}
\gamma_t \geq \tilde{\gamma} :=  \min\left\{ \Gamma, \frac{\beta}{L_f}\right\}. \label{eq:gamma-min-def}
\end{align}
for all $t$. Note that we explicitly include the case of $\Gamma \leq \frac{1}{L_f}$, although in practice $\Gamma$ is often set very large. Clearly,
\[ \Gamma \beta^k \leq \tilde{\gamma}\ \Leftrightarrow \ k \geq \frac{\log \frac{\Gamma}{\tilde{\gamma}}}{\log \frac{1}{\beta}}. \]
Therefore, in Algorithm \ref{alg:ls-subroutine}, the backtracking iteration index satisfies $k_t \leq \frac{\log \frac{\Gamma}{\tilde{\gamma}} }{\log \frac{1}{\beta}}$ for all $t$. Note that if the loop breaks at $k_t$, the number of $\prox$ evaluations is exactly $k_t + 1$.

%Suppose that the Proximal-PL inequality \ref{eq:prox-PL} holds with $\bar{\mu}\leq L_f$. 
%Since $\frac{1}{\tilde{\gamma}} \geq \frac{L_f}{\beta} \geq L_f$, by the monotonicity of $\cD(x, \cdot)$, we have
%\[ \frac{1}{2}\cD\left( x, \frac{1}{\tilde{\gamma}} \right) \geq \frac{1}{2}\cD(x, L_f) \geq \bar{\mu} (F(x) - F^*). \]
Let 
\begin{align}
\bar{L} = \max\left\{\frac{1}{\tilde{\gamma}}, \Lambda\right\} = \max\left\{ \frac{1}{\Gamma}, \frac{L_f}{\beta}, \Lambda \right\}. \label{eq:def-bar-L-triple-max}
\end{align}
Then, monotonicity of $D(x, \cdot)$ implies, for all $x\in \dom g$ such that $F(x)\leq F(x^0)$, 
\[\frac{1}{2} \cD(x, \bar{L}) \geq \frac{1}{2}\cD(x, \Lambda) \geq \lambda\left( F(x) - F^* \right). \]

Following the proof of Theorem \ref{thm:pg-lin-conv} (or that of \cite[Theorem 5]{karimi2016linear}), we have
\begin{align*}
F(x^{t+1}) & \leq F(x^t) + \langle  \nabla f(x^t), x^{t+1}-x^t\rangle + \frac{L_f}{2}\|x^{t+1}-x^t\|^2 + g(x^{t+1}) - g(x^t) \\
& \leq F(x^t) + \langle  \nabla f(x^t), x^{t+1}-x^t\rangle + \frac{\bar{L}}{2}\|x^{t+1}-x^t\|^2 + g(x^{t+1}) - g(x^t) \\
& \leq F(x^t) - \frac{1}{2\bar{L}}\cD\left(x^t, \bar{L}\right) \\
& \leq F(x^t) - \frac{\lambda}{\bar{L}}(F(x^t) - F^*) \\
\Rightarrow \ \ & F(x^{t+1}) - F^* \leq \left(1 - \frac{\lambda}{\bar{L}}\right) \left( F(x^t) - F^* \right). 
\end{align*}
Summarizing the above discussion, we have the following convergence guarantee for PG with linesearch.
\begin{theorem}
	Let $\alpha\geq 1$, $\beta\in (0,1)$ and $\Gamma>0$. For problem \eqref{eq:min-F=f+g} satisfying the Proximal-\PL inequality with $\Lambda \geq \lambda > 0$ for all $x\in \dom g$ such that $F(x)\leq F(x^0)$, proximal gradient \eqref{eq:pg-iter} with linesearch subroutine $\mathcal{LS}_{\alpha, \beta, \Gamma}$ described in Algorithm \ref{alg:ls-subroutine} generates iterates $x^t$ such that
	\begin{align}
	F(x^{t+1}) - F^* \leq \left(1 - \frac{\lambda}{\bar{L}} \right)^t \left( F(x^0)  - F^*\right),\ \ t = 1, 2, \dots, \label{eq:prox-grax-f+g-with-ls-lin-conv}
	\end{align}
	where $\bar{L}$ is defined in \eqref{eq:def-bar-L-triple-max}. Furthermore, each iteration requires at most $1 + \frac{\log \frac{\Gamma}{\tilde{\gamma}}}{\log \frac{1}{\beta}}$ number of $\prox$ evaluations.
	\label{thm:prox-grad-f+g-with-ls-lin-conv}
\end{theorem}

\textit{Proof of Theorem \ref{thm:pg-ls-conv}.} In the above discussion, when $g(x) = \delta_\cX(x)$, we can replace the Lipschitz constant $L_f$ by the restricted one $L_f^\cX$ throughout, since Algorithm \ref{alg:ls-subroutine} ensures $x^t \in \cX$ for all $t$. It remains to apply Theorem \ref{thm:prox-grad-f+g-with-ls-lin-conv}. For $q = 0$, $\Lambda = \lambda = \frac{\mu}{H^2}$ and $\bar{L} = \max\left\{ \frac{1}{\Gamma}, \frac{L_f^\cX}{\beta}, \frac{\mu}{H^2} \right\}$. Therefore, the rate is 
\[ 1 - \frac{\lambda}{\bar{L}} = 1 - \frac{\mu}{\max\{\mu,  H^2/\Gamma,  H^2L_f^\cX/\beta  \}}. \]
For $q\neq 0$, $\Lambda = \lambda = \frac{1}{2 \kappa}$ and $\bar{L} = \max\left\{ \frac{1}{\Gamma}, \frac{L_f^\cX}{\beta}, \frac{1}{2\kappa} \right\}$. Therefore, the rate is
\[1 - \frac{1}{\max\{1, 2\kappa L_f^\cX /\beta, 2\kappa/\Gamma \}}. \]
\qed

\subsection{Other utility functions}\label{app:CD-and-CES}
%Let $a \circ b$ denote element-wise product, $a^\rho$ denote element-wise power and ${\rm Diag}(u)$ denote diagonal matrix with diagonal elements $u = (u_i)$.
Recall that, by Theorem \eqref{thm:eg-gives-me-for-certain-ui}, for any CCNH utilities $u_i$, optimal solutions to the EG convex program \eqref{eq:eisenberg-gale-primal} correspond to equilibrium allocation and prices.

\textbf{CES utilities} are parametrized by a nondegenerate $v$ and exponent $\rho\in (-\infty, 1]\backslash \{ 0\}$:
\[ u_i(x_i) = \left(\sum_{j=1}^m v_{ij}x_{ij}^\rho \right)^{1/\rho}.  
\]
Clearly, $\rho=1$ gives linear utilities. For $\rho < 1$, it has been shown that Proportional Response dynamics achieves linear convergence in prices and utilities \cite[Theorem 4]{zhang2011proportional} under their notion of $\epsilon$-approximate market equilibrium \cite[pp. 2693]{zhang2011proportional}. 
%Let the EG objective (in minimization form) be $f(x) = -\sum_i B_i \log u_i(x_i)$ \eqref{eq:eisenberg-gale-primal}, which is clearly twice differentiable on $\interior(\cX)$. 
%Let $z_i = \sum_j v_{ij}x_{ij}^\rho$. 
%Straightforward calculation gives
%\[ f'_{z_i} = -\frac{B_i}{\rho z_i},\ \ f^{''}_{z_i} = \frac{B_i}{\rho z_i^2},\ \ \nabla_{x_i} z_i = \rho\cdot v_i \circ x_i^{\rho - 1}, \ \ \nabla^2_{x_i} z_i = \rho (\rho - 1) \cdot {\rm Diag}(v_i \circ x_i^{\rho -2 }). \]
%Therefore, 
%\begin{align*}
%	\nabla^2_{x_i} f(x) &= f''_{z_i} \cdot ( \nabla_{x_i}z_i)(\nabla_{x_i}z_i)^\top + f'_{z_i} \cdot \nabla^2_{x_i}z_i \\
%	& = \frac{\rho B_i}{z_i^2}\cdot ( v_i \circ x_i^{\rho - 1})(v_i \circ x_i^{\rho - 1})^\top + \frac{B_i (1 - \rho)}{z_i}\cdot {\rm Diag}(v_i \circ x_i^{\rho -2 }). 
%\end{align*}
%In this case, the equilibrium utilities $u^*$ are unique and the proportionality lower bound (c.f. Lemma \ref{lemma:eg-linear-u_min-u_max} and Appendix \ref{app:eg-linear-u_min-u_max}) holds:
%\[ 
%u_i^* \geq  u_i\left( \frac{B_i}{\|B\|_1}\ones\right) = \left(\frac{v\circ B^\rho}{\|B\|_1}\right)^{1/\rho}. 
%\]
%Therefore, we can use quadratic extrapolation \S \ref{sec:linear-utilities} to obtain an objective function with 
%

%when $v$ is nondegenerate, the objective of  \eqref{eq:eisenberg-gale-primal} is strongly convex. Therefore, PG converges at a linear rate by classical analysis. 
\textbf{Cobb-Douglas utilities} represent substitutive items and take the following form, for parameters $\lambda = (\lambda_i)$, $\lambda_i \in \Delta_m$:
\[ u_i(x_i) = \Pi_j x_{ij}^{\lambda_{ij}}. \]  
In this case, EG \eqref{eq:eisenberg-gale-primal} decomposes item-wise into simple problems with explicit solutions. Specifically, for each item $j$, the minimization problem is 
\[\min_{x_{:, j} \in \Delta_n} - \sum_i B_i \lambda_{ij} \log x_{ij}. \]
Let $p_j$ be the Lagrangian multiplier associated with constraint $\sum_i x_{ij} = 1$. The Lagrangian is
\[ \mathcal{L}(x_{:, j}, p_j) = -\sum_i B_i \lambda_{ij} \log x_{ij} + p_j\left( \sum_i x_{ij} - 1\right). \]
By first-order stationarity condition, for any $p_j\in \RR$, $\mathcal{L}(x_{:, j}, p_j)$ is minimized when 
\begin{align}
	x_{ij} =\frac{B_i \lambda_{ij}}{p_j}. \label{eq:cobb-douglas-x-in-p}
\end{align}
Substituting it into $\mathcal{L}$ and discarding the constants w.r.t. $p_j$, we have
\[ g(p_j) = \left(\sum_i B_i \lambda_{ij}\right) \log p_j - p_j, \]
which is maximized at equilibrium prices
\[ p^*_j  = \sum_i B_i \lambda_{ij}. \]
Therefore, by \ref{eq:cobb-douglas-x-in-p}, the equilibrium $x^*$ under Cobb-Douglas utilities is given by
\[ x^*_{ij} = \frac{B_i \lambda_{ij}}{\sum_i B_i \lambda_{ij}},\ \forall\, i,j. \]

\section{Linear utilities}
\subsection{Shmyrev's convex program} \label{app:shmyrev-cp}
Under linear utilities, it turns out that we can also compute market equilibrium via the following convex program due to Shmyrev \cite{shmyrev2009algorithm, birnbaum2011distributed}. In this convex program, the variables are the \textit{bids} $b_{ij}$, $i\in [n]$, $j\in [m]$ and prices $p_j$, $j\in [m]$. 
\begin{align}
\max\, \sum_{i, j} b_{ij} \log v_{ij} - \sum_j p_j \log p_j\ \ {\rm s.t.}\ \sum_i b_{ij} = p_j,\ j \in [m],\ \sum_j b_{ij} = B_i,\ i\in [n], \ b\geq 0. \label{eq:shmyrev}
\end{align}
Given an optimal solution $b^*$, equilibrium prices and allocations are then given by $p^*_j = \sum_i b^*_{ij}$ and $x^*_{ij} = \frac{b^*_{ij}}{p^*_j}$, respectively.

\subsection{Proof of Lemma \ref{lemma:eg-linear-u_min-u_max}}\label{app:eg-linear-u_min-u_max}
Any $x\in \cX$ satisfies $x\leq 1$. Therefore, $\langle v_i, x_i\rangle \leq \|v_i\|_1 \|x^*_i\|_\infty \leq \|v_i\|_1 = \bar{u}_i$. For the lower bound, recall that at an equilibrium allocation $x^*$ ensures that every buyer gets at least the utility of the proportional share, that is, 
\[ \langle v_i, x^*_i\rangle \geq \left \langle v_i, \frac{B_i}{\|B\|_1} \ones\right \rangle = \frac{B_i\|v_i\|_1}{\|B\|_1} = \ubar{u}_i. \]

\subsection{Uniqueness of equilibrium quantities and convergence of $u^t$, $p^t$} \label{app:eg-convergence-of-simplex-prices}

Convergence of $u^t$ to $u^*$ can be easily seen as follows. Let $x^t$ be the PG iterates and $\tilde{h}$, $A$, $f = \tilde{h}(Ax)$, $\mu$ be defined as in \S \ref{sec:linear-utilities} and $f^* = \min_{x\in \cX} f(x)$. Since $\tilde{h}$ is $\mu$-strongly convex, we have
\[ \frac{\mu}{2}\|u^t - u^*\|^2\leq \tilde{h}(u^t) - \tilde{h}(u^*) \leq \tilde{h}(Ax^t) - f^*, \]
which converges linearly. Next, we show uniqueness of $p^*$ via simple arguments and construct a sequence of linearly convergent prices $p^t$.

\begin{lemma}
	Assume that $v$ is nondegenerate. Then, the equilibrium prices $p^*$ under linear utilities are unique. \label{lemma:unique-p*}
\end{lemma}
\textit{Proof.} By Theorem \ref{thm:eg-gives-me-for-certain-ui} and \cite[Lemma 3]{cole2017convex}, $p^*$ is an optimal solution (together with some $\beta^*$) to the following problem (dual of \eqref{eq:eisenberg-gale-primal} with linear utilities): \qed
\begin{align}
	 \min_{p,\, \beta}\ \sum_j p_j - \sum_i B_i \log \beta_i \ \ {\rm s.t.}\  p \geq 0,\, \beta\geq 0,\, p_j \geq v_{ij}\beta_i,\, \forall\, i,j.  \label{eq:eg-dual-p-beta}
\end{align}
Here, strong duality holds since there clearly exist primal and dual strictly feasible solutions with finite objective values given nondegenerate $v$ (c.f. Theorem \ref{thm:eg-gives-me-for-certain-ui} and Appendix \ref{app:thm-eg-and-mkt-equi}). We can eliminate $p$ by letting $p_j = \max_i v_{ij}\beta_i$ for all $j$ and rewrite \eqref{eq:eg-dual-p-beta} as 
\[ \min_\beta\, \sum_j \max_i v_{ij}\beta_j - \sum_i B_i \log \beta_i\ \ {\rm s.t.}\ \beta\geq 0. \]
In the above, since the objective is strongly convex and the feasible region is $\beta\geq 0$, the optimal solution $\beta^*$ is clearly unique. Furthermore, it must hold that $\beta^* >0$ (since the optimal objective value is finite and strong duality holds). For $p^*$ optimal to \eqref{eq:eg-dual-p-beta}, it must hold that $p^*_j = \max_i v_{ij}\beta^*_i$. In fact, $p^*_j \geq \max_i v_{ij}\beta^*_i$ by feasibility and, for any strict inequality, decreasing the corresponding $p^*_j$ strictly decreases the objective. \qed

The following lemma provides simple upper and lower bounds on feasible and equilibrium prices, respectively. The lower bounds are slightly strengthened over the existing one \cite[Lemma 17]{birnbaum2011distributed}.
\begin{lemma}
	Let $p^*$ be equilibrium prices under linear utilities with nondegenerate valuations $v$. Then, $\ubar{p}_j \leq p^*_j \leq \bar{p}_j$ for all $j$, where $\ubar{p}_j = \max_i \frac{v_{ij} B_i}{\|v_i\|_1}$ and $\bar{p}_j = \|B\|_1$.  \label{lemma:linear-pmin-pmax}
\end{lemma}
\textit{Proof.} It is essentially the same as the proof of Lemma \ref{lemma:ql-shmyrev-upper-lower-bounds}, except that, at optimality, $u_i \leq \|v_i\|_1 + B_i$ can be strengthened to $u_i \leq \|v_i\|_1$ (utility of each buyer is at most that of having a unit of every item). \qed

\paragraph{A linearly convergent sequence of $p^t$.} Here, all norms are vector norms. Note that each step of PG is of the form $x^{t+1} = \Pi_\cX(\bar{x}^t)$, where $\bar{x}^t = x^t - \gamma \nabla f(x^t)$. Since $\nabla f$ is $L_f$-Lipschitz, the mapping \[\phi_1: x \mapsto x- \gamma \nabla f(x^t)\] 
is Lipschitz continuous (w.r.t. $\|\cdot\|_2$) with constant $1 + \gamma L_f = 2$ (where $\gamma = \frac{1}{L \|A\|^2}$ is the fixed stepsize). Meanwhile, we have the following. 
\begin{lemma}
	Let $y \in \RR^n$ and $y^* = \Pi_{\Delta^n}(y)$. There exists a unique multiplier $\lambda \in \RR$, which can be computed in $O(n \log n)$ time, such that
	\begin{align}
	\sum_{i=1}^n (y_i - \lambda)_+ = 1. \label{eq:simplex-unique-multi}
	\end{align}
	Moreover, the mapping $\phi_2: y \mapsto \lambda$ is 
	piecewise linear and 
	$1$-Lipschitz continuous w.r.t. $\|\cdot \|_1$. \label{simplex-projection-multiplier-Lipschitz}
\end{lemma}
\textit{Proof.} 
By the KKT conditions for simplex projection (see, e.g., \cite[\S 3]{wang2013projection}), it holds that there exists unique $\lambda$ such that
\[ y^* = (y - \lambda \ones)_+.  \]
%Note that $\Pi_{\Delta_n}(\cdot)$ is nonexpansive ($1$-Lipschitz) \cite[Theorem 6.4.2. (b)]{beck2017first}. Meanwhile, since $y^* = (y - \lambda \ones)_+$, there exists $i$ such that $y^*_i = y_i - \lambda$ (otherwise $y_i < \lambda$ for all $i$ and $y^*_i = 0$, contradicting to $y^*\in \Delta_n$). In fact, we can take $i = \argmax_{\ell} y_\ell$. 
Suppose there exists $\lambda_1 < \lambda_2$ that satisfy \eqref{eq:simplex-unique-multi}. Then, since the left-hand side of \eqref{eq:simplex-unique-multi}, denoted as $w(\lambda)$, is monotone decreasing in $\lambda$, it must hold that $w(\lambda)=1$ for all $\lambda\in [\lambda_1, \lambda_2]$. In other words, $w(\cdot)$ is \textit{constant} on $[w_1, w_2]$. This further implies $w(\lambda) = 0$ for all $w\in [w_1, w_2]$, a contradiction. Therefore, $\lambda = \phi_2(y)$ is uniquely defined. Let $I^+(y)$, $I^0(y)$, $I^-(y)$ denote the set of indices $i\in [n]$ such that $y_i > \lambda$, $y_i = \lambda$, $y_i < \lambda$, respectively (where $\lambda = \phi_2(y)$). We have 
\[ \lambda = \frac{\sum_{i\in I^+(y)} y_i - 1}{ |I^+(y)|} = \frac{\sum_{i\in I^+(y)\cup I^0(y)} y_i - 1}{|I^+(y)| + |I^0(y)|}, \]
which is piecewise linear in $y$ since there are only finitely many index possible sets of indices and $I^+(y)$ is always nonempty (otherwise $\sum_i (y_i - \lambda)_+ = 0$). To see Lipschitz continuity, let $y'$ be such that $\|y'-y\|_1 \leq \epsilon$, where $0<\epsilon < \min \{|y_i - y_j|: i,j\in [n],\, y_i \neq y_j \}$. It must hold that $I^+(y)\subseteq I^+(y')$. In other words, $\lambda' = \phi_2(y')$ does not deactivate any $i\in I^+(y)$, only bringing new $i\in I^0(y)$. Hence, it holds that $|\lambda' - \lambda| \leq \frac{\|y - y'\|_1}{|I^+(y)|} \leq \|y-y'\|_1$. In other words, $\phi_2$ is $1$-Lipschitz continuous w.r.t. $\|\cdot \|_1$. 

Finally, \cite[Algorithm 1]{wang2013projection}) computes $\lambda$ and $y^*$ in $O(n \log n)$ time. 
\qed

Abusing the notation, let $\phi_2$ also denote the mapping from $x\in \RR^{n\times m}$ to $\lambda\in \RR^m$, that is, $\lambda_j = \varphi_2(x_{1j}, \dots, x_{nj})$. 
Let \[\phi(x) = \phi_2 (\phi_1(x))/\gamma\] 
and $p^t = \phi(x^t)$. Here, $\phi_1$ is $2$-Lipschitz continuous and $\phi_2$ is $1$-Lipschitz continuous w.r.t. $\|\cdot \|_1$. For any optimal solution $x^* \in \cX^*$, by $x^* = \Pi_\cX (x^*)$ and KKT conditions for \eqref{eq:eisenberg-gale-primal} and \eqref{eq:eg-dual-p-beta}, it can be seen that 
\[p^* = \phi(x^*).\] 
%Since $h$ is strongly convex, it can be expressed as $\cX^* = \cX \cap \cS$, where $\cS = \{ x: Ax = u^*\}$ for some unique $u^*$ (i.e., $\langle v_i, x_i \rangle = u^*_i$). Therefore, by Theorem \ref{thm:pg-for-eg-lin-conv}, we have
Using the Lipschitz continuity properties of $\phi_1, \phi_2$ and Theorem \ref{thm:h(Ax)+poly_ind}, we have
\begin{align*}
\|p^t - p^*\|_1 &= \|\phi(x^t) - \phi( \Pi_{\cX^*}(x^t))\|_1 \leq \frac{1}{\gamma}\| \phi_1(x^t)  - \phi_1(\Pi_{\cX^*}(x^t)) \|_1
\\ & \leq \frac{n}{\gamma} \|\phi_1(x^t) - \phi_1(\Pi_{\cX^*}(x^t))\|  \leq \frac{2n}{\gamma} \cdot \|x^t - \Pi_{\cX^*}(x^t)\| \\
& \leq \frac{2n }{\gamma} \cdot \sqrt{\frac{2 H_\cX(A)}{\mu} \left( f(x^t) - f^* \right) } \\
& \leq \frac{2n}{\gamma} \sqrt{ \frac{2 H_\cX(A)}{\mu}}\cdot  \left( 1 - \frac{\mu}{2 H L \|A\|^2}\right)^{t/2} \cdot \sqrt{f(x^0) - f^*}. 
%& \leq \frac{2n H_\cX(A) }{\gamma} \cdot \| Ax^t - u^* \| \leq \frac{2n H_\cX(A)}{\gamma} \sqrt{\frac{2}{}}
\end{align*}
Therefore, we can take $C = \frac{2n}{\gamma} \sqrt{ \frac{2 H_\cX(A)}{\mu}} \cdot \sqrt{f(x^0) - f^*}$ and $\rho = \sqrt{1 - \frac{\mu}{2 H L \|A\|^2}} \in (0,1)$.

Since $p^* \geq \ubar{p} > 0$, we can bound the maximum relative price error $\eta^t = \max_j \frac{|p^t_j - p^*_j|}{p^*_j}$ as follows, where $\ubar{p}_{\min} = \min_j \ubar{p}_j$. 
\[ \eta^t \leq \frac{\|p^t - p^*\|_1}{\ubar{p}_{\min}} \leq \frac{C}{\ubar{p}_{\min}} \cdot \rho^t. \]
In other words, $\eta^t$ converges (R-)linearly to zero.  

%Since $f(x^t) - f^*$ converges linearly, 
%where $\mu = \min_i \frac{B_i}{\|v_i\|_1^2}$ is the strong convexity modulus of $h$ on $\cX$ and $\rho = 1 - \frac{1}{H^2 L\|v\|_F^2}$ is the linear convergence rate. Combining the above two inequalities, we have 
%\[ \|p^t - p^*\| \leq \frac{2^{3/2} n HL  \|v\|_F^2 \left(1 - \frac{1}{H^2 L \|v\|_F^2} \right)^{t/2} }{\sqrt{\mu}}(f(x^0) - f^*),\ t = 1,2, \dots \]
\section{QL utilities}
\subsection{Derivation of the QL-Shmyrev convex program \eqref{eq:ql-shmyrev}} \label{app:ql-shmyrev-dual}
In \cite[Lemma 5]{cole2017convex}, the convex program for the equilibrium prices is as follows:
\begin{align}
\min \, \sum_j p_j - \sum_i B_i \log \beta_i \ \ 
{\rm s.t.} \ v_{ij}\beta_i \leq p_j,\ \forall\, i,j,\ \ 0 \leq \beta \leq 1. \label{eq:ql-dual-p-beta}
\end{align}
Note that it is simply the dual of EG under linear utilities \eqref{eq:eg-dual-p-beta} with additional constraints $\beta\leq 1$.
Assuming $v$ is nondegenerate, by a change of variable and Lagrange duality, we can derive the dual of \eqref{eq:ql-dual-p-beta}. First, at optimality, it must holds that $\beta_i>0$ for all $i$. Therefore, by nondegeneracy of $v$, $p_j > 0$ for all $j$ at optimality. Let $p_j = e^{q_j}$ and $\beta_i = e^{-\gamma_i}$. The above problem is equivalent to 
\begin{align}
\begin{split}
	\min \  & \sum_j e^{q_j} + \sum_i B_i \gamma_i \\
	\ {\rm s.t.}& \quad q_j + \gamma_i \geq \log v_{ij}, \ \forall\, i,j, \\
	& \quad \gamma \geq 0. 
\end{split} \label{eq:ql-dual-exponentiated}
\end{align}
Let $b_{ij} \geq 0$ be the dual variable associated with constraint $q_j + \gamma_i \geq \log v_{ij}$. The Lagrangian is
\begin{align*}
	L(q, \gamma, b) & := \sum_j e^{q_j} + \sum_i B_i \gamma_i - \sum_{i, j} b_{ij}\left( q_j + \gamma_i - \log v_{ij} \right) \\
	& =  \sum_j \left( e^{q_j} -\left(\sum_i b_{ij}\right) q_j \right) + \sum_i (B_i - \sum_j b_{ij}) \gamma_i + \sum_{i, j} (\log v_{ij}) b_{ij}. 
\end{align*}
Clearly, when $\sum_j b_{ij} \leq B_i$ for all $i$, $\gamma\geq 0$, $L(q, \gamma, b)$ is minimized at $q_j = \log \sum_i b_{ij}$ and $\gamma = 0$. When $\sum_j b_{ij} > B_i$ for some $i$, $L \rightarrow -\infty$ as $\gamma_i \rightarrow \infty$. Therefore, when $\sum_j b_{ij} \leq B_i$ for all $i$, we have
\[ g(b) = \sum_j \left[\sum_{i, j} b_{ij} - \left( \sum_i b_{ij}\right) \log \sum_i b_{ij}  \right] + \sum_{i, j} (\log v_{ij}) b_{ij}. \]
Therefore, the dual is
\[ \max\, g(b) \ \ {\rm s.t.}\ b\geq 0,\ \sum_j b_{ij} \leq B_i,\ \forall\, i. \]
Adding slack variables $\delta = (\delta_1 \dots, \delta_n)$ and writing it in minimization form yield \eqref{eq:ql-shmyrev}. 
\paragraph{Remark.} As mentioned in \S \ref{sec:ql-utilities}, when some $v_{ij} = 0$ (but $v$ is still nondegenerate), by the above derivation, the first summation in \eqref{eq:ql-shmyrev} should be replaced by $\sum_{(i,j)\in \mathcal{E}}$, where $\mathcal{E} = \{(i,j): v_{ij}>0\}$. The dual remains the same otherwise.

\subsection{Proof of Lemma \ref{lemma:ql-shmyrev-upper-lower-bounds}}\label{app:proof-ql-shmyrev-bounds}
Similar to the proof of Lemma \ref{lemma:unique-p*}, this can be seen via the uniqueness of the optimal solution $(p^*, \beta^*)$ of \eqref{eq:ql-dual-p-beta}, that is, from uniqueness of $\beta^*$ to that of $p^*_j = \max_i v_{ij}\beta^*_i$. 

Let $(b^*, \delta^*)$ be an optimal solution to \eqref{eq:ql-shmyrev}. Note that strong duality holds for \eqref{eq:ql-dual-exponentiated} and \eqref{eq:ql-shmyrev}, since there exit simple strictly feasible solutions. By the derivation in Appendix \ref{app:ql-shmyrev-dual}, it holds that $q^*_j = \log \sum_i b^*_{ij}$ gives an optimal solution to \eqref{eq:ql-dual-exponentiated} (the first-order optimality condition). Therefore, 
\[p^*_j = e^{q^*_j} = \sum_i b^*_{ij}. \]
%By Lemma \ref{lemma:cole-duality-lemma}, $p^*_j = \sum_i b^*_{ij}$ (the first-order optimality condition).  

Next we establish the upper and lower bounds on $p^*$. By the derivation in Appendix \ref{app:ql-shmyrev-dual} and Lagrange duality, for any optimal solution $b^*$ to \eqref{eq:ql-shmyrev}, it holds that $p^*_j:= \sum_i b^*_{ij}$ and $\beta^*_j = \min_{j\in J_i}\frac{p^*_j}{v_{ij}}$ give the (unique) optimal solution to \eqref{eq:ql-dual-p-beta}. Clearly, $\beta^*\leq 1$ and therefore
\[ p^*_j = \max_i v_{ij}\beta^*_i \leq \max_i v_{ij} = \bar{p}_j. \]
By \cite[Lemma 5]{cole2017convex}, the dual of \eqref{eq:ql-dual-p-beta} is (c.f. the original EG primal \ref{eq:eisenberg-gale-primal})
\begin{align}
\begin{split}
	 \max_{u,\, x,\, s} & \sum_i B_i \log u_i - s_i \\
	\st & u_i \leq v_i^\top x_i + s_i,\ \forall\, i,\ \\ 
	& \sum_i x_{ij} \leq 1,\ \forall\, j, \\
	&  x, s\geq 0.
\end{split} \label{eq:ql-eg-form}
\end{align}
Clearly, strong duality holds for \eqref{eq:ql-dual-p-beta} and \eqref{eq:ql-eg-form}. Furthermore, notice the following. 
\begin{itemize}
	\item $\beta^*_i = \frac{B_i}{u^*_i}$ at optimality, where $u^*_i$ is the amount of utility of buyer $i$. This is by the stationarity condition in the KKT optimality conditions.
	\item $u^*_i \leq \|v_i\|_1 + B_i$, where the right hand side is the amount of utility of all items and the entire budget. This can also be seen as follows. When $s_i > B_i$, decreasing $s_i$ strictly increases the objective of \eqref{eq:ql-eg-form}. Therefore, the optimal $s^*$ must satisfy $s^*_i \leq B_i$. It then follows from the constraint $u_i \leq v_i^\top x_i + s_i$. 
\end{itemize}
Therefore, 
\[ p^*_j \geq \max_i v_{ij} \beta^*_i \geq \max_i \frac{v_{ij} B_i }{\|v_i\|_1 + B_i} = \ubar{p}_j. \] 
%For the lower bound, consider the following feasible solution of \eqref{eq:ql-dual-p-beta}: $\beta_j = 1$, $p_j = \bar{p}_j = \max_i v_{ij}$. It gives an objective value of $f^0:= \sum_j \bar{p}_j$. Note that 
%\[ -B_i \log \beta_i > f^0 \Leftrightarrow \beta_i <  e^{-f^0/B_i}. \]
%In other words, if $\beta_i < e^{-f^0/B_i}$ for any $i$, the solution is certainly not optimal. Therefore, an optimal solution must satisfy 
%\[ \beta_i = \min_{j\in J_i} \frac{p_j}{v_{ij}} \geq e^{-f^0/B_i} \]
%for all $i$. This means $p_j \geq v_{ij} \cdot e^{-f^0/B_i}$ for all $i,j$. Therefore, 
%\[ p_j \geq \max_{i\in I_j} \left\{v_{ij} \cdot e^{-f^0/B_i}\right\}. \]

\subsection{Proof of Theorem \ref{thm:pr-ql-shmyrev-1/T}} \label{app:proof-MD-1/T}
Similar to \cite[Lemma 7]{birnbaum2011distributed}, we first establish the following ``generalized Lipschitz condition'' for $\varphi$, which is key to the claimed last-iterate convergence.
\begin{lemma}
	For all $(b, \delta), (b', \delta')\in \cB$, it holds that 
	\begin{align}
	\varphi(b') \leq \varphi(b) + \langle \nabla \varphi(b), b' - b \rangle + D(b', \delta'\| b, \delta). \label{eq:generalized-Lipschitz}
	\end{align}
	\label{lemma:generalized-Lipschiz}
\end{lemma}
\textit{Proof.} 
Recall that $p_j(b) = \sum_i b_{ij}$, $\frac{\partial}{\partial b_{ij}} \varphi(b) = \log \frac{p_j(b)}{v_{ij}}$. For $(a, \delta^a), (b, \delta^b)\in \cB$, we have
\begin{align}
& \varphi(b) - \varphi(a) - \langle \nabla \varphi(a), b - a \rangle \nnnl
&= -\sum_{i,j} (1+\log v_{ij})(b_{ij} - a_{ij}) + \sum_j p_j(b)\log p_j(b) - \sum_j p_j(a)\log p_j(a) \nnnl
& \quad \quad - \sum_{i,j} (b_{ij} - a_{ij})\log \frac{p_j(a)}{v_{ij}} \nnnl
&= -\sum_{i,j} (b_{ij} - a_{ij}) + \sum_{j} p_j(b)\log \frac{p_j(b)}{p_j(a)} \nnnl
& = \sum_i (\delta^b_i - \delta^a_i) + \sum_j p_j(b)\log \frac{p_j(b)}{p_j(a)}. \label{eq:bounding-phi(b)-phi(a)-...}
\end{align}
Note that convexity and smoothness of $x \mapsto x\log \frac{x}{y}$ ($y>0$) implies
\begin{align}
\delta^b_i - \delta^a_i \leq \delta_i^b \log \frac{\delta^b_i}{\delta^a_i}. \label{eq:delta(b)-delta(a)<=...}
\end{align}
As in the proof of \cite[Lemma 7]{birnbaum2011distributed}, by convexity of $q(x, y) = x\log \frac{x}{y}$, it holds that
\begin{align}
\sum_j p_j(b)\log \frac{p_j(b)}{p_j(a)} \leq \sum_{i,j} b_{ij}\log \frac{b_{ij}}{a_{ij}}. \label{eq:D(p(b)||p(a))<=D(b||a)}
\end{align}
By \eqref{eq:delta(b)-delta(a)<=...} and \eqref{eq:D(p(b)||p(a))<=D(b||a)}, the right hand side of \eqref{eq:bounding-phi(b)-phi(a)-...} can be bounded by $D(b, \delta^b\| a, \delta^a)$. Therefore, \eqref{eq:generalized-Lipschitz} holds. \qed

Next, we prove the inequality on the right. Clearly, $(b^0, \delta^0)\in \cB$. By \cite[Theorem 3]{birnbaum2011distributed} (with objective $f = \varphi$, constraint set $C = \cB$ and stepsize $\gamma$), we have 
\[ \varphi(b^t) - \varphi(b^*) \leq \frac{D(b^*, \delta^*\|b^0, \delta^0)}{t}. \]
Similar to the proof of \cite[Lemma 13]{birnbaum2011distributed}, we can bound the Bregman divergence on the right hand side as follows, where $b_{ij} = \delta_i = \frac{B_i}{m+1}$. 
\begin{align*}
D(b^*, \delta^*\|b^0, \delta^0) &= \sum_{i,j} b^*_{ij} \log \frac{b^*_{ij} }{B_i} + \sum_i \delta^*_i \log \frac{\delta^*_i}{B_i} + \sum_{i,j} b^*_{ij}\log (m+1) + \sum_i \delta^*_i \log (m+1) \\ 
& \leq \sum_{i,j} b^*_{ij}\log (m+1) + \sum_i \delta^*_i \log (m+1) \\
& \leq \|B\|_1 \log (m+1),
\end{align*}
where the first inequality is because $\frac{b^*_{ij}}{B_i} \leq 1$. 
Combining the above yields the desired inequality.\footnote{In fact, the bound $\log(mn)$ in  \cite[Lemma 13]{birnbaum2011distributed} (which assumes $\|B\|_1 = 1$) can be easily strengthened to $\log m$ via the above derivation. In other words, it does not depend explicitly on the number of buyers (but implicitly through $\|B\|_1$ in general).}

Finally, we show the inequality on the left. By optimality of $(b^*, \delta^*)$, we have \[\langle \nabla \varphi(b^*), b - b^*\rangle \geq 0,\ \ \forall \, (b, \delta)\in \cB. \] 
Recall that $p_j(b) = \sum_i b_{ij}$. By \eqref{eq:bounding-phi(b)-phi(a)-...}, we have
\[  D(p^t\|p^*) = -\sum_{i,j} (b^t_{ij} - b^*_{ij}) + \sum_{j} p_j(b^t)\log \frac{p_j(b^t)}{p_j(b^*)}  \leq \varphi(b^t) - \varphi^*. \]\qed

\subsection{Details from MD \eqref{eq:md-udpate} to PR \eqref{eq:pr-ql-shmyrev-final}} \label{app:details-md-to-pr-ql}
Note that \eqref{eq:md-udpate} is buyer-wise separable: for each $i$, we have (where $\frac{\partial}{\partial b_{ij}} \varphi_b(b) = \log \frac{p_j(b)}{v_{ij}}$ and $\cB_i = B_i \cdot \Delta_{m+1}$)
\begin{align}
(b_i^{t+1}, \delta_i^{t+1}) & = \argmin_{(b_i, \delta_i)\in \cB_i}\, \sum_j \left( \log \frac{p_j(b^t)}{v_{ij}} - \log b^t_{ij}\right) b_{ij} - (\log \delta^t_i)\delta_i + \sum_j b_{ij} \log b_{ij} + \delta_i \log \delta_i \nonumber \\
& = \argmin_{(b_i, \delta_i)\in \cB_i}\, - \sum_j (\log b^t_{ij}) b_{ij} - (\log \delta^t_i)\delta_i + \sum_j b_{ij} \log b_{ij} + \delta_i \log \delta_i.
\label{eq:md-update-each-i}
\end{align}
By Lemma \ref{lemma:md-subproblem}, for all $i,j$, 
\begin{align}
b_{ij}^{t+1} = B_i\cdot \frac{\frac{v_{ij}b^t_{ij}}{p_j(b^t)}}{\sum_\ell \frac{v_{i\ell}b^t_{i\ell}}{p_\ell(b^t)} + \delta^t_i}
\label{eq:formula-b(t+1),delta(t+1)},\ \  \delta^{t+1}_j = B_i\cdot \frac{\delta^t_i}{\sum_\ell \frac{v_{i\ell}b^t_{i\ell}}{p_\ell(b^t)} + \delta^t_i}.
\end{align}
Let $p^t_j = p_j(b^t)$. Then, \eqref{eq:formula-b(t+1),delta(t+1)} can be written in terms of the allocations $x^t_{ij} = b^t_{ij} /p^t_j$ (which sum up to $1$ over buyers $i$ for any item $j$) and leftover $\delta^t_i$, thus giving \eqref{eq:pr-ql-shmyrev-final}. 
%Comparing \eqref{eq:pr-ql-shmyrev-final} with \cite[\S1.2]{birnbaum2011distributed}, it can be viewed as a natural generalization of PR dynamics to QL utilities, in which case the leftover $\delta_i$ is maintained.

%\section{Proof of Theorem \ref{thm:ql-price-unique-converge}} \label{app:proof-ql-price-unique-convergence}

\subsection{Convergence of prices} \label{app:ql-price-conv}
Let $\eta^t = \max_j \frac{|p^t_j - p^*_j|}{p^*_j}$ be the relative price error, which can clearly be bounded by $\frac{\|p^t - p^*\|_1}{\ubar{p}_{\min}}$, where $\ubar{p}_{\min} = \min_j \ubar{p}_j >0$ is given in Lemma \ref{lemma:ql-shmyrev-upper-lower-bounds}. By Theorem \ref{thm:pr-ql-shmyrev-1/T} and strong convexity of KL divergence (w.r.t. $\|\cdot \|_1$), for $b^t$ and $p^t = p(b^t)$ generated by either PG or PR,
\begin{align}
	\frac{1}{2} \|p^t - p^*\|_1^2 \leq D(p^t\|p^*) \leq \varphi(b^t) - \varphi^*. \label{eq:price-1-norm-KL-obj-error}
\end{align}
Therefore, for PG, the quantities $\eta^t$, $\|p^t - p^*\|$ and $D(p^t\|p^*)$ all converge linearly to $0$. For PR, they converge at $O(1/T)$. 
%By Theorem \ref{thm:h(Ax)+<q,x>-lin-conv}, we have $\|\bar{b}^t - \Pi_{\cB^*}(\bar{b}^t)\|^2 \leq \kappa\left( \tilde\varphi(b^t) - \varphi^*\right)$ for all $\bar{b} = (b, \delta)\in \cB$ with $\tilde \varphi(b) \leq \varphi(b^0)$, where $\cB^* = \argmin_{\bar{b}\in \cB} \varphi(b) = \{ \bar{b} = (b, \delta)\in \cB: p(b) = p^* \}$. Since $\tilde\varphi(b^t) - \varphi^*$ converges linearly, so is the distance to equilibrium bids $\|b^t - \Pi_{\cB^*}(b^t)\|$.

%\paragraph{Convergence of $b^t$ and $p^t$}  Let $p^t = p(b^t)$. Linear convergence of $D(p^t\|p^*)$ can be seen from the inequality $D(p^t\|p^*) \leq \varphi(b^t) - \varphi^*$ in Theorem \ref{thm:pr-ql-shmyrev-1/T} and that of $\eta^t$ can be seen from the next paragraph.

We can further bound $\varphi(b^t) - \varphi^*$ by the duality gap. Specifically, given $b^t$, $p^t = p(b^t)$, let 
\[b^t_i = \min\left\{ \min_j \frac{p^t_j}{v_{ij}}, 1 \right\}.\] 
Then, $(p^t, \beta^t)$ is feasible to \eqref{eq:ql-dual-p-beta}. By weak duality,
\begin{align}
	\varphi(b^t) - \varphi^* \leq \varphi(b^t) + g(p^t, \beta^t), \label{eq:ql-shmyrev-dgap-expre}
\end{align}
where $g(p, \beta)$ is the (minimization) objective of \eqref{eq:ql-dual-p-beta}. Combining the above, we have
\[ \eta^t \leq \frac{\sqrt{ 2\left( \varphi(b^t) + g(p^t, \beta^t) \right) }}{\ubar{p}_{\min}}. \]
Note that the above holds for $b^t$ from either PG or PR. Although neat in theory, numerical experiments suggest that the above bound can be loose and is not suitable as a termination criteria.

%
%Under QL utilities, it can be bounded by computable quantities including the duality gap and a lower bound on $p^*$ \cite[Eq. (18)]{birnbaum2011distributed} and \cite[Lemma 9]{zhang2011proportional}, we can also bound $\eta^t = \max_j \frac{|p^t_j - p^*_j|}{p^*_j}$ using computable quantities under QL utilities. First, . We know a uniform lower bound $\ubar{p}_j$ on $p^*_j$ by Lemma \ref{lemma:ql-shmyrev-upper-lower-bounds}. Meanwhile, , where the right hand side can be further bounded, due to weak duality, by the duality gap  Therefore, 
%\begin{align*}
%\eta^t \leq \frac{\sqrt{ 2(\varphi(b^t, \delta^t) - \varphi^*)}}{\min_j \ubar{p}_j} \leq  \frac{\sqrt{ 2(\varphi(b^t, \delta^t)+g(p^t, \beta^t))}}{\min_j \ubar{p}_j}. % \label{eq:bound-ql-price-diff}
%\end{align*}
%Note that it holds for any iterates $(b^t, \delta^t)$ feasible to \eqref{eq:ql-shmyrev} and $(p^t, \beta^t)$ defined as above, regardless of the algorithm used. 

\section{Leontief utilities}
\subsection{Derivation of \eqref{eq:eg-Leontief-dual-reform}}\label{app:leontief-dual}
The primal EG \eqref{eq:eisenberg-gale-primal} under Leontief utilities $u_i(x_i) = \min_{j\in J_i} \frac{x_{ij}}{a_{ij}}$ can be written in both $x$ and $u$:
\begin{align*}
\min_{u,\, x}\ &  -\sum_i B_i \log u_i \\
{\rm s.t.} \ \ &  u_i \leq \frac{x_{ij}}{a_{ij}} ,\, \forall j\in J_i,\ \forall i\in [n],  \\
& \sum_i x_{ij} \leq 1,\ \forall\, j\in [m],\\
& x\geq 0,\ u\geq 0.
\end{align*}
Clearly, it can also be written in terms of $u_i$ only as follows:
\begin{align}
\min\, -\sum_i B_i \log u_i\ \ {\rm s.t.}\ \sum_{i \in I_j} a_{ij} u_i \leq 1,\ \forall\, j,\ u\geq 0. \label{eq:eg-Leontief}
\end{align}
Let $p_j\geq 0$ be the dual variable associated with constraint $ \sum_{i\in I_j} a_{ij} u_i \leq 1 $. The Lagrangian is 
\begin{align*}
\mathcal{L}(u, p) & = -\sum_i B_i \log u_i + \sum_j p_j \left( \sum_{i\in I_j} a_{ij} u_i - 1 \right) \\
&= - \sum_j p_j + \sum_i \left[ -B_i \log u_i + \langle a_i, p \rangle u_i \right].
\end{align*}
Note that minimizing $\mathcal{L}$ w.r.t. $u$ can be performed separably for each $u_i$. For any $i$ such that $\sum_{j\in J_i} p_j>0$, by first-order stationarity condition, the term $-B_i \log u_i + \langle a_i, p \rangle u_i $ is minimized at $u^*_i(p) = \frac{B_i}{\langle a_i, p \rangle}$
with minimum value $B_i(1 - \log B_i) + B_i \log \langle a_i, p \rangle $.
If $\sum_{j\in J_i} p_j = 0$, the term approaches $-\infty$ as $u_i \rightarrow \infty$. Therefore, the dual objective is
\[ g(p) = \begin{cases}
-\sum_j p_j + 
\sum_i B_i \log \langle a_i, p \rangle + \sum_i B_i (1 - \log B_i) & {\rm if}\ p\geq 0\ {\rm and}\ \sum_{j\in J_i} a_{ij} p_j > 0 \\
-\infty & {\rm o.w.}
\end{cases} \]
Hence the (Lagrangian) dual problem is $\max_p g(p)$. Its minimization form, up to the constant $-\sum_i B_i (1 - \log B_i)$, is
\begin{align}
\min \left[\sum_j p_j - \sum_i B_i \log \langle a_i, p \rangle \right] \ \ {\rm s.t.}\ p \geq 0. \label{eq:eg-Leontief-dual}
\end{align}
%In order to make use of Theorem \ref{thm:pg-lin-conv} in solving \eqref{eq:eg-Leontief-dual} to achieve linear convergence, we can reformulate Problem \eqref{eq:eg-Leontief-dual} into the form $h(Ap) + g(p)$, where $h$ is a strongly convex function with Lipschitz continuous gradient, $A$ is a linear map and $g$ is an indicator function of a polyhedral set. We next describe the necessary reformulations.
By Theorem \ref{thm:eg-gives-me-for-certain-ui}, we have the following.
\begin{itemize}
	\item  An optimal solution to \eqref{eq:eg-Leontief-dual} gives equilibrium prices.
	\item A market equilibrium $(x^*, p^*)$ satisfies $\langle p^*, x_i\rangle = B_i$ for all $i$ and $\sum_i x^*_{ij} = 1$ for all $j$. Therefore, we have $\sum_j p^*_j = \|B\|_1$.
\end{itemize}
Therefore, we can add the constraint $\sum_j p_j = \|B\|_1$ to \eqref{eq:eg-Leontief-dual} without affecting any optimal (equilibrium) solution. This leads to \eqref{eq:eg-Leontief-dual-reform}.

\subsection{Proof of Lemma \ref{lemma:r-ubar-r-bar}} \label{app:leontief-bounds-<a(i),p>}
let $p$ be any feasible solution to \eqref{eq:eg-Leontief-dual-reform}. Since $\sum_j p_j = \|B\|_1$, we have $\langle a_i, p \rangle \leq \|a_i\|_\infty \|p\|_1 = \|a_i\|_\infty \|B\|_1 $ for all $i$. By Appendix \ref{app:leontief-dual}, at equilibrium, $p^*$ and primal variables $u^*_i$ satisfy $u^*_i = \frac{B_i}{\langle a_i, p^* \rangle}$ (the stationarity condition) and 
\[u^*_i \leq \textnormal{utility of getting one unit of every item} = \min_{j\in J_i} \frac{1}{a_{ij}} = \frac{1}{\|a_i\|_\infty}\] 
for all $i$. Therefore $\langle a_i, p^* \rangle = \frac{B_i}{u^*_i} \leq \|a_i\|_\infty \|B\|_1$.

\subsection{Linear convergence of utilities} \label{app:leontief-utilities-lin-conv}
Note that the equilibrium utilities $u^*$ are clearly unique by \eqref{eq:eg-Leontief}. By the KKT stationary condition,  
\[ u^*_i = \frac{B_i}{\langle a_i, p^*\rangle},\ \ \forall\, i \]
for equilibrium prices $p^*$. Therefore, an intuitive construction of $u^t$ is as follows. Let $p^t$ be the current iterate, $r_i^t = \langle a_i, p\rangle$. First compute $\tilde{u}^t_i = \frac{B_i}{r^t_i}$. Then, to satisfy the primal constraints $\sum_i u_i a_{ij} \leq 1$, take 
\[ u^t = \frac{\tilde{u}^t}{\max_j \sum_i u_i a_{ij}  } = \frac{\tilde{u}^t}{\|a^\top \tilde{u}\|_\infty}. \]
Let $r^* = \langle a_i, p^*\rangle = \frac{B_i}{u^*_i}$ and $f^* = \argmin_{p\in \cP} \tilde{h}(ap) = \tilde{h}(r^*) = h(r^*)$. Strong convexity of $\tilde{h}$ implies $\frac{\mu}{2}\|r^t - r^*\|^2\leq h(r^t) - f^*$.
Furthermore, the mapping $r^t\mapsto \tilde{u}^t \mapsto u^t$ is Lipschitz continuous on $r^t \in [\ubar{r}, \bar{r}]$. Therefore, $\|u^t - u^*\|$ converges to $0$ linearly as well. 

\section{Additional details on numerical experiments} \label{app:details-experiment}

For linear utilities, we generate market data $v = (v_{ij})$ where $v_{ij}$ are i.i.d. from standard Gaussian, uniform, exponential, or lognormal distribution. For each of the sizes $n=50, 100, 150, 200$ (on the horizontal axis) and $m=2n$, we generate $30$ instances with unit budgets $B_i=1$ and random budgets $B_i = 0.5 + \tilde{B}_i$ (where $\tilde{B}_i$ follows the same distribution as $v_{ij}$). See \S \ref{sec:experiments} for plots under random budgets and below for those under uniform budgets.

The termination conditions (on the vertical axis) are
\[\epsilon(p^t, p^*) \leq \eta,\ \eta = 10^{-2}, 10^{-3},\] 
where $p^*$ is the optimal Lagrange multipliers of \eqref{eq:eisenberg-gale-primal} computed by CVXPY+Mosek. Then, for $n = 100, 200, 300, 400$ and $n=2m$, we repeat the above with termination conditions
\[{\rm dgap}_t/n \leq \eta,\ \eta = 10^{-3}, 10^{-4}, 10^{-5}, 5\times 10^{-6}.\] 

For QL utilities, we repeat the above (same random $v$, same sizes and termination conditions) using budgets $B_i = 5( 1+\tilde{B}_i)$. This is to make buyers have nonzero bids and leftovers (i.e., $0< \delta^*_i < B_i$) at equilibrium in most scenarios. In this case, $p^* = p(b^*)$, where $b^*$ is the optimal solution to \eqref{eq:ql-shmyrev} computed by CVXPY+Mosek. For QL, FW does not perform well in initial trials and is excluded in subsequent experiments.

For the linesearch subroutine $\mathcal{LS}_{\alpha,\beta,\Gamma}$ in PG (see Appendix \ref{app:pg-ls-details}), we use parameters $\alpha=1.02$, $\beta = 0.8$ and $\Gamma = 100 L\|A\|^2$ throughout.

For Leontief utilities, in addition to ${\rm dgap}_t/n\leq \eta$, we also use the termination condition $\epsilon(u^t, u^*) = \max_j \frac{|u^t_j - u^*_j|}{u^*_j} \leq \eta$, where $u^*$ is the optimal solution to EG under Leontief utilities \eqref{eq:eg-Leontief} computed by CVXPY+Mosek.

\paragraph{Computing the duality gap.} For linear utilities, the objective of the original Shmyrev's convex program \eqref{eq:shmyrev} is
\[ \varphi(b) = -\sum_{i,j} (\log v_{ij})b_{ij} +  \sum_j p_j(b)\log p_j(b) \]
where $p_j(b) = \sum_i b_{ij}$. Recall the objective of the (EG) dual \eqref{eq:eg-dual-p-beta}, equivalent to the dual of Shmyrev's \eqref{eq:shmyrev},
\[g(p, \beta) = \sum_j p_j  - \sum_i B_i \log \beta_i. \]
Given iterate $b^t$, let $p^t_j = p_j(b^t)$ and $\beta^t_i = \min_j \frac{p_j}{v_{ij}}$, which is finite since $v$ is nondegenerate and $p^t>0$. The duality gap is computed via
\[ {\rm dgap}_t = \varphi(b^t) + g(p^t, \beta^t). \] 
For QL utilities, it is computed similarly, that is, through \eqref{eq:ql-shmyrev-dgap-expre}. For Leontief utilities, it is computed using the construction in Appendix \ref{app:leontief-utilities-lin-conv}.

%\paragraph{Wall-clock time saving} The experiments are conducted on a Linux compute server with Intel(R) Xeon(R) CPU E5-2620 v4 @ 2.10GHz and 128GB ECC RAM memory. The numerical software is CVXPY 1.0.31 and Mosek 9.2.4 in Python 3.7.6. Consider the set up $n = 500$, $m=1000$, $v_{ij} \sim $ standard Gaussian, $B_i = 0.5 + \tilde{B}_i$ (where $\tilde{B}_i\sim $ standard Gaussian as described above). In table \ref{table:wall-clock-time}, we report the wall-clock solving times of CVXPY+Mosek, PG with linesearch, PR and FW with exact linesearch (averages and standard errors in seconds across 10 repeats). The termination criteria for the FOMs are ${\rm dgap}_t/n\leq 10^{-4}$, which is more than sufficient for practical purposes. Note that the simplex projection algorithm for PG is implemented in Python, which incurs some overhead compared to the highly efficient matrix-vector multiplications in PR. 
%\highlight{Meanwhile, CVXPY also incurs significant processing time on top of Mosek's solving time.}
%
%\begin{table}
%	\centering
%	\begin{tabular}{||c c c c||} 
%		\hline
%		CVXPY+Mosek & PG & PR & FW \\ [0.5ex] 
%		\hline
%		94.1 (2.82) & 38.1 (1.94) & 25.6 (1.30) & 46.2 (0.71) \\ 
%		\hline
%	\end{tabular}
%	\vspace{2.5pt}
%	\caption{Wall-clock times of each algorithm in seconds}
%	\label{table:wall-clock-time}
%\end{table}

\paragraph{Additional plots.} In \S \ref{sec:experiments}, the plots for linear utilities are generated under random $B_i$. Here we present an augmented set of plots under different random market data, different utilities, unit v.s. varying, random budgets $B_i$ and different termination conditions (${\rm dgap}_t/n \leq \eta$ or $\epsilon(p^t, p^*)\leq \eta$). All horizontal axes are market sizes $n$ (with $m=2n$) and all vertical axes are number of iterations. The error vertical bars are the standard errors of the number of iterations across $30$ repeats. The legends are in the subplot ``Linear utilities, dgap$/n\leq$1e-5''.
\begin{center}
	% linear
	\includegraphics[width=0.8\linewidth]{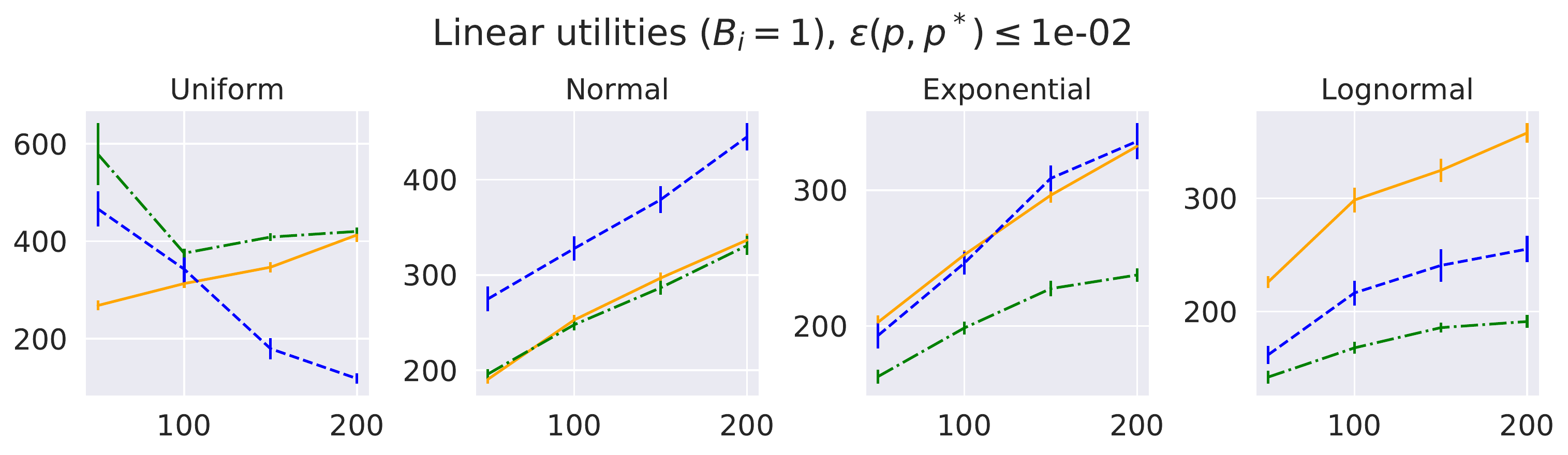}
	\includegraphics[width=0.8\linewidth]{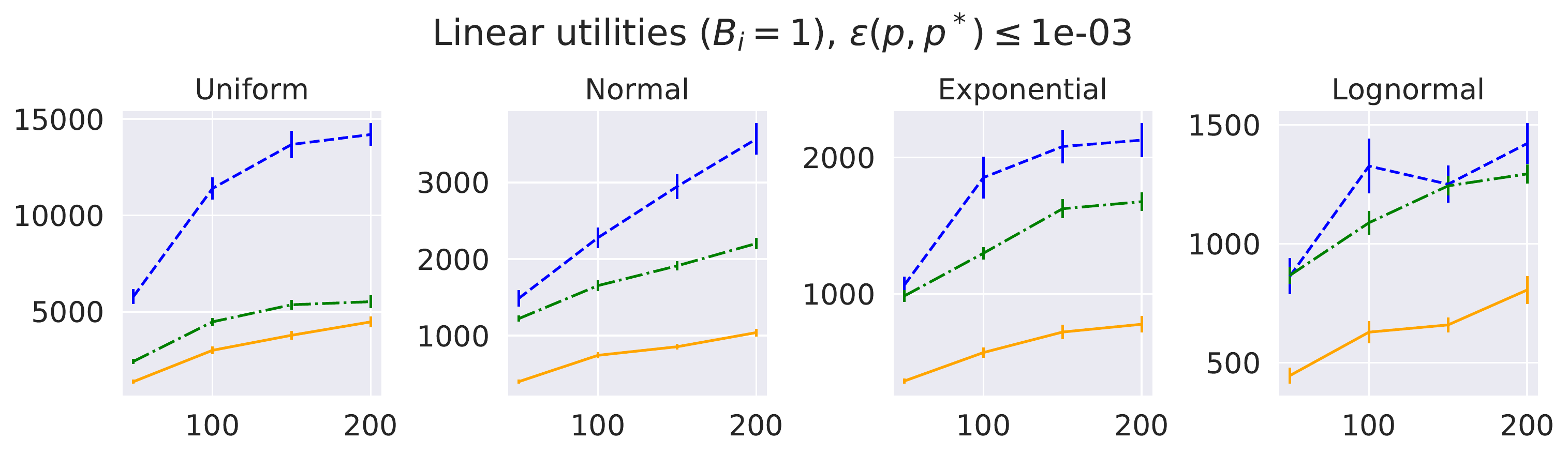}
	\includegraphics[width=0.8\linewidth]{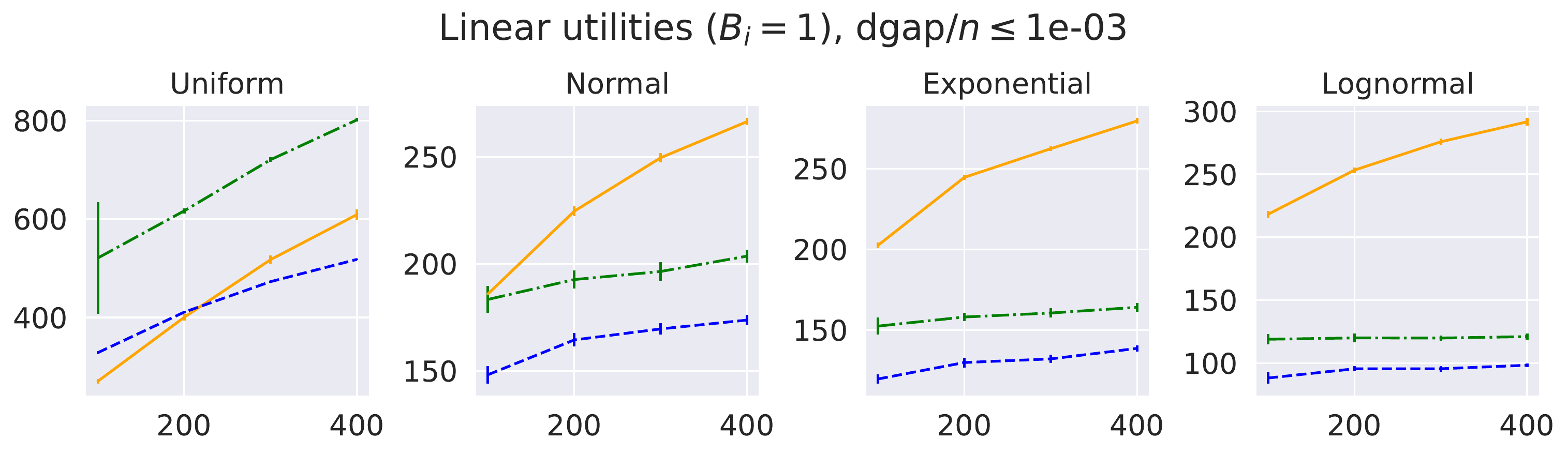}
	\includegraphics[width=0.8\linewidth]{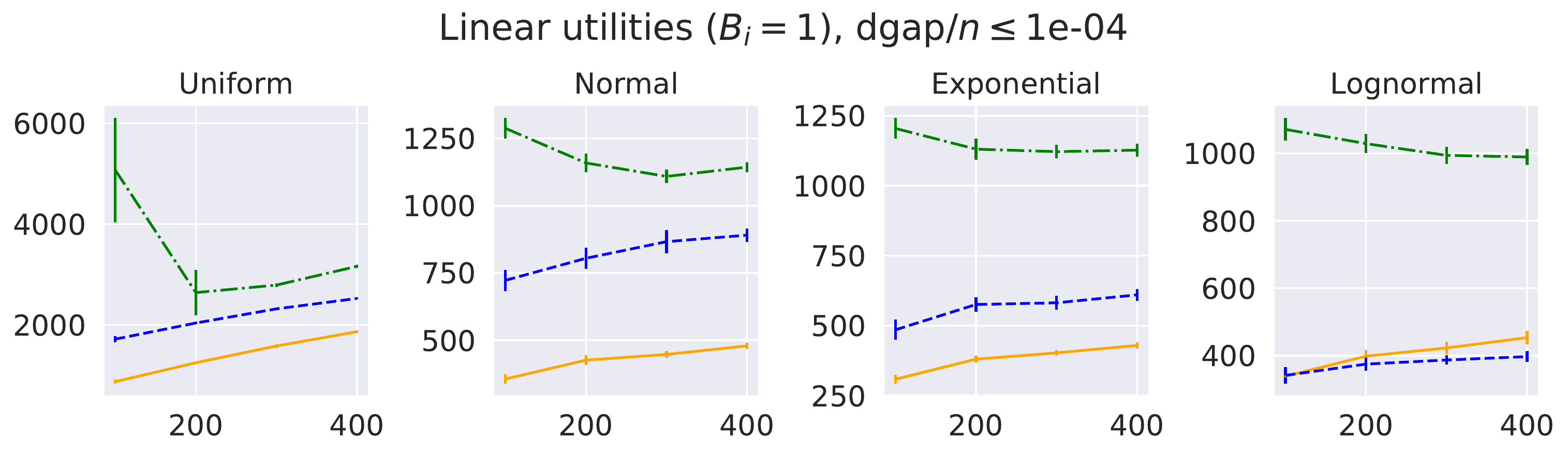} 
	\includegraphics[width=0.8\linewidth]{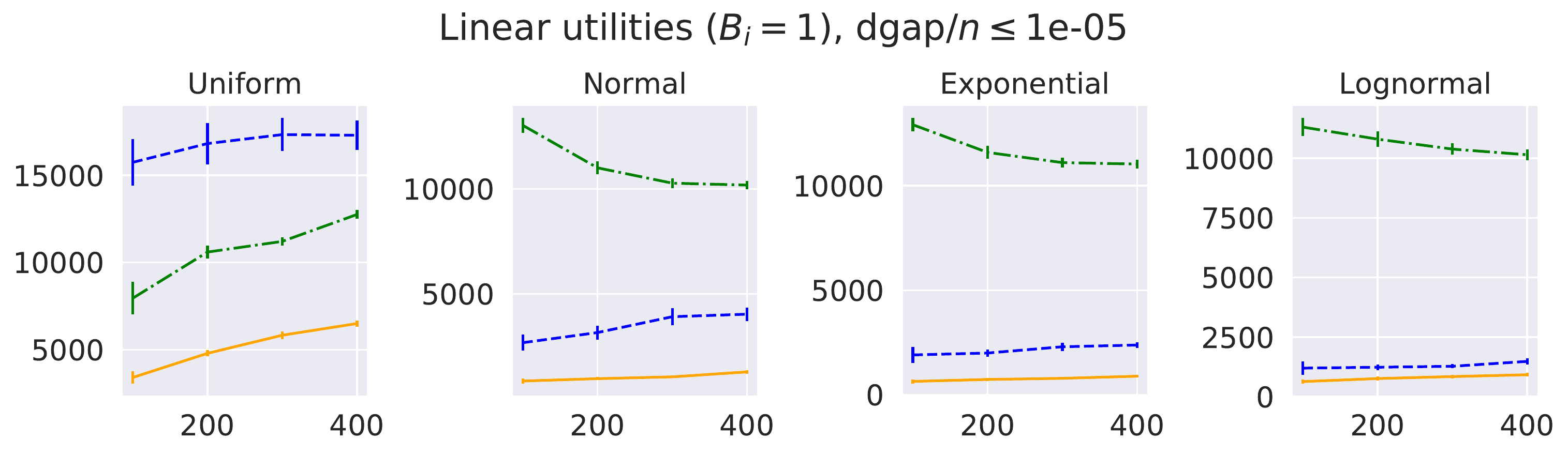}
	\includegraphics[width=0.8\linewidth]{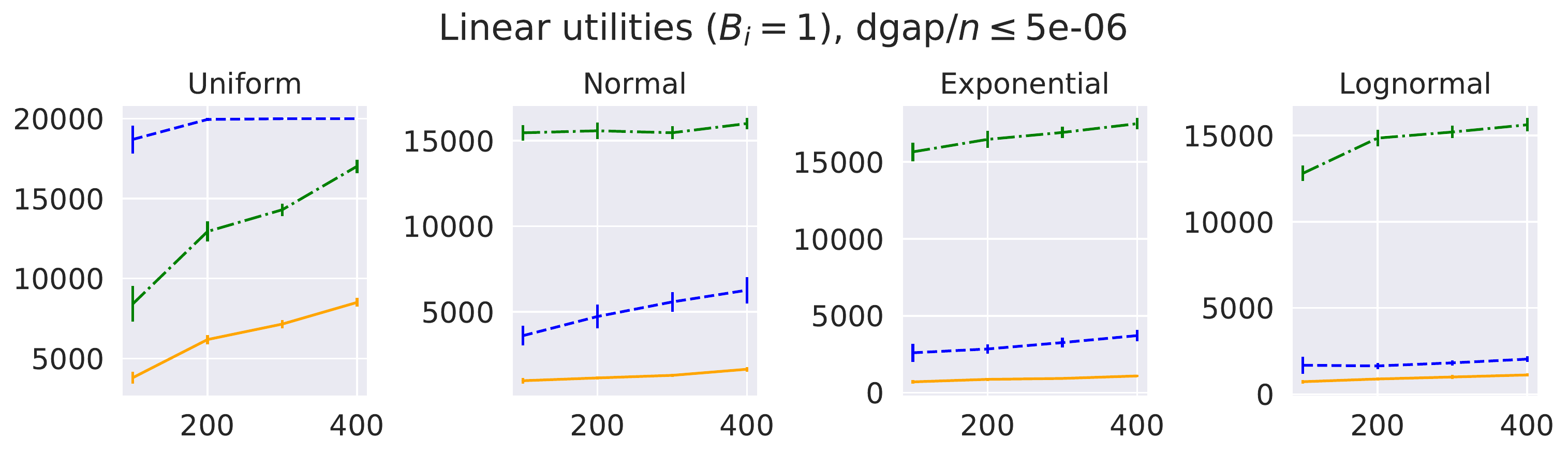}
	\includegraphics[width=0.8\linewidth]{plots/iters_to_ave_dgap/linear-varying-budgets/iters-to-normalized-dgap-1e-03.pdf}
	\includegraphics[width=0.8\linewidth]{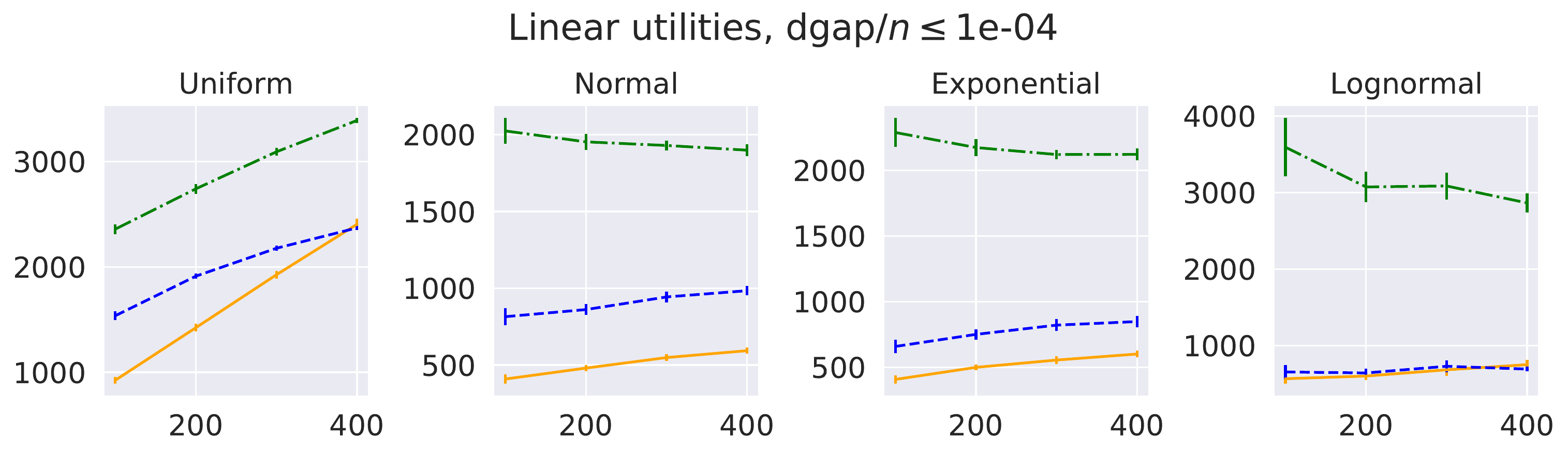} 
	\includegraphics[width=0.8\linewidth]{plots/iters_to_ave_dgap/linear-varying-budgets/iters-to-normalized-dgap-1e-05.pdf}
	\includegraphics[width=0.8\linewidth]{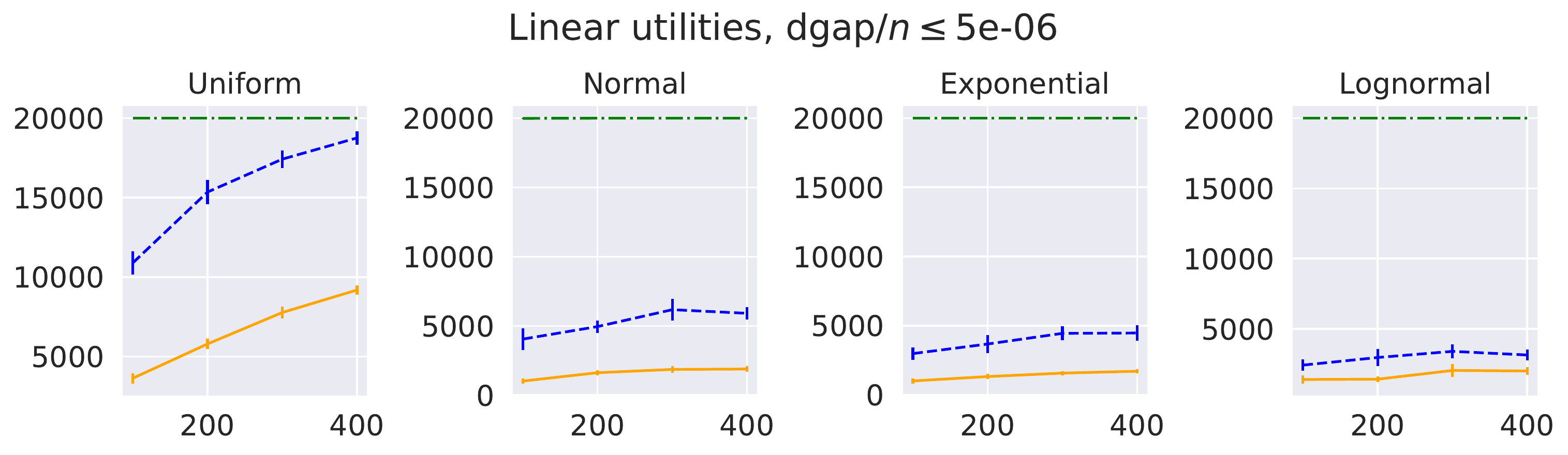}
	% ql
	\includegraphics[width=0.8\linewidth]{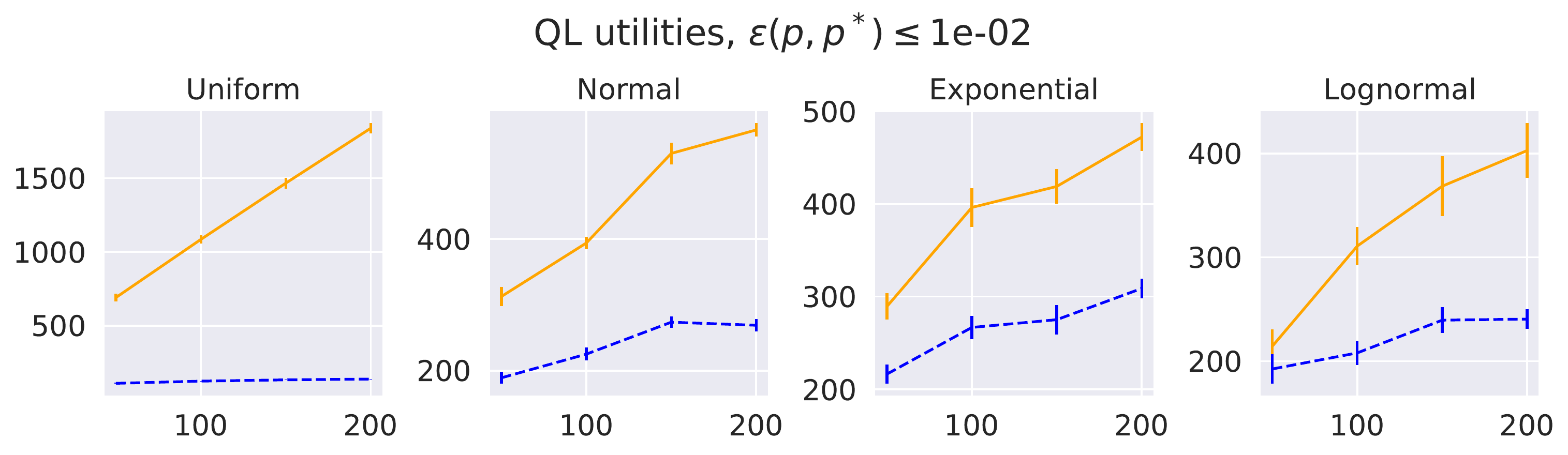}
	\includegraphics[width=0.8\linewidth]{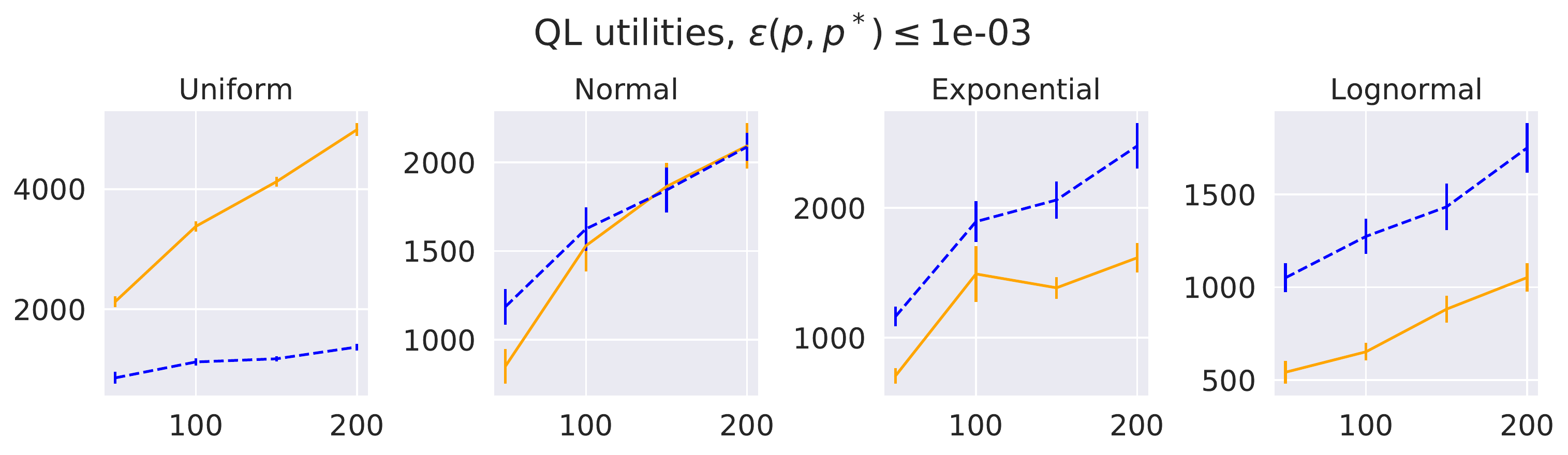}
	\includegraphics[width=0.8\linewidth]{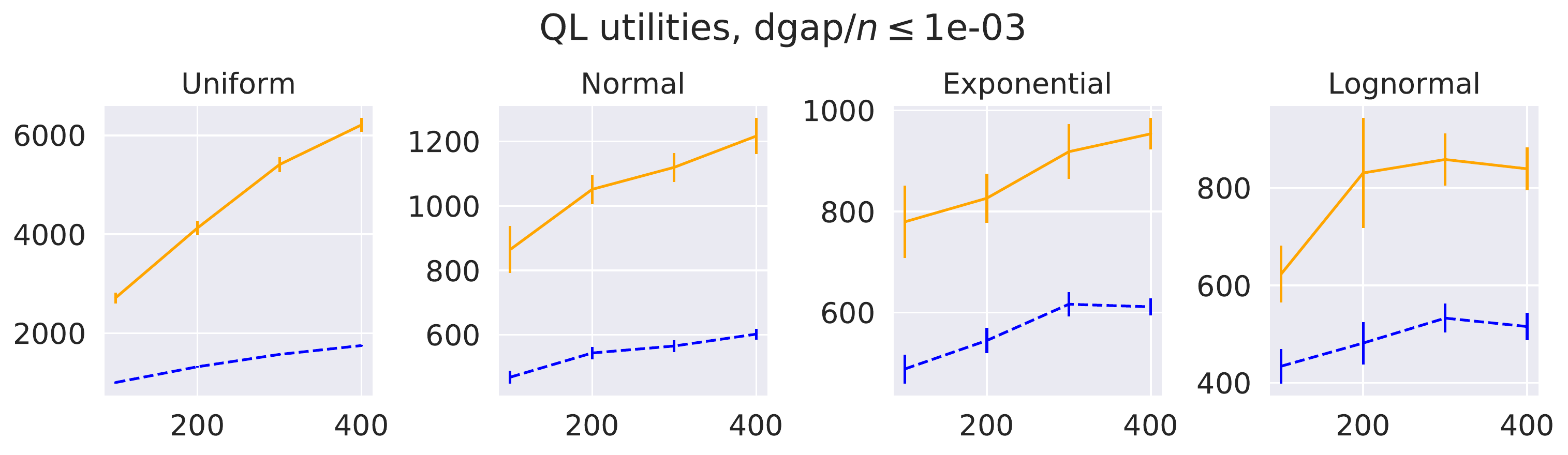}
	\includegraphics[width=0.8\linewidth]{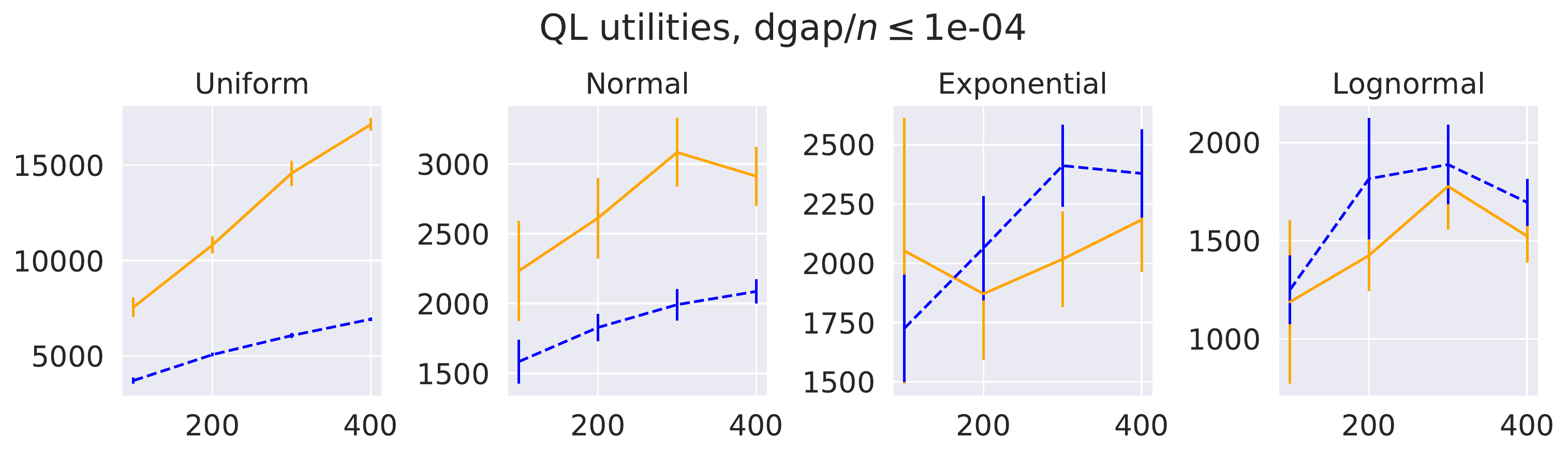} 
	\includegraphics[width=0.8\linewidth]{plots/iters_to_ave_dgap/ql-varying-budgets-pg-pr/iters-to-normalized-dgap-1e-05.pdf}
	\includegraphics[width=0.8\linewidth]{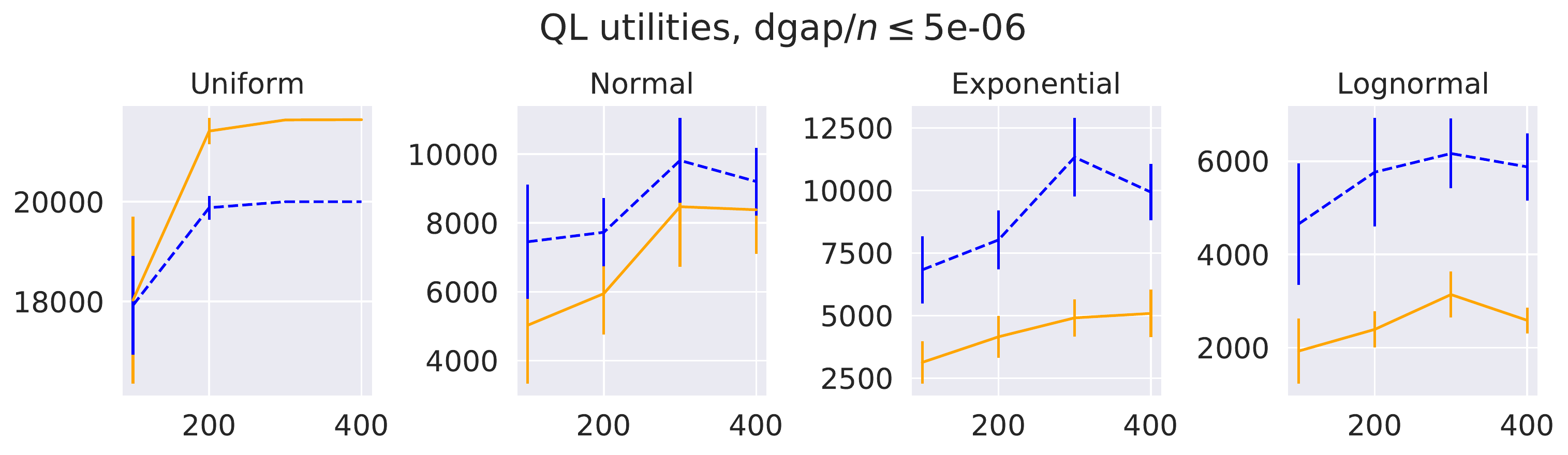}
	\includegraphics[width=0.8\linewidth]{plots/iters_to_ave_dgap/ql-varying-budgets-pg-pr/iters-to-normalized-dgap-1e-03.pdf}
	\includegraphics[width=0.8\linewidth]{plots/iters_to_ave_dgap/ql-varying-budgets-pg-pr/iters-to-normalized-dgap-1e-04.pdf} 
	\includegraphics[width=0.8\linewidth]{plots/iters_to_ave_dgap/ql-varying-budgets-pg-pr/iters-to-normalized-dgap-1e-05.pdf}
	\includegraphics[width=0.8\linewidth]{plots/iters_to_ave_dgap/ql-varying-budgets-pg-pr/iters-to-normalized-dgap-5e-06.pdf}
	% leontief
%	\includegraphics[width=0.8\linewidth]{plots/iters_to_p_diff_vs_mosek/leontief-budgets-all-1/iters-to-normalized-dgap-1e-02.pdf}
%	\includegraphics[width=0.8\linewidth]{plots/iters_to_p_diff_vs_mosek/leontief-budgets-all-1/iters-to-normalized-dgap-1e-03.pdf}
	\includegraphics[width=0.8\linewidth]{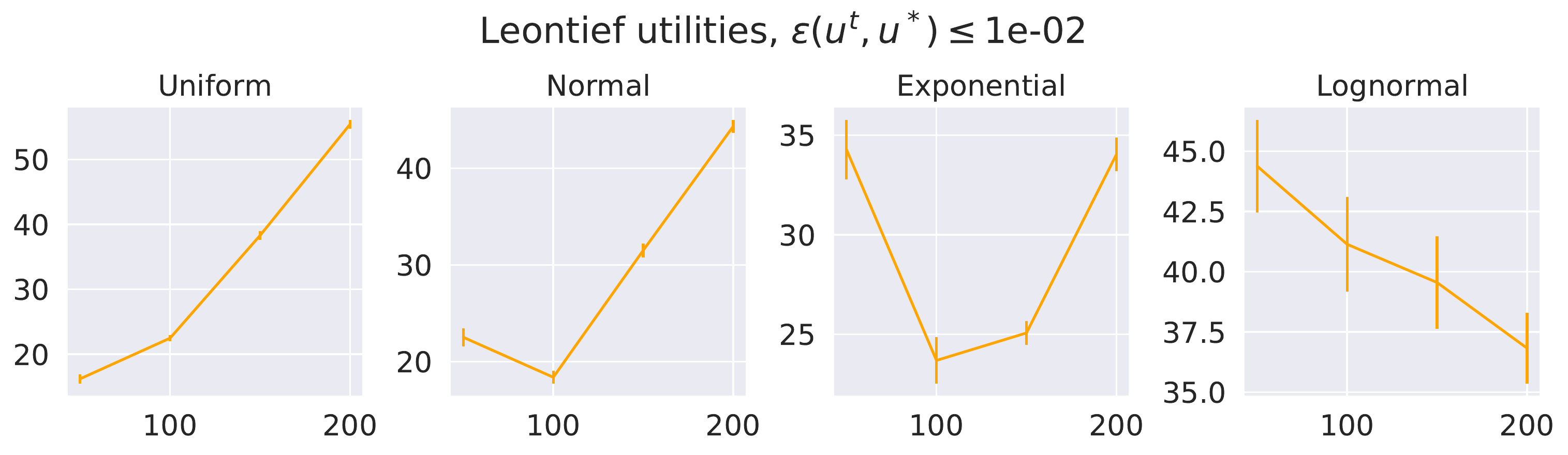}
	\includegraphics[width=0.8\linewidth]{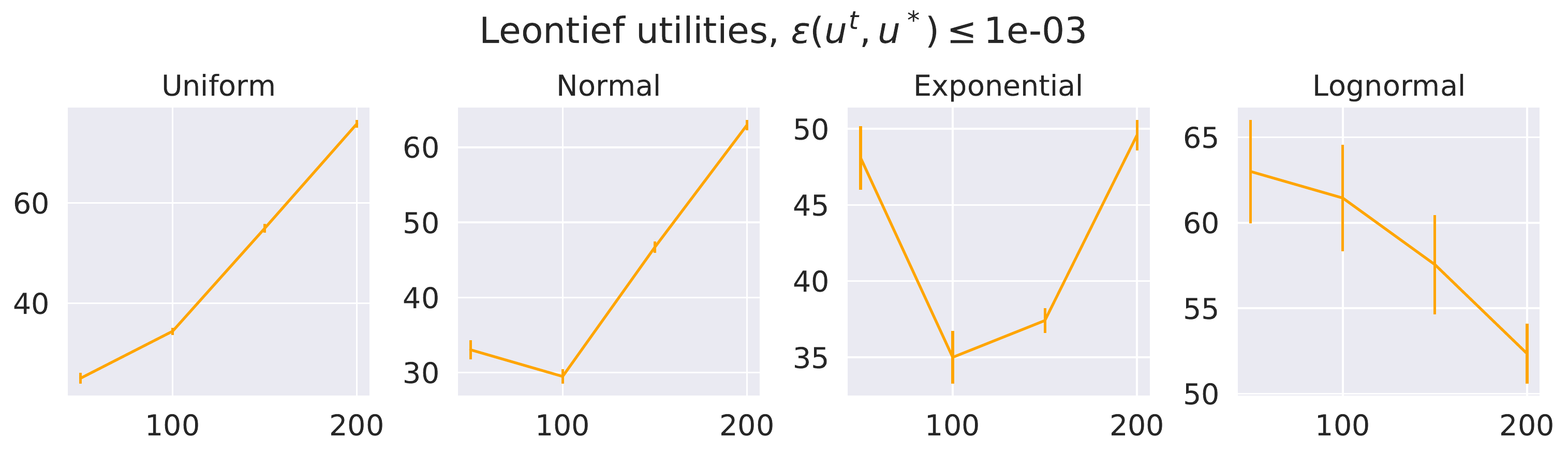}
	\includegraphics[width=0.8\linewidth]{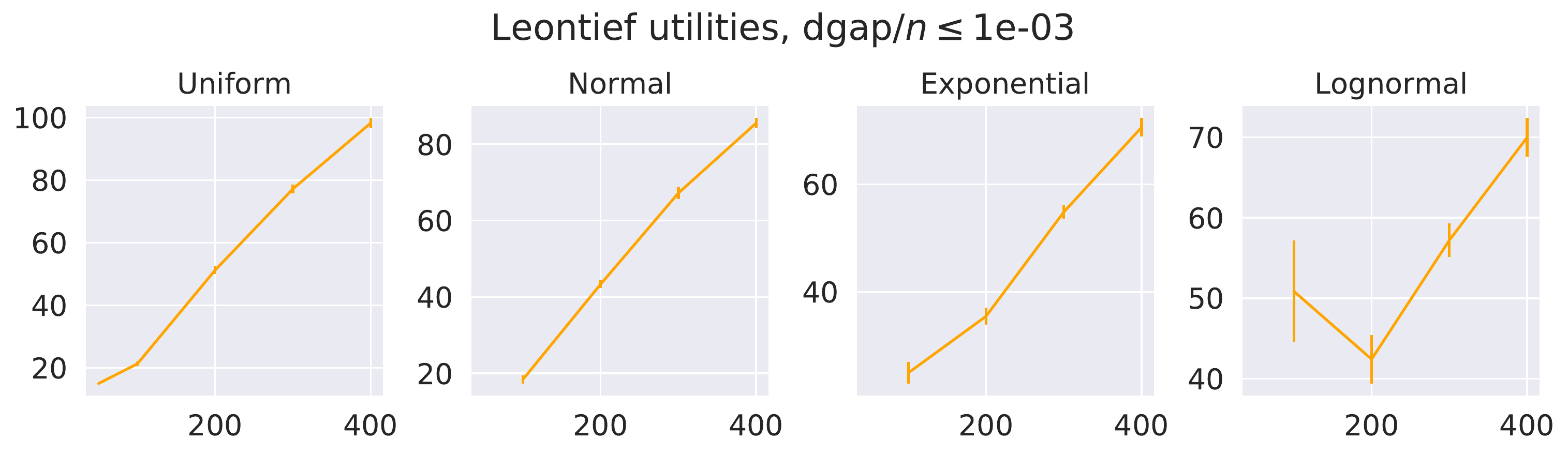}
	\includegraphics[width=0.8\linewidth]{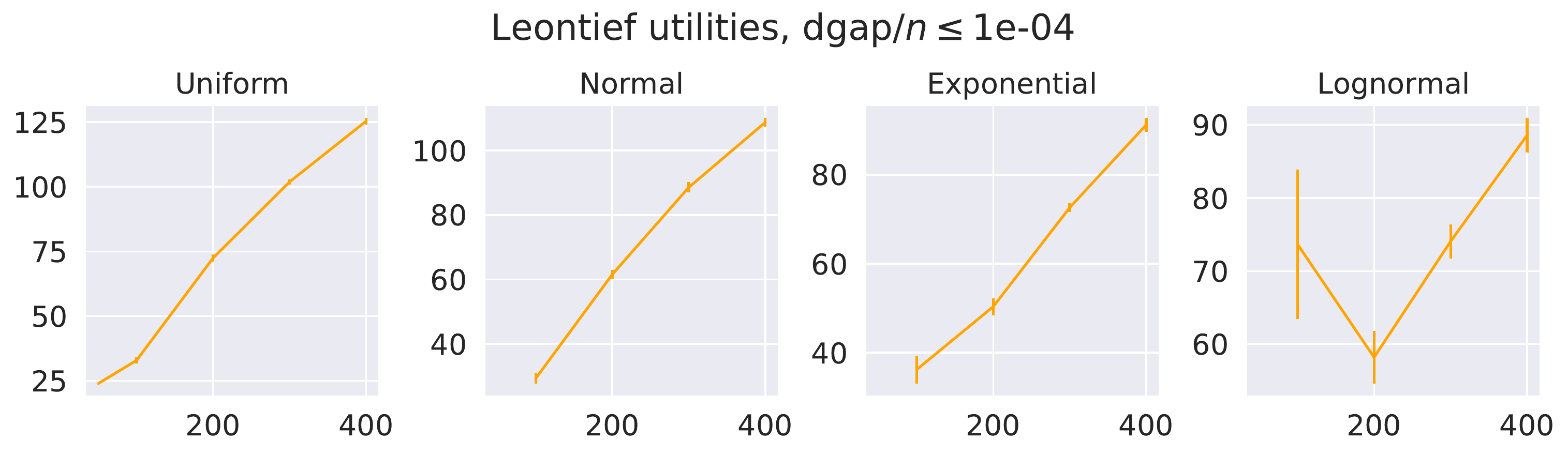}
	\includegraphics[width=0.8\linewidth]{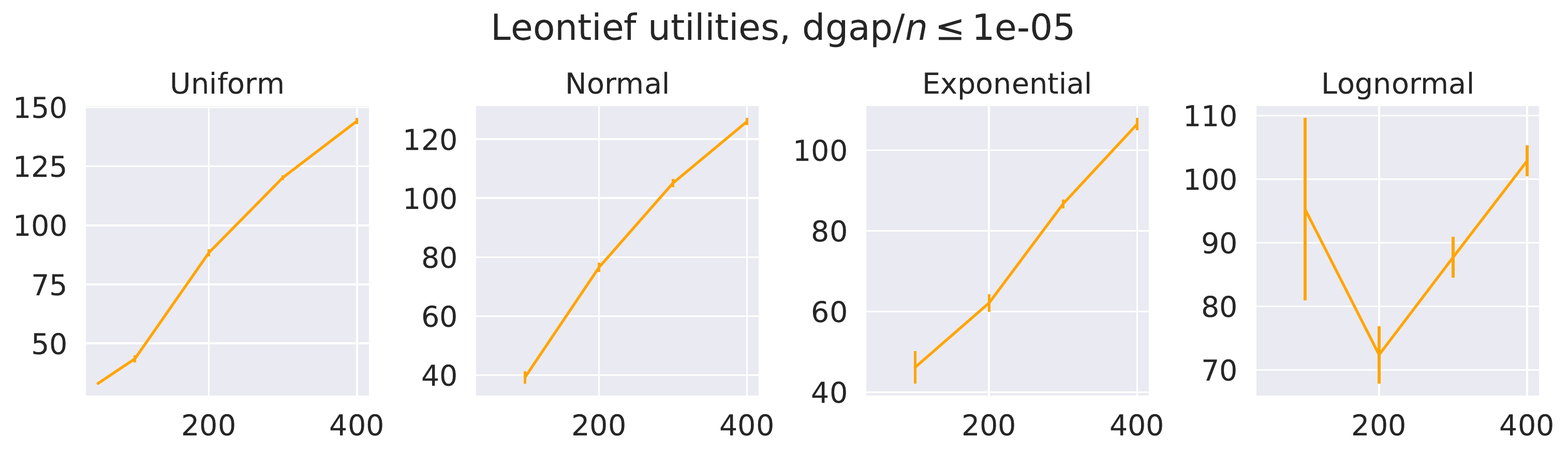}
	\includegraphics[width=0.8\linewidth]{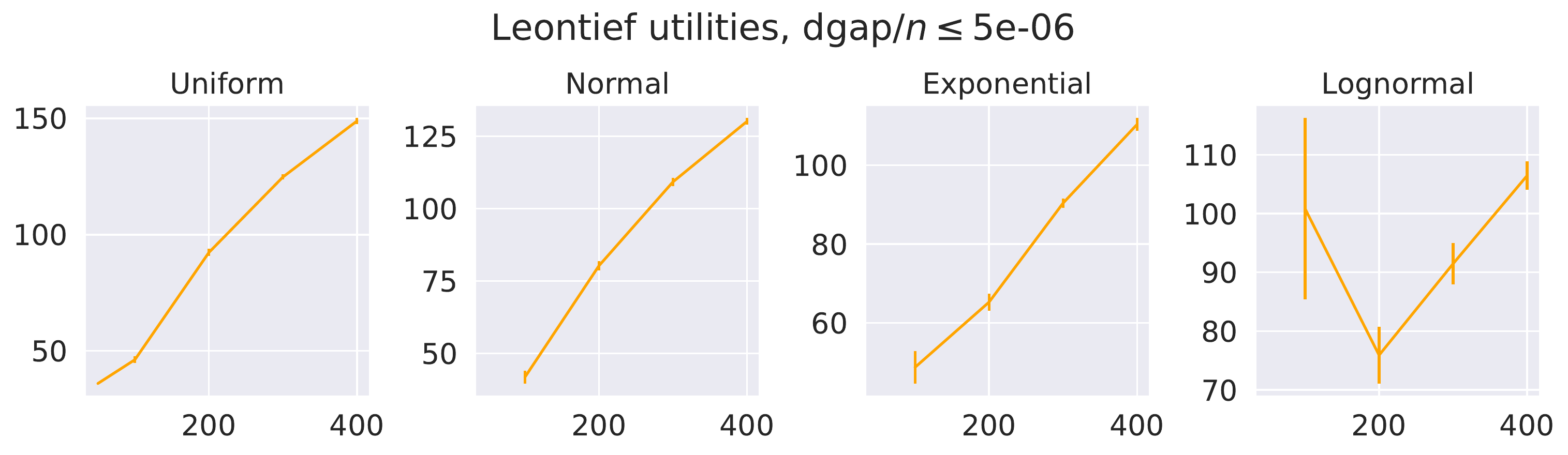}
\end{center}

%\textbf{Heavy-tail distributions.} For $v_{ij} \sim {\rm Cauchy}$

%
% Each instance is solved using PG with linesearch (see Appendix \ref{app:pg-ls-details} for more details), PR and FW with exact linesearch (implemented via one-dimensional bisection search).  Let $p^*$ denote the ``true'' optimal prices computed using CVXPY+Mosek \cite{diamond2016cvxpy, mosek2010mosek, orabona2019modern}. We record the number of iterations to reach $\epsilon(p^t, p^*) = \max_i \frac{|p^t_i - p^*_i|}{p^*_i} \leq \eta$ for $\eta = 10^{-2}, 10^{-3}$. Next, for larger sizes, we repeat the same using termination criteria $\frac{{\rm dgap}_t}{n}\leq \eta$ for $\eta = 10^{-3}, 10^{-4}, 10^{-5}, 5\times 10^{-6}$. For each setup, we report the average number of iterations before reaching the desired termination threshold and standard error $\frac{1.96 \hat{\sigma}}{\sqrt{k}}$, where $\hat{\sigma}$ is the sample standard deviation of the $30$ repeats. For PG, we report the total number of projections $\Pi_\cX$. Each experiment is terminated at $20000$ iterations regardless of whether termination threshold is reached.  As can be seen, PR is often more efficient, except very high accuracy (${\rm dgap}/n\leq 10^{-6}$) is required. For \textbf{Leontief} utilities, we repeat the similar with random $a \in \RR_+^{n\times m}$ having a random number of (approximately half) nonzeros.
%

%%% Local Variables:
%%% mode: latex
%%% TeX-master: "../main"
%%% End:

\end{document}